\documentclass[fleqn,10pt]{wlscirep}
\usepackage[utf8]{inputenc}
\usepackage[T1]{fontenc}
\usepackage{nameref}
\usepackage{hyperref}       
\usepackage{url}            
\usepackage{booktabs}       
\usepackage{multirow}       
\usepackage{multicol}       
\usepackage{subcaption}
\usepackage{amsmath}
\usepackage{amsfonts}       
\usepackage{nicefrac}       
\usepackage{microtype}      
\usepackage{graphicx}
\usepackage{xspace}
\usepackage{fontawesome}
\usepackage{xcolor,colortbl}

\definecolor{bg}{HTML}{f2f2f2}

\makeatletter
\DeclareRobustCommand\onedot{\futurelet\@let@token\@onedot}
\def\@onedot{\ifx\@let@token.\else.\null\fi\xspace}

\def\ie{\emph{i.e}\onedot} 
\makeatother

\title{Rethinking model prototyping through the MedMNIST+ dataset collection}

\author[1*]{Sebastian Doerrich}
\author[1]{Francesco Di Salvo}
\author[1,2]{Julius Brockmann}
\author[1]{Christian Ledig}
\affil[1]{University of Bamberg, xAILab Bamberg, Bamberg, 96047, Germany}
\affil[2]{Ludwig Maximilian University of Munich, Munich, 80539, Germany}

\affil[*]{sebastian.doerrich@uni-bamberg.de}


\begin{abstract}
The integration of deep learning based systems in clinical practice is often impeded by challenges rooted in limited and heterogeneous medical datasets. In addition, the field has increasingly prioritized marginal performance gains on a few, narrowly scoped benchmarks over clinical applicability, slowing down meaningful algorithmic progress. This trend often results in excessive fine-tuning of existing methods on selected datasets rather than fostering clinically relevant innovations. In response, this work introduces a comprehensive benchmark for the MedMNIST+ dataset collection, designed to diversify the evaluation landscape across several imaging modalities, anatomical regions, classification tasks and sample sizes. We systematically reassess commonly used Convolutional Neural Networks (CNNs) and Vision Transformer (ViT) architectures across distinct medical datasets, training methodologies, and input resolutions to validate and refine existing assumptions about model effectiveness and development. Our findings suggest that computationally efficient training schemes and modern foundation models offer viable alternatives to costly end-to-end training. Additionally, we observe that higher image resolutions do not consistently improve performance beyond a certain threshold. This highlights the potential benefits of using lower resolutions, particularly in prototyping stages, to reduce computational demands without sacrificing accuracy. Notably, our analysis reaffirms the competitiveness of CNNs compared to ViTs, emphasizing the importance of comprehending the intrinsic capabilities of different architectures. Finally, by establishing a standardized evaluation framework, we aim to enhance transparency, reproducibility, and comparability within the MedMNIST+ dataset collection. Code is available at \href{https://github.com/sdoerrich97/rethinking-model-prototyping-MedMNISTPlus}{github.com/sdoerrich97/rethinking-model-prototyping-MedMNISTPlus}.
\end{abstract}

\begin{document}

\flushbottom
\maketitle
%
%
\thispagestyle{empty}


\section*{Introduction}
\label{sec:introduction}
In recent years, significant strides in deep learning (DL) have reshaped various domains, from image classification to natural language processing \cite{Wang2023}. This progress was driven by the development of increasingly sophisticated models, exemplified by architectures like the Transformer \cite{Vaswani2017} for text or Vision Transformer (ViT) \cite{dosovitskiy2021} for images. Moreover, advanced training methodologies, including self-supervised contrastive methods such as CLIP \cite{Radford2021} for image and text pairs, and DINO \cite{Caron2021,oquab2024dinov2} for image pairs, have enabled the training of complex models without the need for exhaustive labeling efforts.
Simultaneously, there is an accumulating interest in integrating machine learning techniques into medical imaging, where DL models are approaching performance comparable to medical experts on specific tasks \cite{LIU2019e271} and software applications are beginning to receive clinical certifications \cite{Sendak2020APF}. Despite this progress and the exponential growth of DL-related publications across various medical fields in the past few years \cite{Kocak2023}, the adoption of DL algorithms in daily clinical practice has been comparatively slow \cite{Stacke2021}.

One major obstacle is the scarcity of appropriate datasets, often characterized by limited sample sizes and heterogeneous image acquisition conditions \cite{Lafarge2017, Oksuz2020, Khan2022}, thereby posing challenges to the generalizability of supervised DL algorithms. Ongoing progress in domain adaptation (DA) and domain generalization (DG) techniques aims to increase algorithmic robustness through aligning feature distributions \cite{Li2021} or acquiring domain-invariant features \cite{LiDa2018}. However, the generalizability of these methods across diverse domains remains a significant challenge, constraining their real-world applicability \cite{Eche2021}.

In addition, there is a concerning trend in DL research towards prioritizing the adaptation and scaling of existing methodologies in order to achieve incremental performance improvements on influential benchmarks \cite{raji2021ai} rather than addressing clinically relevant needs \cite{Varoquaux2022}. This trend is particularly pronounced in academic research, where the incentive structure often prioritizes quantity over relevance, leading to the incorporation of additional complexity into existing methodologies often at the expense of increased computational requirements \cite{Janiesch2021}. While benchmark datasets play a crucial role in coordinating machine learning research and facilitating standardized evaluations \cite{Koch2021ReducedRA}, overreliance on a handful of influential yet narrowly scoped benchmarks may stifle innovation and exacerbate inherent biases within these datasets such as the underrepresentation of certain demographic groups \cite{Crawford2021, Birhane2021}. The latter, in particular, limits the applicability of current DL techniques across diverse patient populations, thereby impeding their real-world deployment \cite{Norori2021}. Instead, research endeavors should focus more on proposing new benchmarks to diversify the landscape, mitigate bias-induced challenges, and cover a broader range of real-world tasks. Rather than solely determining a winner based on state-of-the-art performance, benchmarking should promote understanding to drive impactful algorithmic development and alternative evaluation methods \cite{raji2021ai}.

Furthermore, the limitations of scaling alone are becoming increasingly evident, as larger models start to falter in model trustworthiness \cite{Rae2021ScalingLM,thoppilan2022lamda} or performance on well-specified tasks \cite{mckenzie2022inverse}. Nonetheless, there is a paramount trend of increasing hardware and compute requirements estimated via the total number of FLOPs, and the number of trainable parameters in deep learning architectures \cite{Sevilla2022}. This further impedes the application of these approaches in the clinical environment. Therefore, it is imperative to explore qualitative enhancements alongside quantitative scaling in DL research as called for by A. Goyal and Y. Bengio \cite{Goyal2022}, particularly in the context of real-world medical applications.

Research endeavors should prioritize the creation of larger and more diverse datasets and benchmarks, with a focus on incorporating additional inductive biases and fostering the continuous development of more sophisticated approaches. The recent emergence of foundation models exemplifies this direction. These models, pre-trained on extensive datasets, offer the potential to enhance performance by capturing intricate patterns and serving as a foundational basis for further fine-tuning \cite{Bommasani2021}. Existing works, building on top of these models, can be readily evaluated across diverse benchmarks due to their high transferability to new datasets and tasks, as well as their remarkable zero- or few-shot performance \cite{Kirillov2023,girdhar2023imagebind}. This facilitates a more comprehensive assessment of these methods without necessitating extensive retraining. 

In this work, we aim to contribute to this effort by reassessing traditional DL models and training schemes, and presenting a new benchmark in the context of medical image classification. Our objective is to reevaluate commonly held assumptions regarding these methodologies, thereby enhancing and confirming comprehension of their inherent strengths and limitations as well as diversifying the benchmark landscape. Consequently, we will offer recommendations and insights for prototyping, model development, and deployment. To this end, we extend upon the existing MedMNIST v2 classification benchmark \cite{medmnistv2}, using the recently introduced MedMNIST+ database \cite{yang2024dataset}. MedMNIST v2 offers a collection of 12 distinct biomedical 2D datasets, ranging from Chest-X-ray to Dermatology, in a MNIST-like \cite{Deng2012} resolution of $28 \times 28$ pixels for medical image analysis. Its limitation to $28 \times 28$ images represented a critical constraint for comprehensive method evaluation. However, this has been addressed with the introduction of MedMNIST+, extending the previous dataset collection with resolutions: $64 \times 64$, $128 \times 128$, and $224 \times 224$ pixels.
By systematically benchmarking a diverse array of baseline models and training paradigms, including selective convolutional and Transformer-based models using both end-to-end training and linear probing on this distinct multi-dimensional database, our goal is to provide critical insights into the strengths and weaknesses of these techniques. Furthermore, we investigate the integration of the k-nearest neighbors ($k$-NN) classifier into the feature space of these models, aiming to enhance computational efficiency and interpretability. Given that most clinically validated and regulated systems currently rely on supervised deep learning, in particular CNN-based architectures \cite{Jones2020,FDAAIMLDevices}, we prioritize these widely adopted models in our analysis. Our primary aim is to re-investigate whether compute-intensive architectures are always necessary, whether higher resolution input consistently improves model performance, and whether end-to-end training is always optimal. Additionally, we want to foster greater transparency, reproducibility, and comparability in future research endeavors within the domain of medical image analysis. Key contributions of our work include:
\begin{itemize}
\item Systematic benchmarking of a wide range of commonly used models across diverse medical datasets, accounting for variations in resolutions, tasks, sample sizes, and class distributions.
\item Identification of systematic strengths and weaknesses inherent in traditional models within the context of medical image classification.
\item Reevaluation of prevalent assumptions with respect to model design, training schemes and input resolution requirements.
\item Presentation of a solid baseline performance for MedMNIST+ and a standardized evaluation framework for assessing future model performance in medical image classification.
\item Formulation of recommendations and take-aways for model development and deployment.
\end{itemize}
\begin{figure}
    \centering
    \includegraphics[width=0.81\linewidth]{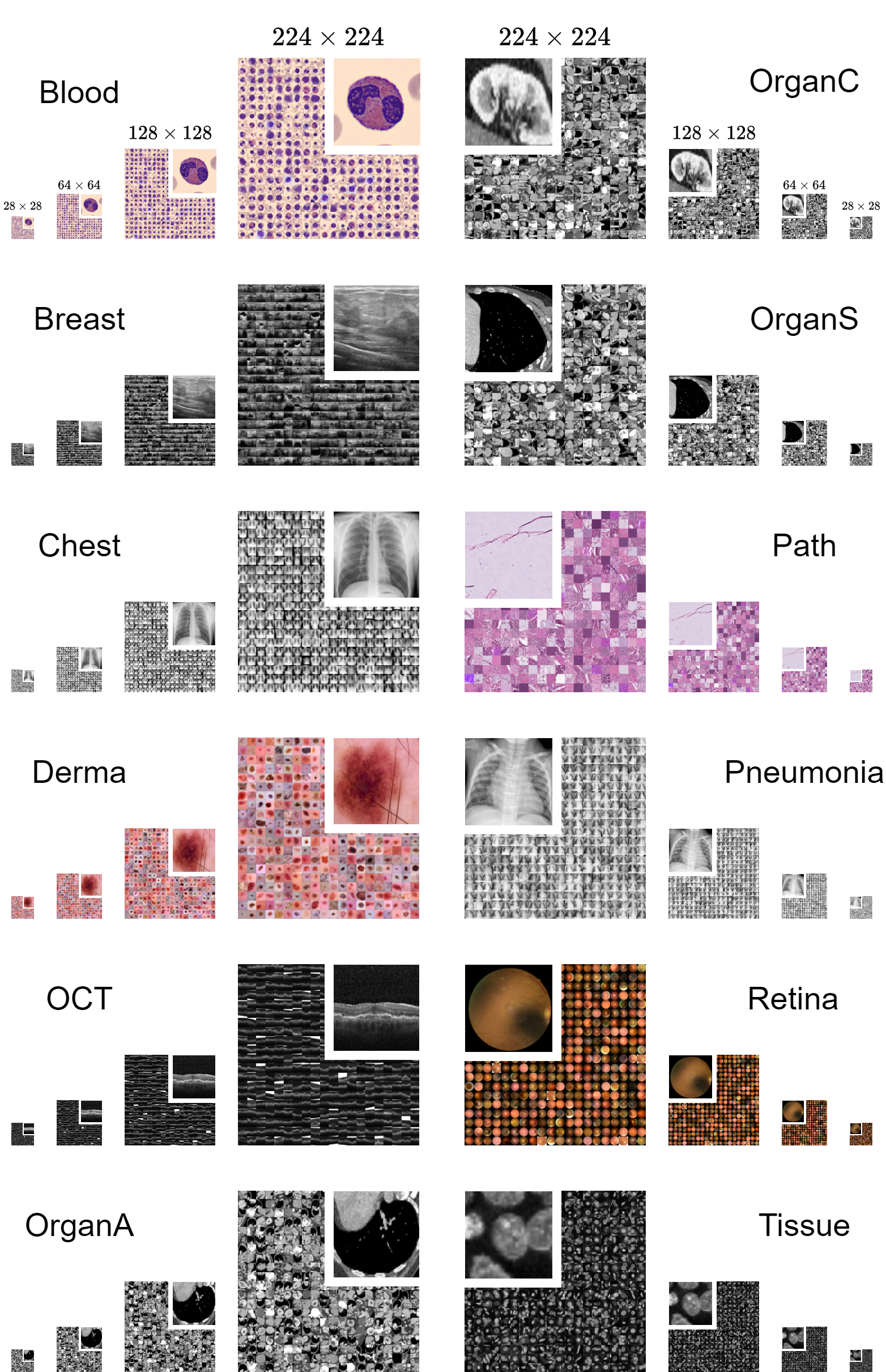}
    \caption{Side-by-side comparison of the 12 2D datasets included in MedMNIST+, showcasing diverse primary data modalities and classification tasks across four image resolutions.}
    \label{fig:dataset}
\end{figure}
\section*{Experiments and Results}
\subsection*{Datasets}
The dataset selection employed in this work originates from MedMNIST v2 \cite{medmnistv2}, initially introduced at a resolution of $28 \times 28$ pixels and recently expanded by MedMNIST+ \cite{yang2024dataset} to four distinct image resolutions, namely $28 \times 28$, $64 \times 64$, $128 \times 128$, and $224 \times 224$. The collection comprises twelve 2D datasets that are curated from carefully selected sources, encompassing primary data modalities such as X-ray, OCT, ultrasound, CT, and electron microscope. Furthermore, these datasets cater to diverse classification tasks, including binary/multi-class, ordinal regression, and multi-label classification, spanning a wide range of dataset scales, ranging from 780 samples for \textit{Breast} up to 236,386 for \textit{Tissue}. The details of each dataset including data source, imaging modality, type of classification task (along with the number of classes), and the publicly available data splits, provided by MedMNIST \cite{yang2024dataset} and forming a one-to-one correspondence with our benchmark, are described in \tablename~\ref{tab:dataset}. Additionally, we report Shannon's Equitability for each split to quantify class imbalance. A visual comparison of all datasets across the four evaluated image resolutions is presented in \figurename~\ref{fig:dataset}.

\begin{table}
\caption{Dataset details including data source, imaging modality, type of classification task (with number of classes), predefined data splits, and class imbalance measured using Shannon's Equitability (ranging from 0 for total imbalance to 1 for perfect balance) for each split. (ML: Multi-Label, MC: Multi-Class, BC: Binary-Class, OR: Ordinary Regression).}
\label{tab:dataset}
    \centering
    \setlength{\tabcolsep}{5.5pt}
    {\small
    \begin{tabular}{l l l r r r }
        \toprule
        \multirow{2}{*}{Dataset} & \multirow{2}{*}{Source} & \multirow{2}{*}{Imaging Modality} & Task & Number of Samples & Class Imbalance \\
         &  &  & (\# Classes) & Train / Val / Test & Train / Val / Test \\
        \midrule \\
        Blood & A. Acevedo et al. \cite{Acevedo2020} & Blood Cell Microscope & MC (8) & $11,959$ / $1,712$ / $3,421$ & 0.96 / 0.96 / 0.96 \\
        Breast & W. Al-Dhabyani et al. \cite{Al-Dhabyani2020} & Breast Ultrasound & BC (2) & $546$ / $78$ / $156$ & 0.84 / 0.84 / 0.84 \\
        Chest & X. Wang et al. \cite{Wang2017ChestXRay8HC} & Chest X-Ray & ML-BC (2) &  78,468 / 11,219 / 22,433 & 0.26 / 0.26 / 0.26 \\
        \multirow{2}{*}{Derma} & P. Tschandl et al. \cite{Tschandl2018} & \multirow{2}{*}{Dermatoscope} & \multirow{2}{*}{MC (7)} & \multirow{2}{*}{7,007 / 1,003 / 2,005} & \multirow{2}{*}{0.58 / 0.58 / 0.58} \\
         & N. Codella et al. \cite{Codella2019} & & & \\
        OCT & D. S. Kermany et al. \cite{Kermany2018} & Retinal OCT & MC (4) & 97,477 / 10,832 / 1,000 & 0.84 / 0.84 / 1.00 \\      
        \multirow{2}{*}{OrganA} & P. Bilic et al. \cite{BILIC2023102680}  & \multirow{2}{*}{Abdominal CT} & \multirow{2}{*}{MC (11)} & \multirow{2}{*}{34,561 / 6,491 / 17,778} & \multirow{2}{*}{0.96 / 0.95 / 0.96} \\
        & X. Xu et al. \cite{Xu2019} & & & \\
        \multirow{2}{*}{OrganC} & P. Bilic et al. \cite{BILIC2023102680}  & \multirow{2}{*}{Abdominal CT} & \multirow{2}{*}{MC (11)} & \multirow{2}{*}{12,975 / 2,392 / 8,216} & \multirow{2}{*}{0.95 / 0.95 / 0.95} \\
        & X. Xu et al. \cite{Xu2019} & & & \\
        \multirow{2}{*}{OrganS} & P. Bilic et al. \cite{BILIC2023102680}  & \multirow{2}{*}{Abdominal CT} & \multirow{2}{*}{MC (11)} & \multirow{2}{*}{13,932 / 2,452 / 8,827} & \multirow{2}{*}{0.93 / 0.96 / 0.94} \\
        & X. Xu et al. \cite{Xu2019} & & & \\
        Path & J. N. Kather et al. \cite{Kather2019} & Colon Pathology & MC (11) & 89,996 / 10,004 / 7,180 & 0.99 / 0.99 / 0.96 \\
        Pneumonia & D. S. Kermany et al. \cite{Kermany2018} & Chest X-Ray & BC (2) & 4,708 / 524 / 624 & 0.82 / 0.82 / 0.95 \\
        Retina & R. Liu et al. \cite{Liu2022} & Fundus Camera & OR (5) & 1,080 / 120 / 400 & 0.87 / 0.86 / 0.87 \\
        Tissue & V. Ljosa et al. \cite{Ljosa2012} & Kidney Cortex Microscope & MC (8) & 165,466 / 23,640 / 47,280 & 0.87 / 0.87 / 0.87 \\
        \bottomrule
    \end{tabular}
    }
\end{table}
\subsection*{Benchmark Evaluation}
\label{subsec:benchmarkeval}
We report the performance assessment of each model selected from our diverse pool, following the methodology outlined in Section \hyperref[subsec:training_pipeline]{Training Pipeline}. Our evaluation encompasses all datasets, image resolutions, and training schemes described earlier. The summary of average accuracy (ACC) and area under the receiver operating characteristic curve (AUC) across all datasets is provided in \tablename~\ref{tab:average_benchmark}. In addition, detailed performance evaluations for each dataset individually can be found in the appendix, spanning Supplementary Tables~\ref{tab:bloodmnist benchmark} to \ref{tab:tissuemnist benchmark}. To ensure the reliability of our assessments for all datasets, we report the mean and standard deviation of ACC and AUC for three random seeds.

It is important to note that the $k$-NN approach does not furnish  an AUC score. This peculiarity arises from its distinct classification methodology, which involves a majority vote on the labels associated with the $k$-closest training embeddings (lowest cosine distance). Although the possibility exists to convert this voting into a probability distribution by normalizing the vote through a division by $k$, such a metric would lack reliability and accuracy as it fails to provide a comprehensive probability distribution across all classes, but solely among neighboring ones. Consequently, the AUC score would be significantly influenced by factors such as the choice of $k$, local density, and data imbalance, prompting us to exclude this in our evaluation.

\begin{figure}
    \centering
    \includegraphics[width=\textwidth]{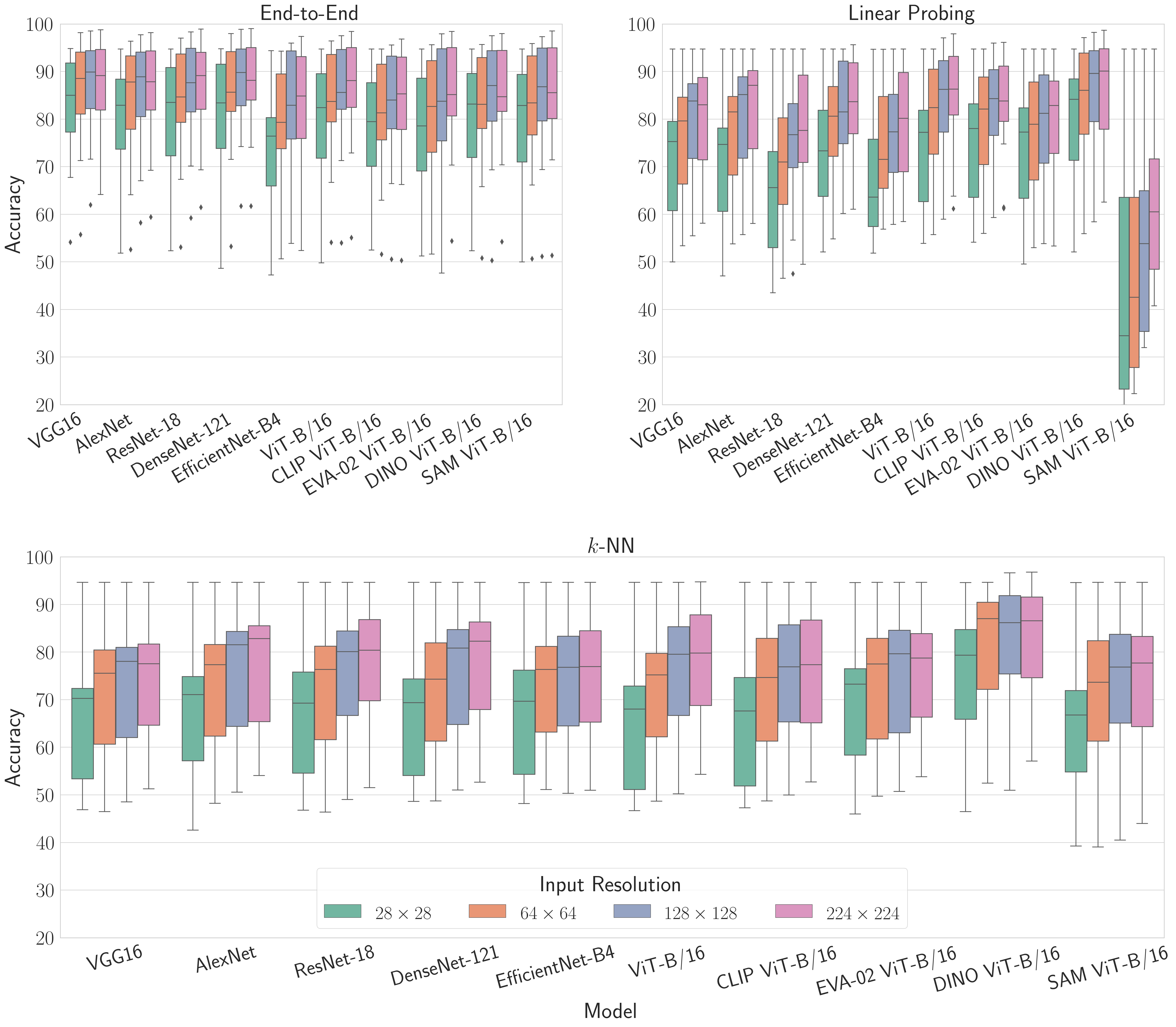}
    \caption{Illustrating the accuracy (ACC) distributions exhibited by each model averaged across all 12 datasets, delineated by training scheme and input resolution. Each subplot within the figure illustrates the performance distributions pertaining to distinct training schemes, with color coding employed to signify the associated input resolution.}
    \label{fig:performance_box}    
\end{figure}

As anticipated, end-to-end training yields the highest overall performance for all training schemes, and higher resolutions appear to enable all models across all training schemes to exhibit performance enhancements compared to lower resolutions. However, these enhancements begin to saturate when transitioning from inputs of $128 \times 128$ pixels to $224 \times 224$ pixels. Despite the increased data information, characterized by a fourfold increase in pixel count, all models and training schemes across all datasets show only marginal improvements or, in some cases, even worse overall performance. This correspondence is visually depicted in Figure \ref{fig:performance_box}, where the accuracy distributions of each model across all datasets are displayed for each training scheme and input resolution, respectively. Furthermore, the performance relationships of the models to each other for a specific training scheme and input resolution remain largely consistent. This observation challenges the common assumption that evaluations solely on higher image resolutions (\textit{e.g.} above $200 \times 200$ pixels) are deemed valid, while evaluations on lower input resolutions are generally considered less meaningful, since the performance trends appear to be resolution independent. This in turn supports the utilization of lower resolution inputs, particularly during the prototyping phase of model development, as they generally allow for faster processing speeds while demanding fewer computational resources.

Moreover, our analysis reveals that more extensive self-supervised pretraining strategies such as CLIP and DINO, when compared to ImageNet pretraining, do not necessarily lead to improved performance for end-to-end trained models. However, they do demonstrate enhanced performance for linear probing and the integration of $k$-NN. Particularly notable is the performance of DINO, which achieves results close to the end-to-end trained baseline while requiring minimal training (linear probing) or no training at all ($k$-NN). This suggests that the latter two training schemes benefit from extensive pretraining even if conducted on unrelated images. This raises questions about whether more sophisticated foundation models can further narrow the performance gap between expensive end-to-end training and more computationally efficient schemes such as $k$-NN or even close it entirely. On the contrary, the results for SAM illustrate that a model pretrained for a distant task (\ie segmentation) may not readily adapt to a new task (\ie classification) and may require complete retraining to perform effectively.
%
\begin{table}[!htpb]
\setlength{\tabcolsep}{4.9pt}
\renewcommand{\arraystretch}{1.1}
\caption{Benchmark outcomes summarizing the average mean and standard deviation of accuracy (ACC), for a fixed operating point of $0.5$, and area under the receiver operating characteristic curve (AUC) across all datasets for all training scheme-model-image resolution combinations, derived from three independent random seeds. Notably, the \mbox{$k$-NN} algorithm, devoid of a training phase, remains unaffected by the stochasticity inherent in model training, thus reporting only the total ACC value without standard deviation. Moreover, owing to its direct utilization of embeddings and labels for classification, \mbox{$k$-NN} does not furnish a reliable AUC score. The overall best result across all training schemes, models, and resolutions is highlighted with a \colorbox{bg}{background color}; the best result per resolution across all training schemes and models is highlighted with \underline{underline}; and the best result per training scheme and resolution is highlighted in \textbf{bold}.}
\label{tab:average_benchmark}
\centering
\begin{tabular}{lcccccccccc}
& \\
\toprule
\multirow{2.5}{*}{Methods} & \multicolumn{4}{c}{Accuracy (ACC)} &  & \multicolumn{4}{c}{Area Under the ROC Curve (AUC)} \\
\cmidrule(r){2-5}\cmidrule(l){7-10}
& $28 \times 28$ & $64 \times 64$ & $128 \times 128$ & $224 \times 224$ &  & $28 \times 28$ & $64 \times 64$ & $128 \times 128$ & $224 \times 224$\\ 
\midrule
{\textsc{End-to-End}} & & & & & \\        
\; VGG16           & \underline{\textbf{82.34{\scriptsize$\pm$0.88}}} & \underline{\textbf{85.33{\scriptsize$\pm$1.16}}} & 86.64{\scriptsize$\pm$0.82} & 86.70{\scriptsize$\pm$0.93} &  & \underline{\textbf{92.66{\scriptsize$\pm$0.41}}} & \underline{\textbf{94.24{\scriptsize$\pm$0.28}}} & \underline{\textbf{95.16{\scriptsize$\pm$0.27}}} & \cellcolor{bg}\underline{\textbf{95.30{\scriptsize$\pm$0.22}}} \\ 
\; AlexNet         & 78.92{\scriptsize$\pm$0.81} & 82.94{\scriptsize$\pm$0.78} & 85.04{\scriptsize$\pm$0.74} & 85.74{\scriptsize$\pm$0.64} &  & 91.14{\scriptsize$\pm$0.43} & 92.72{\scriptsize$\pm$0.33} & 94.29{\scriptsize$\pm$0.30} & 94.90{\scriptsize$\pm$0.23} \\ 
\; ResNet-18       & 79.66{\scriptsize$\pm$0.74} & 83.42{\scriptsize$\pm$0.65} & 85.73{\scriptsize$\pm$0.66} & 86.22{\scriptsize$\pm$0.58} &  & 90.92{\scriptsize$\pm$0.28} & 92.49{\scriptsize$\pm$0.50} & 93.91{\scriptsize$\pm$0.27} & 94.51{\scriptsize$\pm$0.24} \\ 
\; DenseNet-121    & 80.32{\scriptsize$\pm$0.93} & 84.62{\scriptsize$\pm$0.80} & \cellcolor{bg}\underline{\textbf{87.13{\scriptsize$\pm$0.56}}} & \underline{\textbf{87.11{\scriptsize$\pm$0.64}}} &  & 91.75{\scriptsize$\pm$0.55} & 93.59{\scriptsize$\pm$0.23} & 94.57{\scriptsize$\pm$0.21} & 95.03{\scriptsize$\pm$0.23} \\ 
\; EfficientNet-B4 & 73.18{\scriptsize$\pm$1.61} & 79.37{\scriptsize$\pm$1.10} & 82.52{\scriptsize$\pm$0.79} & 82.44{\scriptsize$\pm$1.11} &  & 87.04{\scriptsize$\pm$0.82} & 90.07{\scriptsize$\pm$0.65} & 91.89{\scriptsize$\pm$0.39} & 91.64{\scriptsize$\pm$0.73} \\ 
\; ViT-B/16        & 78.23{\scriptsize$\pm$0.88} & 83.17{\scriptsize$\pm$0.92} & 84.94{\scriptsize$\pm$0.93} & 86.06{\scriptsize$\pm$0.92} &  & 90.54{\scriptsize$\pm$0.47} & 92.53{\scriptsize$\pm$0.69} & 93.25{\scriptsize$\pm$0.35} & 94.08{\scriptsize$\pm$0.38} \\ 
\; CLIP ViT-B/16   & 76.73{\scriptsize$\pm$0.80} & 80.39{\scriptsize$\pm$0.99} & 82.33{\scriptsize$\pm$1.02} & 82.75{\scriptsize$\pm$1.01} &  & 89.22{\scriptsize$\pm$1.11} & 90.91{\scriptsize$\pm$0.51} & 91.51{\scriptsize$\pm$0.31} & 91.83{\scriptsize$\pm$0.58} \\ 
\; EVA-02 ViT-B/16 & 76.69{\scriptsize$\pm$1.44} & 80.77{\scriptsize$\pm$0.97} & 82.76{\scriptsize$\pm$0.97} & 84.72{\scriptsize$\pm$1.09} &  & 88.91{\scriptsize$\pm$1.01} & 90.53{\scriptsize$\pm$0.54} & 91.59{\scriptsize$\pm$0.75} & 92.60{\scriptsize$\pm$0.94} \\ 
\; DINO ViT-B/16   & 78.51{\scriptsize$\pm$1.09} & 82.13{\scriptsize$\pm$1.02} & 84.31{\scriptsize$\pm$1.05} & 84.84{\scriptsize$\pm$1.08} &  & 91.02{\scriptsize$\pm$0.48} & 91.94{\scriptsize$\pm$0.32} & 92.91{\scriptsize$\pm$0.36} & 93.90{\scriptsize$\pm$0.73} \\ 
\; SAM ViT-B/16    & 78.26{\scriptsize$\pm$0.84} & 82.13{\scriptsize$\pm$1.12} & 84.19{\scriptsize$\pm$1.06} & 84.30{\scriptsize$\pm$0.82} &  & 89.36{\scriptsize$\pm$0.79} & 90.79{\scriptsize$\pm$1.01} & 91.94{\scriptsize$\pm$1.08} & 91.91{\scriptsize$\pm$0.55} \\ 
\midrule
{\textsc{Linear Probing}} & & & & & \\         
\; VGG16           & 71.18{\scriptsize$\pm$0.13} & 75.14{\scriptsize$\pm$0.26} & 78.58{\scriptsize$\pm$0.15} & 79.62{\scriptsize$\pm$0.18} &  & 87.70{\scriptsize$\pm$0.06} & 89.55{\scriptsize$\pm$0.06} & 91.94{\scriptsize$\pm$0.04} & 92.47{\scriptsize$\pm$0.05} \\ 
\; AlexNet         & 69.63{\scriptsize$\pm$0.35} & 76.11{\scriptsize$\pm$0.25} & 79.08{\scriptsize$\pm$0.17} & 81.02{\scriptsize$\pm$0.18} &  & 85.91{\scriptsize$\pm$0.13} & 89.06{\scriptsize$\pm$0.11} & 91.78{\scriptsize$\pm$0.07} & 93.18{\scriptsize$\pm$0.05} \\ 
\; ResNet-18       & 64.41{\scriptsize$\pm$0.05} & 70.49{\scriptsize$\pm$0.08} & 74.94{\scriptsize$\pm$0.07} & 76.89{\scriptsize$\pm$0.08} &  & 82.95{\scriptsize$\pm$0.18} & 85.24{\scriptsize$\pm$0.22} & 87.97{\scriptsize$\pm$0.76} & 90.37{\scriptsize$\pm$0.17} \\ 
\; DenseNet-121    & 72.10{\scriptsize$\pm$0.28} & 78.01{\scriptsize$\pm$0.19} & 80.77{\scriptsize$\pm$0.19} & 82.22{\scriptsize$\pm$0.23} &  & 86.76{\scriptsize$\pm$0.94} & 90.11{\scriptsize$\pm$0.29} & 92.00{\scriptsize$\pm$0.19} & 93.02{\scriptsize$\pm$0.11} \\ 
\; EfficientNet-B4 & 67.29{\scriptsize$\pm$0.54} & 73.95{\scriptsize$\pm$0.51} & 76.39{\scriptsize$\pm$0.25} & 77.91{\scriptsize$\pm$0.62} &  & 83.67{\scriptsize$\pm$0.64} & 86.48{\scriptsize$\pm$0.45} & 88.49{\scriptsize$\pm$0.55} & 89.57{\scriptsize$\pm$0.22} \\ 
\; ViT-B/16        & 73.21{\scriptsize$\pm$0.24} & 79.62{\scriptsize$\pm$0.33} & 83.08{\scriptsize$\pm$0.70} & 84.01{\scriptsize$\pm$0.27} &  & 88.25{\scriptsize$\pm$0.14} & 91.33{\scriptsize$\pm$0.22} & 93.57{\scriptsize$\pm$0.09} & 94.31{\scriptsize$\pm$0.13} \\ 
\; CLIP ViT-B/16   & 74.17{\scriptsize$\pm$0.24} & 78.67{\scriptsize$\pm$0.32} & 81.54{\scriptsize$\pm$0.24} & 82.24{\scriptsize$\pm$0.18} &  & 88.66{\scriptsize$\pm$0.08} & 91.48{\scriptsize$\pm$0.20} & 93.28{\scriptsize$\pm$0.10} & 93.66{\scriptsize$\pm$0.10} \\ 
\; EVA-02 ViT-B/16 & 73.04{\scriptsize$\pm$0.19} & 76.50{\scriptsize$\pm$0.16} & 78.48{\scriptsize$\pm$0.11} & 79.30{\scriptsize$\pm$0.10} &  & 88.41{\scriptsize$\pm$0.04} & 90.66{\scriptsize$\pm$0.03} & 92.04{\scriptsize$\pm$0.04} & 92.41{\scriptsize$\pm$0.04} \\ 
\; DINO ViT-B/16   & \textbf{78.23{\scriptsize$\pm$0.22}} & \textbf{82.74{\scriptsize$\pm$0.36}} & \textbf{84.46{\scriptsize$\pm$0.24}} & \textbf{85.11{\scriptsize$\pm$0.55}} &  & \textbf{90.94{\scriptsize$\pm$0.14}} & \textbf{93.29{\scriptsize$\pm$0.12}} & \textbf{94.50{\scriptsize$\pm$0.11}} & \textbf{94.99{\scriptsize$\pm$0.14}} \\ 
\; SAM ViT-B/16    & 43.69{\scriptsize$\pm$0.01} & 48.10{\scriptsize$\pm$0.03} & 54.14{\scriptsize$\pm$0.04} & 61.20{\scriptsize$\pm$0.03} &  & 66.51{\scriptsize$\pm$1.38} & 74.68{\scriptsize$\pm$0.80} & 80.46{\scriptsize$\pm$0.13} & 81.75{\scriptsize$\pm$0.11} \\ 
\midrule
{\textsc{$k$-NN ($k = 11$)}} & & & & & \\       
\; VGG16           & 66.07 & 70.65 & 72.26 & 73.78 & & - & - & - & - \\ 
\; AlexNet         & 67.47 & 72.14 & 74.56 & 76.22 & & - & - & - & - \\ 
\; ResNet-18       & 66.98 & 71.42 & 74.20 & 76.60 & & - & - & - & - \\ 
\; DenseNet-121    & 67.12 & 71.29 & 74.97 & 76.97 & & - & - & - & - \\ 
\; EfficientNet-B4 & 68.15 & 72.45 & 73.91 & 74.17 & & - & - & - & - \\ 
\; ViT-B/16        & 65.92 & 70.90 & 75.45 & 77.51 & & - & - & - & - \\ 
\; CLIP ViT-B/16   & 66.40 & 71.61 & 74.83 & 75.52 & & - & - & - & - \\ 
\; EVA-02 ViT-B/16 & 69.63 & 72.64 & 74.64 & 75.52 & & - & - & - & - \\ 
\; DINO ViT-B/16   & \textbf{73.61} & \textbf{79.41} & \textbf{81.17} & \textbf{81.90} & & - & - & - & - \\ 
\; SAM ViT-B/16    & 65.05 & 70.40 & 71.58 & 71.95 & & - & - & - & - \\ 
\bottomrule
\end{tabular}
\end{table}
\subsection*{Input Resolution Impact}
\label{subsec:inputres}
\begin{figure}
    \centering
    \includegraphics[width=\textwidth]{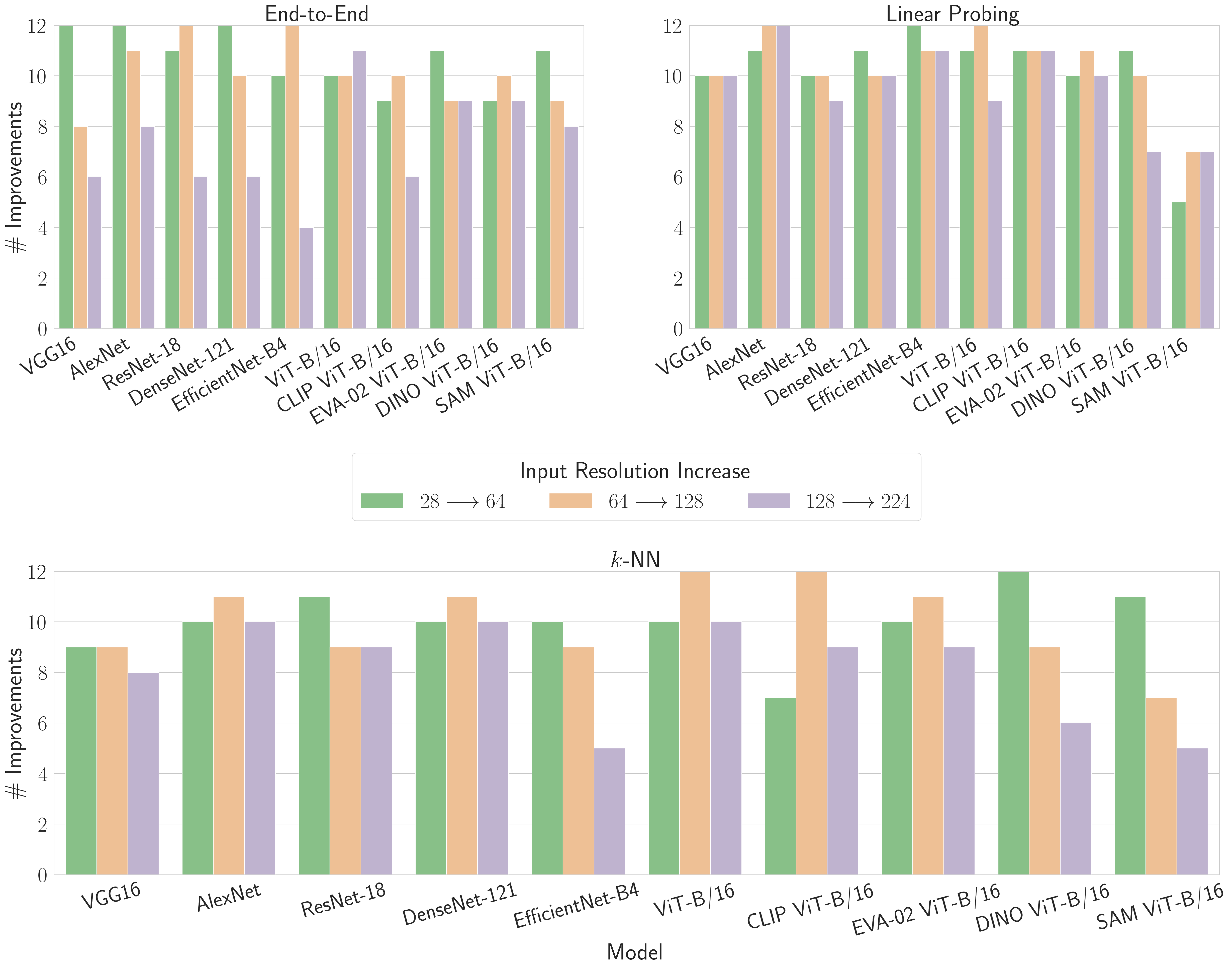}
    \caption{Analysis of model performance (ACC) improvement with increasing input resolution across all $12$ datasets. The figure illustrates the frequency of performance enhancements as input resolutions progress from $28 \times 28$ to $64 \times 64$, $64 \times 64$ to $128 \times 128$, and $128 \times 128$ to $224 \times 224$, encompassing all models and training schemes. Each bar signifies for how many datasets the model performance, in terms of the mean accuracy across the three random seeds, is superior at the next higher resolution compared to the preceding lower one, with a maximum of 12 improvements per transition.}
    \label{fig:performance_improvement}    
\end{figure}

We further investigate the effect of input resolution on model performance. Our objective is to determine how often model performance enhances with incremental increases in input resolution. Intuitively, higher input resolutions are expected to enable models to capture more intricate features, potentially leading to improved overall performance. However, as demonstrated in Section \hyperref[subsec:benchmarkeval]{Benchmark Evaluation}, this trend reaches a saturation point around an input resolution of $128 \times 128$ pixels for our underlying setting. To validate this observation, we analyze the instances where a model's performance surpasses that of the previous, lower resolution for each training scheme individually. Specifically, we compare the mean accuracy values across three different random seeds for the same model and training scheme between two resolutions. Figure \ref{fig:performance_improvement} depicts this analysis for transitions from $28 \times 28$ to $64 \times 64$, $64 \times 64$ to $128 \times 128$, and $128 \times 128$ to $224 \times 224$ resolutions. We calculate the frequency of performance improvements per increase in input resolution across all 12 datasets, resulting in a maximum improvement count of 12 for each transition.

Our findings substantiate the observations from Section \hyperref[subsec:benchmarkeval]{Benchmark Evaluation} to some extent. Overall, higher input resolutions lead to performance improvements across all models and training schemes. However, this improvement diminishes notably when transitioning from a $128 \times 128$ to $224 \times 224$ resolution, with superior performance observed only for a limited number of dataset instances. This trend is particularly pronounced for end-to-end trained convolutional models, whereas ViT-based models demonstrate less sensitivity to input resolution variations. This discrepancy could be attributed to the specific design of the ViT architecture, which is tailored for $224 \times 224$ pixel images, unlike convolutional models. Additionally, we observe that linear probing benefits the most from higher resolution images, with slight differences noted for the $k$-NN approach, potentially due to the pretraining with images of the same size. These results underscore that while input resolution impacts model performance, the effect is less significant than initially anticipated, with slight variations depending on the architecture used. This supports the utility of lower input resolutions at least during the prototyping phase.
\subsection*{Model Ranking}
\label{subsec:ranking}
\begin{figure}
    \centering
    \begin{subfigure}[b]{0.85\textwidth}
        \includegraphics[width=\textwidth]{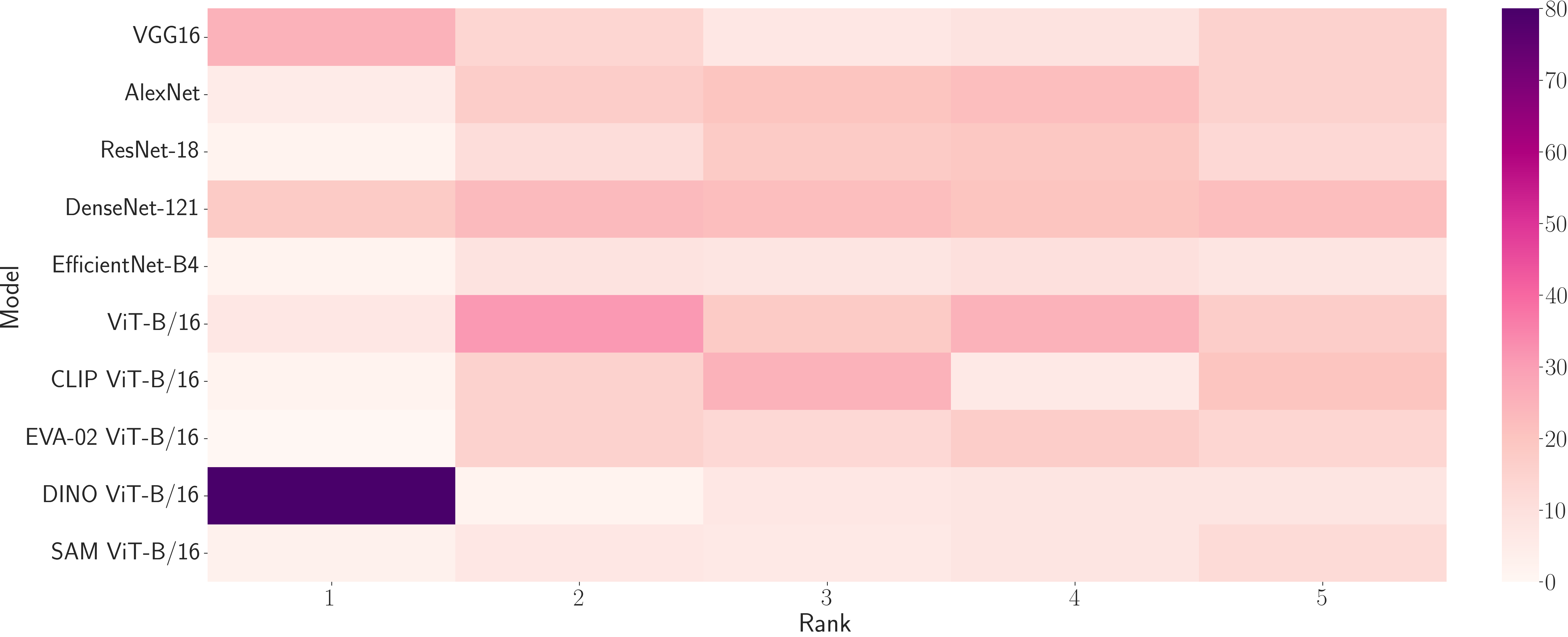}
        \caption{Frequency of top-5 performance placement per model counted for each dataset, training scheme and resolution.}
        \label{fig:ranking_total}
    \end{subfigure}

    \vspace{0.4cm}
    
    \centering
    \begin{subfigure}[b]{0.85\textwidth}
        \includegraphics[width=\textwidth]{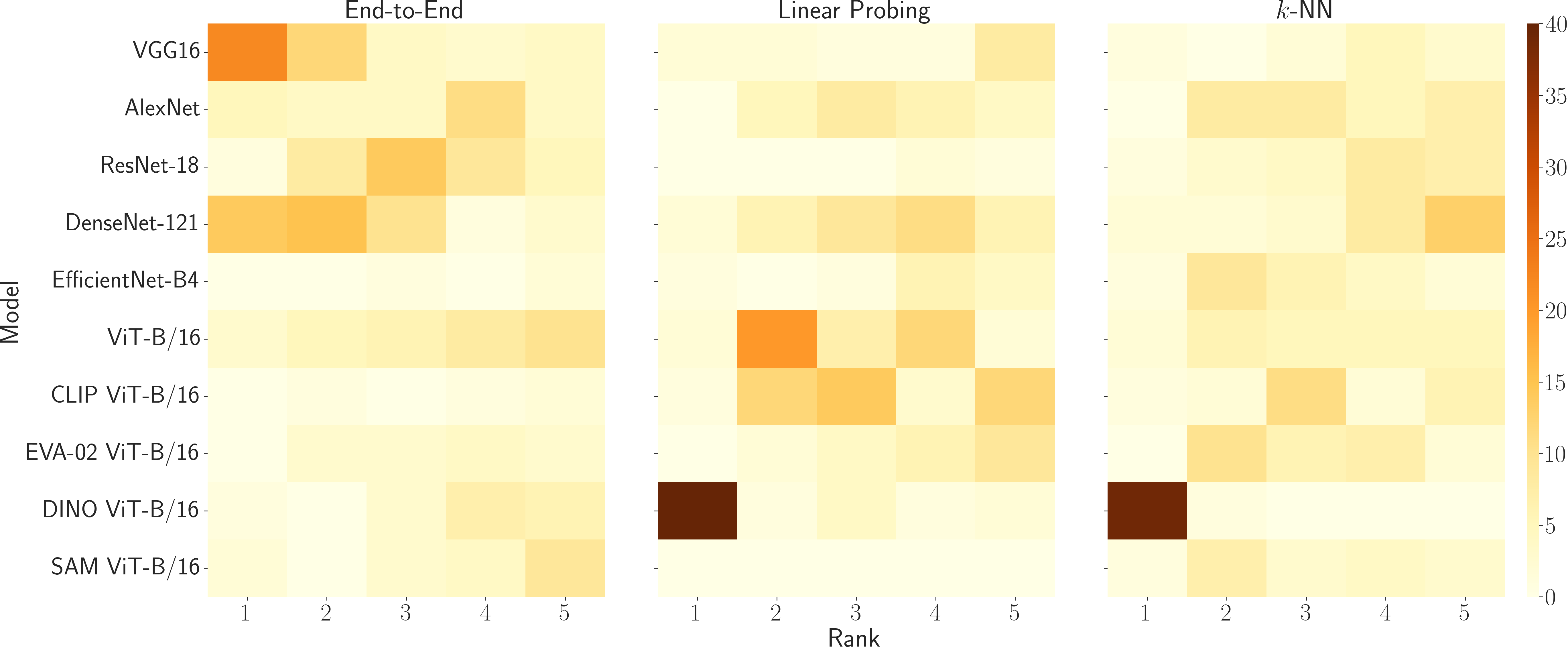}
        \caption{Frequency of top-5 performance placement per model and training scheme counted for each dataset and resolution.}
        \label{fig:ranking_trainigscheme}
    \end{subfigure}

    \vspace{0.4cm}
    
    \centering
    \begin{subfigure}[b]{0.85\textwidth}
        \includegraphics[width=\textwidth]{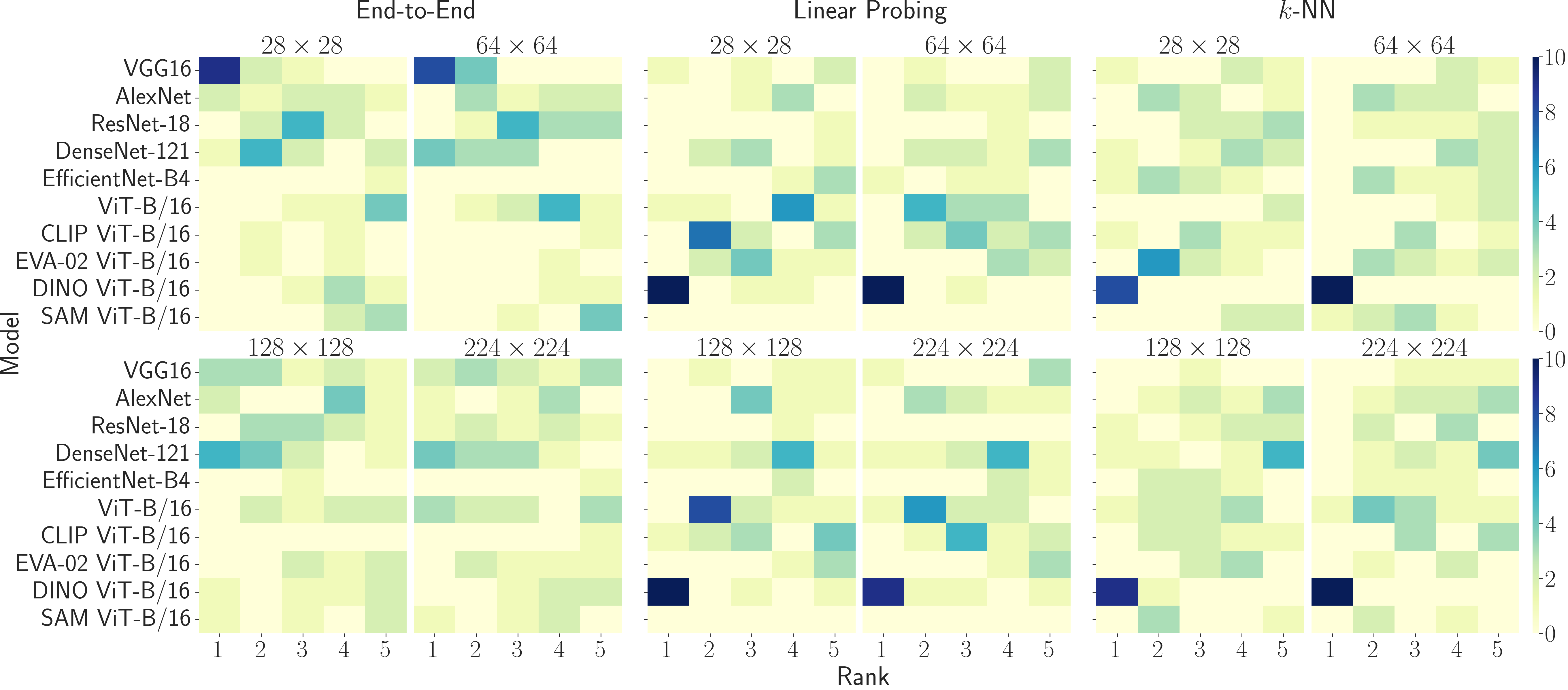}
        \caption{Frequency of top-5 performance placement per model, training scheme and resolution counted for each dataset.}
        \label{fig:ranking_trainigscheme_resolution}    
    \end{subfigure}
    
    \vspace{0.1cm}
    
    \caption{Ranking analysis showcasing the frequency of model placements among the top-5 performers in terms of accuracy (ACC) across all training schemes and resolutions (a), for each training scheme separately (b), and for both training schemes and resolutions, respectively (c) across all datasets.}
    \label{fig:ranking}
\end{figure}

We further assess how frequently a model's performance ranks among the top-5 performers concerning accuracy (ACC). Figure \ref{fig:ranking} visually portrays this as heatmaps, illustrating the total count of top-5 rank appearances for each model across all datasets, training schemes and image resolutions. Sub-figure (a) consolidates the overall ranking across all training schemes and resolutions, while sub-figure (b) presents the ranking for each training scheme separately. At last, sub-figure (c) provides the ranking broken down on both training schemes and resolutions collectively.

Our observations unveil that convolutional models consistently outperform ViT-based models concerning ACC for end-to-end training, regardless of their pretraining strategy. Notably, VGG16 and DenseNet-121 emerge as the top performers in this aspect. The performance of the DenseNet-121 backbone is particularly intriguing, given its relatively low number of parameters and activations compared to almost all other models. This finding challenges the prevailing assumption that a more complex architecture invariably outperforms a simpler, smaller one given sufficient training samples.

In contrast, ViT-based architectures, especially those pretrained with DINO, exhibit superior performance for linear probing and the $k$-NN approach compared to Convolutional Neural Networks. This is likely due to their enhanced representational capacity within the feature space. Moreover, sub-figure (c) illustrates the consistency of these observations across different input resolutions, with minimal variations observed. This underscores the significance of exhaustive pretraining for linear probing and the $k$-NN approach, highlighting ViTs' suitability as foundation models compared to their convolutional counterparts. Additionally, it emphasizes that the complexity of a model architecture does not necessarily align directly with its suitability for this purpose.
\subsection*{Quantitative Evaluation}
In order to substantiate the qualitative findings presented earlier, we conduct a quantitative evaluation using non-parametric statistical tests. The aim is to discern the potential impact of input resolution (specifically, $28 \times 28$, $64 \times 64$, $128 \times 128$, and $224 \times 224$) and training scheme (comprising end-to-end training, linear probing, and $k$-NN integration) on the model performance, measured in terms of accuracy. Additionally, we aim to assess whether there are notable variations in performance across datasets depending on the model architecture. To this end, Friedman tests, with a significance level set at $p = 0.05$, are initially employed to ascertain if statistically significant differences exist among the experimental conditions. Post-hoc two-tailed Wilcoxon signed-rank tests are then conducted to identify specific group differences, with a Bonferroni adjustment applied to mitigate the risk of type I errors resulting from multiple comparisons. Despite the limited number of observations ($n = 10$ for model-related evaluations and $n = 12$ for dataset-related analyses), we decided to assume an approximately normally distributed sampling distribution of the sample mean, following the Central Limit Theorem, in order to allow the computation of $Z$-values for the Wilcoxon tests. Finally, we asses the effect size of each test based on Cohen’s interpretation guidelines \cite{Cohen1992} for Kendall's Coefficient of Concordance $W$, and Pearson's correlation coefficient $r$ for the Friedman and Wilcoxon tests, respectively.

First, we want to investigate whether the input resolution significantly influences model performance. For this, we average the accuracy results per model and resolution across all datasets and training schemes. The Friedman test reveals a statistically significant difference in accuracy between the different image resolutions $\left( \chi^2(dof=3) = 30.0,\;p = 0.000001\right)$ with a perfect agreement among the rankings of model performance across the range of input resolutions ($W = 1.0$) for 3 degrees of freedom ($dof = \text{number of input resolutions} - 1 = 3$). Median interquartile range (IQR) perceived accuracy for resolution $28 \times 28$, $64 \times 64$, $128 \times 128$, and $224 \times 224$ are $72.44$ ($70.76$ to $73.17$), $76.97$ ($75.60$ to $77.69$), $79.36$ ($78.38$ to $80.61$), and $80.10$ ($79.86$ to $81.82$), respectively. Post-hoc analysis with two-tailed Wilcoxon signed-rank tests and a Bonferroni correction, adjusted significance level of $p < 0.008$, shows significant differences between all resolution pairs $\left(Z = -2.803,\;p = 0.002,\;r = 0.886\right)$. Notably, the results affirmed the trend observed in previous sections, indicating that higher resolutions generally lead to improved accuracy, albeit with diminishing returns at higher resolution levels (\ie the transition from $128 \times 128$ to $224 \times 224$).

Next, we explore potential disparities in model performance based on the employed training schemes (end-to-end, linear probing, and $k$-NN). For this, we average the accuracy results per model and training scheme across all datasets and input resolutions. The Friedman test (with $dof = \text{number of training schemes} - 1 = 2$) reveals overall significant differences suggesting a strong effect of the used training scheme on the model performance, $\left(\chi^2(\text{dof}=2) = 14.6,\;p = 0.0007,\;W = 0.73\right)$. Furthermore, post-hoc Wilcoxon signed-rank tests with Bonferroni correction, adjusted significance level of $p < 0.0167$, confirm these findings between end-to-end training and linear probing $\left(Z = -2.701,\;p = 0.0039,\;r = 0.854\right)$ as well as end-to-end training and $k$-NN $\left(Z = -2.803,\;p = 0.002,\;r = 0.886\right)$. These results, in conjunction with median IQR perceived accuracy values for end-to-end training of $82.77$ ($81.48$ to $83.61$), linear probing of $76.65$ ($74.45$ to $78.94$) and $k$-NN of $72.11$ ($72.37$ to $72.60$), respectively, affirm the superiority of end-to-end training in achieving the highest overall performance from Section \hyperref[subsec:benchmarkeval]{Benchmark Evaluation}. Interestingly, no significant differences are observed between linear probing and $k$-NN integration $\left(Z = -1.682,\;p = 0.1055\right)$, despite a large effect size $\left(r = 0.532\right)$. This suggests comparable efficacy of these training schemes despite variations in accuracy, thereby highlighting the potential of training-free strategies such as $k$-NN.

Finally, we assess performance variations across datasets based on the model architecture. Through averaging accuracy results per dataset and model across all training schemes and input resolutions, we discern distinct performance trends. The IQR perceived accuracy for all models is visualized in \tablename~\ref{tab:percentiles_models} and the existence of statistical differences in dataset performance depending on the model architecture is illustrated in \tablename~\ref{tab:model_differences}. Notably, DenseNet-121 and DINO ViT-B/16 emerged as the top-performing models, aligning with our previous findings of Section \hyperref[subsec:ranking]{Model Ranking}. Except of these, ViT-based models generally exhibit superior performance compared to convolutional models, which may seem contradictory to earlier findings. However, considering that we average the performance results across all training schemes, ViT models benefit from their higher performance for linear probing and $k$-NN integration compared to convolutional models, which only demonstrate higher performance in end-to-end training. Thus, these findings are indeed in line with our earlier observations. Furthermore, the results for SAM confirm its inferior performance compared to all other models as depicted in Section \hyperref[subsec:benchmarkeval]{Benchmark Evaluation}, thus contributing to a comprehensive understanding of model capabilities across various training contexts and tasks. Despite its similar performance for the end-to-end training, its task-foreign pretraining for segmentation rather than classification seems to limit its capabilities for training-less (linear probing) or training-free ($k$-NN) approaches.

Further details regarding dataset-specific performance and model differences are elaborated in the appendix throughout Supplementary Tables~\ref{tab:bloodmnist benchmark} to \ref{tab:tissuemnist benchmark}.

\begin{table}
    \caption{Percentile statistics for each model performance in terms of averaged accuracy (ACC) across all training schemes and input resolutions across all 12 datasets. The highest overall value per percentile is highlighted in \underline{underline} and the highest value per architecture type (convolution vs ViT, separated by a line) is highlighted in \textbf{bold}.}
    \label{tab:percentiles_models}
    \centering
    \begin{tabular}{lccc}
        \toprule
        \multirow{2.5}{*}{Model} & \multicolumn{3}{c}{Percentile} \\
        \cmidrule(r){2-4}
        & 25th & 50th (Median) & 75th \\ 
        \midrule
        VGG16 & 69.78 & 82.54 & 85.45 \\
        AlexNet & 69.87 & \textbf{83.88} & 86.00 \\
        ResNet-18 & 68.82 & 80.53 & 84.24 \\
        DenseNet-121 & \textbf{72.04} & 82.63 & \textbf{87.73} \\
        EfficientNet-B4 & 67.96 & 76.33 & 84.27 \\
        \midrule
        ViT-B/16 & 71.87 & 82.53 & 87.57 \\
        CLIP ViT-B/16 & 69.77 & 80.34 & 86.18 \\
        EVA-02 ViT-B/16 & 69.46 & 80.89 & 86.81 \\
        DINO ViT-B/16 & \underline{\textbf{75.99}} & \underline{\textbf{86.08}} & \underline{\textbf{91.17}} \\
        SAM ViT-B/16 & 57.83 & 69.64 & 75.01 \\
        \bottomrule
    \end{tabular}
\end{table}

\begin{table}[htbp]
    \renewcommand{\arraystretch}{2}
    \caption{Illustration of pair-wise significant differences between model performance in terms of averaged accuracy across all training schemes, input resolutions, and all 12 datasets using the results of the pair-wise Wilcoxon signed-rank tests with a Bonferroni correction (adjusted significance level of $p < 0.0011$). (\colorbox{green!50}{\faDiamond}: significant difference favoring the model in the \textbf{row}, \colorbox{green!50}{\faTrophy}: significant difference favoring the model in the \textbf{column}, \colorbox{red!50}{\faTimes}: no significant difference).}
    \label{tab:model_differences}
    \centering
    \begin{tabular}{l|cccccccccc}
        \toprule
        Model & VGG16 & AlexNet & ResNet & DenseNet & EfficientNet & ViT-B/16 & CLIP & EVA-02 & DINO & SAM \\
        \midrule
        VGG16 
        	& \cellcolor{bg} 
        	& \cellcolor{red!50} \faTimes 
        	& \cellcolor{green!50} \faDiamond
        	& \cellcolor{red!50} \faTimes 
        	& \cellcolor{red!50} \faTimes 
        	& \cellcolor{red!50} \faTimes 
        	& \cellcolor{red!50} \faTimes 
        	& \cellcolor{red!50} \faTimes 
        	& \cellcolor{red!50} \faTimes 
        	& \cellcolor{green!50} \faDiamond \\
        AlexNet 
        	& \cellcolor{red!50} \faTimes 
        	& \cellcolor{bg} 
        	& \cellcolor{red!50} \faTimes 
        	& \cellcolor{red!50} \faTimes 
        	& \cellcolor{red!50} \faTimes 
        	& \cellcolor{red!50} \faTimes 
        	& \cellcolor{red!50} \faTimes 
        	& \cellcolor{red!50} \faTimes 
        	& \cellcolor{green!50} \faTrophy 
        	& \cellcolor{green!50} \faDiamond \\
        ResNet 
        	& \cellcolor{green!50} \faTrophy 
        	& \cellcolor{red!50} \faTimes 
        	& \cellcolor{bg} 
        	& \cellcolor{green!50} \faTrophy 
        	& \cellcolor{red!50} \faTimes 
        	& \cellcolor{green!50} \faTrophy 
        	& \cellcolor{red!50} \faTimes 
        	& \cellcolor{red!50} \faTimes 
        	& \cellcolor{green!50} \faTrophy 
        	& \cellcolor{green!50} \faDiamond \\
        DenseNet 
        	& \cellcolor{red!50} \faTimes 
        	& \cellcolor{red!50} \faTimes 
        	& \cellcolor{green!50} \faDiamond 
        	& \cellcolor{bg} 
        	& \cellcolor{green!50} \faDiamond 
        	& \cellcolor{red!50} \faTimes 
        	& \cellcolor{red!50} \faTimes 
        	& \cellcolor{red!50} \faTimes 
        	& \cellcolor{red!50} \faTimes 
        	& \cellcolor{green!50} \faDiamond \\
        EfficientNet 
        	& \cellcolor{red!50} \faTimes 
        	& \cellcolor{red!50} \faTimes 
        	& \cellcolor{red!50} \faTimes 
        	& \cellcolor{green!50} \faTrophy 
        	& \cellcolor{bg} 
        	& \cellcolor{green!50} \faTrophy 
        	& \cellcolor{red!50} \faTimes 
        	& \cellcolor{red!50} \faTimes 
        	& \cellcolor{green!50} \faTrophy 
        	& \cellcolor{red!50} \faTimes \\
        ViT-B/16 
        	& \cellcolor{red!50} \faTimes 
        	& \cellcolor{red!50} \faTimes 
        	& \cellcolor{green!50} \faDiamond 
        	& \cellcolor{red!50} \faTimes 
        	& \cellcolor{green!50} \faDiamond 
        	& \cellcolor{bg} 
        	& \cellcolor{red!50} \faTimes 
        	& \cellcolor{green!50} \faDiamond 
        	& \cellcolor{green!50} \faTrophy 
        	& \cellcolor{green!50} \faDiamond \\
        CLIP 
        	& \cellcolor{red!50} \faTimes 
        	& \cellcolor{red!50} \faTimes 
        	& \cellcolor{red!50} \faTimes 
        	& \cellcolor{red!50} \faTimes 
        	& \cellcolor{red!50} \faTimes 
        	& \cellcolor{red!50} \faTimes 
        	& \cellcolor{bg} 
        	& \cellcolor{red!50} \faTimes 
        	& \cellcolor{green!50} \faTrophy 
        	& \cellcolor{green!50} \faDiamond \\
        EVA-02 
        	& \cellcolor{red!50} \faTimes 
        	& \cellcolor{red!50} \faTimes 
        	& \cellcolor{red!50} \faTimes 
        	& \cellcolor{red!50} \faTimes 
        	& \cellcolor{red!50} \faTimes 
        	& \cellcolor{green!50} \faTrophy 
        	& \cellcolor{red!50} \faTimes 
        	& \cellcolor{bg} 
        	& \cellcolor{green!50} \faTrophy 
        	& \cellcolor{green!50} \faDiamond \\
        DINO 
        	& \cellcolor{red!50} \faTimes 
        	& \cellcolor{green!50} \faDiamond 
        	& \cellcolor{green!50} \faDiamond  
        	& \cellcolor{red!50} \faTimes 
        	& \cellcolor{green!50} \faDiamond 
        	& \cellcolor{green!50} \faDiamond 
        	& \cellcolor{green!50} \faDiamond 
        	& \cellcolor{green!50} \faDiamond 
        	& \cellcolor{bg} 
        	& \cellcolor{green!50} \faDiamond  \\
        SAM 
        	& \cellcolor{green!50} \faTrophy 
        	& \cellcolor{green!50} \faTrophy 
        	& \cellcolor{green!50} \faTrophy  
        	& \cellcolor{green!50} \faTrophy 
        	& \cellcolor{red!50} \faTimes 
        	& \cellcolor{green!50} \faTrophy 
        	& \cellcolor{green!50} \faTrophy 
        	& \cellcolor{green!50} \faTrophy 
        	& \cellcolor{green!50} \faTrophy 
        	& \cellcolor{bg} 
        	\\
        \bottomrule
    \end{tabular}
\end{table}
\section*{Discussion and Conclusion}
This work presents a comprehensive benchmarking analysis of convolutional and Transformer-based networks, for medical image classification across diverse datasets, training schemes, and input resolutions. Through systematic evaluation, we challenge prevailing assumptions regarding model design, training schemes, and input resolution requirements. Our experiments are designed to highlight both general dataset-average findings (see \tablename~\ref{tab:average_benchmark}) and dataset-specific results (see Supplementary Tables~\ref{tab:bloodmnist benchmark} - \ref{tab:tissuemnist benchmark} in the appendix), which are basically coherent. By reassessing these methodologies, our aim is to foster genuine progress in the field and provide insights to inform the development of more efficient and effective models, rather than supporting the current trend of continuous scaling.

Our findings offer valuable insights into the performance of traditional models across various scenarios. End-to-end training consistently delivers the highest overall performance, with higher resolutions generally enhancing performance up to a certain threshold. Notably, we observe diminishing returns beyond $128 \times 128$ to $224 \times 224$ pixels, suggesting the potential viability of lower resolution inputs, particularly during the prototyping phase of model development. Moreover, this implies the existence of an optimal image resolution in terms of performance (accuracy and processing speed), likely lying between these two distinct image resolutions. However, this behavior is expected to be contingent on dataset characteristics, including color space and sample size. Therefore, further investigation is needed to determine the existence and dataset specificity of this optimal image resolution, as well as its contribution to overall model performance.

Furthermore, our analysis highlights the nuanced impact of self-supervised pretraining strategies like CLIP and DINO. While they do not always improve end-to-end trained models, they demonstrate enhanced performance for linear probing and $k$-NN integration. The near-baseline performance of DINO-pretrained models, requiring minimal training for linear probing or none for $k$-NN integration, raises questions about the necessity for full end-to-end training, emphasizing the potential for pretrained models to achieve comparable performance using computationally efficient methodologies. Particularly noteworthy is the fact that CLIP was trained on pairs of images and text, potentially limiting its suitability for image-centric tasks, while DINO exclusively utilizes pairs of natural images from ImageNet, which likely limits its suitability for medical image classification. The remarkable performance of CLIP and DINO despite their domain foreignness underscores the potential of foundation models and emphasizes the need for domain-specific foundation models to further enhance performance and applicability.

Finally, our model ranking analysis underscores the performance disparities between CNNs and ViTs. Convolutional models consistently outperform ViTs in accuracy for end-to-end training, while ViTs excel in linear probing and $k$-NN approaches. This emphasizes the continued competitiveness of convolutional models compared to ViTs and underscores the significance of exhaustive pretraining for the latter, highlighting the particular suitability of ViTs for foundation models.

In addition to performance considerations, robustness to distribution shifts, model interpretability, and explainability are critical factors in clinical applications, where heterogeneous data is prevalent, and trust and transparency are paramount. While our work does not explicitly evaluate these aspects, prior research suggests that ViTs and ViT-based foundation models demonstrate greater robustness to distribution shifts compared to CNNs \cite{Zhang2022CVPR,Guo2023,Zhou2023}. Furthermore, $k$-NN integration enhances interpretability compared to end-to-end training and linear probing by enabling direct, sample-based reasoning within the feature space \cite{Klein2023,doerrich2024integrating,Xiaomeng2024}. Lastly, both CNNs and ViTs can be complemented with existing explainability techniques \cite{Haar2023,Choi2024}, improving model transparency and aiding clinical decision-making. However, future work should further investigate these aspects to ensure that models not only achieve high performance but also provide clinically meaningful and interpretable insights for practitioners.

However, it is crucial to acknowledge the limitations of our study. First, our analysis is limited to the datasets and distinct dataset splits provided by MedMNIST+ \cite{yang2024dataset} making it inherently susceptible to biases such as class imbalance, data inhomogeneities, and demographic underrepresentation. In addition, the dataset collection lacks images from common modalities such as MRI, SPECT, and PET, and encompasses only a limited number of anatomical regions and disease patterns. Therefore, further investigation is warranted to include these modalities and assess the applicability of our findings in these unexplored settings. Second, our evaluation across all datasets simultaneously may overlook dataset-specific nuances. Future studies should explore dataset-specific results, additional datasets with varying characteristics (\ie sample size, noisy labels, corruptions, etc.), and different dataset splits. Furthermore, our benchmark is restricted to Convolutional Neural Networks and Vision Transformer architectures. While these models remain fundamental to medical image classification, emerging architectures, such as Graph Neural Networks (GNNs), warrant further evaluation. GNNs have shown promise, particularly in histopathology, where they effectively capture topological structures in Whole Slide Images \cite{Brussee2025}. Expanding future benchmarks to include such architectures could provide a more comprehensive evaluation of the deep learning landscape for medical image classification. Additionally, we do not analyze the interpretability and explainability of the assessed architectures, which are critical for clinical adoption. Future work should address these aspects to enhance model transparency and trustworthiness. Finally, this work does not focus on the deployment of deep learning models in clinical practice which is a substantially more complex endeavor and requires addressing a broad set of challenges, including model explainability, data privacy, regulatory approval, and human-AI interaction. Instead, our benchmark aims to accelerate model development by providing insights that facilitate more efficient and scalable approaches for eventual real-world integration. In conclusion, our work advocates for the following key takeaways and recommendations for model development:
\begin{itemize}
\item \textbf{Prioritize computational efficiency:} prioritize the development of computationally efficient alternatives to full end-to-end training for faster model development iterations and reduced hardware demands during deployment.
\item \textbf{Utilize lower-resolution images for prototyping:} consider utilizing lower resolution images during prototyping to conserve computational resources and time.
\item \textbf{Benchmark models across diverse datasets:} evaluate methods on multiple distinct benchmarks to cover real-world situations, rather than focusing solely on achieving state-of-the-art performance on a single benchmark.
\item \textbf{Focus on efficiency and robustness:} focus on the development of efficient and robust methods, rather than scaling existing methods to attain state-of-the-art performance.

\end{itemize}
\section*{Method}
\subsection*{Model Selection}
Our selection of model architectures encompasses a diverse array of both convolutional and Transformer-based networks. Among the chosen convolutional models are well-established architectures such as VGG16 \cite{Simonyan2014VeryDC}, AlexNet \cite{Krizhevsky2012}, ResNet-18 \cite{He2015DeepRL}, DenseNet-121 \cite{Huang2016DenselyCC}, and EfficientNet-B4 \cite{Tan2019EfficientNetRM}, all of which were pretrained on the ImageNet1k dataset \cite{Russakovsky2014ImageNetLS}. In the domain of Transformers, we include the Vision Transformer (ViT) \cite{dosovitskiy2021}, renowned for its exceptional performance across various image classification tasks, pretrained on ImageNet1k as well as the CLIP \cite{Radford2021} and DINO \cite{Caron2021} pretrained ViT variants. Acknowledging recent advancements in this area, our selection extends to adaptations of ViT such as EVA-02 \cite{Fang2023} and the Segment Anything Model (SAM) \cite{Kirillov2023}. EVA-02 represents a series of efficiently optimized plain ViTs with moderate model sizes, employing bidirectional visual representations learned from a robust CLIP encoder. Conversely, SAM, initially conceived as a foundational model for image segmentation, is intentionally designed and trained to be promptable, thus facilitating zero-shot transferability to new image distributions and tasks, including image classification. For all ViT architectures, we opted for the base backbone with a patch size of 16 (\ie ViT-B/16). With the exception of the AlexNet model obtained from the torchvision library, all models were sourced from the "Pytorch Image Models (timm)" library at Huggingface \cite{rw2019timm}. Further details regarding the employed backbone architectures, including the number of parameters, activations, Giga Multiply-Add Operations (GMACs), feature dimension before the final classification layer, memory requirements for single precision (fp32), and inference times on both GPU and CPU, are outlined in Table \ref{tab:models}.
\begin{table}
\caption{Details of the evaluated model architectures, including the number of parameters (in million, M), activations (in million, M), Giga Multiply-Add Operations (GMACs), feature dimension before the final classification layer, memory requirements (in megabytes, MB) for single precision (fp32), and inference times on both GPU and CPU (in milliseconds, ms). GMACs, activations, inference times, and memory requirements are reported for processing a single input image of resolution 224 × 224 pixels. Inference times were measured on an Intel(R) Core(TM) i9-14900K processor (CPU) and an NVIDIA RTX 6000 GPU (Ada Generation).}
\label{tab:models}
    \centering
    \begin{tabular}{l r r r r r r r}
        \toprule
        \multirow{3}{*}{Model} & Number & Number & Number & Output & Memory & CPU & GPU \\
        & Parameters & Activations & GMACs & Dimension & Requirements & Inference & Inference \\
        & (M) & (M) & \# & \# & (MB) & (ms) & (ms) \\
        \midrule
        VGG16                    & 138.4 & 13.6 & 15.5 & 4096 & 1454 & 108.0 & 0.4 \\ 
        AlexNet                  & 62.3 & 0.6 & 0.36 & 4096 & 766 & 8.6 & 0.3 \\ 
        ResNet-18                & 11.7 & 2.5 & 1.8 & 512 & 602 & 9.2 & 0.7 \\ 
        DenseNet-121             & 8.0 & 6.9 & 2.9 & 1024 & 690 & 318.0 & 3.9 \\ 
        EfficientNet-B4          & 19.3 & 17.1 & 1.5 & 1792 & 728 & 29.3 & 3.9 \\ 
        ViT-B/16                 & 86.6 & 16.5 & 16.9 & 768 & 900 & 49.0 & 1.3 \\ 
        CLIP ViT-B/16            & 86.6 & 16.5 & 16.9 & 768 & 900 & 48.9 & 1.3 \\ 
        EVA-02 ViT-B/16          & 86.3 & 16.5 & 16.9 & 768 & 902 & 60.0 & 2.8 \\ 
        DINO ViT-B/16            & 85.8 & 16.5 & 16.9 & 768 & 900 & 49.1 & 1.2 \\ 
        SAM ViT-B/16             & 89.7 & 64.3 & 23.3 & 256 & 924 & 71.4 & 13.2 \\ 
        \bottomrule
    \end{tabular}
\end{table}
\subsection*{Training Pipeline}
\label{subsec:training_pipeline}
We adopt diverse training paradigms, encompassing both end-to-end training (\ie training the whole model), and linear probing, where we solely train the classification head, while keeping the encoder frozen. Additionally, we explore the integration of the k-nearest neighbors (\mbox{$k$-NN}) classifier into the feature space of pre-trained models. Following K. Nakata et al.\cite{Nakata2022} and S. Doerrich et al.\cite{doerrich2024integrating}, the pre-trained image encoder initially extracts feature embeddings from the training set, which are then stored in an external database along with their corresponding labels. During inference, the image encoder generates a feature embedding for a given query image. Subsequently, the top-$k$ feature embeddings having highest similarity scores (\ie lowest cosine distance) with the query embedding are retrieved from the training set along with their associated labels. The classification of the query image is afterward determined through a majority vote on these labels. This approach facilitates efficient classification without necessitating retraining of the classification head or the entire encoder, thereby enhancing computational efficiency, interpretability, and generalizability, with reduced dependence on hyperparameters. Given the substantial computational cost associated with training deep neural networks, particularly foundation models, the adoption of \mbox{$k$-NN} methods presents an efficient alternative to traditional training schemes.

The training regimen consisted of 100 epochs with early stopping based on the validation set. We employed the AdamW optimizer \cite{Loshchilov2017DecoupledWD} with a learning rate of $0.0001$, along with a cosine annealing learning rate scheduler \cite{Loshchilov2016SGDRSG} with a single cycle. Each model was trained with a batch size of $64$, allowing for training on a single NVIDIA RTX™ A5000 GPU. For evaluations utilizing the k-nearest neighbors (\mbox{$k$-NN}) approach, we set $k$ to $11$, in line with Z. Zhu et al. \cite{Zhu2021DetectingCL}, who demonstrated the suitability of $k > 10$ for detecting noisy labels. To maintain compatibility with the pretrained models while preserving the inherent properties of individual resolutions, all image resolutions were padded to $224 \times 224$ pixels using zero padding. This choice was motivated by M. Hashemi \cite{Hashemi2019}, who demonstrated that zero-padding has no discernible effect on classification accuracy while significantly reducing training time compared to image resizing. With zero-padding, neighboring zero input units (pixels) do not activate their corresponding convolutional unit in the subsequent layer, resulting in decreased requirements for updating synaptic weights on outgoing links and ensuring robust feature preservation during image reshaping.
\subsection*{Loss Criterion and Evaluation Metrics}
In line with the methodology outlined by J. Yang et al. \cite{medmnistv2}, we select the choice of loss criteria to suit the specific classification tasks associated with each dataset. For binary (BC) and multi-class classification (MC), as well as ordinal regression (OR) tasks, we utilize the Cross-Entropy (CE) loss function applied to the logits:
\begin{equation}
    \text{CE} = -\frac{1}{N} \sum_{n=1}^{N} \log\left(\frac{\exp(z_{n,y_n})}{\sum_{c=1}^{C} \exp(z_{n,c})} \right),
\end{equation}
where $N$ denotes the number of samples in the current batch, $C$ represents the total number of classes, $z_{n,c}$ signifies the logit for class $c$ of the $n$-th sample, and $z_{n,y_n}$ denotes the logit corresponding to the target class for the $n$-th sample. For binary classification (BC), this equation simplifies to Binary Cross-Entropy with $C = 2$.

Additionally, we treat the multi-label classification task of the Chest dataset as a multi-label binary classification (ML-BC) problem. Here, each class label $c$ is addressed as a distinct binary classification task, aiming to predict the presence or absence of each class label $c$ for a given sample $n$. To this end, we employ the Binary Cross-Entropy with Logits (BCEwithLogits) loss function across all class labels $c \in C$:
\begin{equation}
    \text{BCEwithLogits} = -\frac{1}{N} \sum_{n=1}^{N} \sum_{c=1}^{C} \left[ y_{n,c} \log\left(\sigma(z_{n,c})\right) + (1 - y_{n,c}) \log\left(1 - \sigma(z_{n,c})\right) \right],
\end{equation}
where $N$ represents the number of samples in the current batch, $z_{n,c}$ indicates the logit for sample $n$ and class label $c$, $y_{n,c}$ denotes the binary label for sample $n$ and label $c$, representing the presence (1) or absence (0) of class $c$, and $\sigma(\cdot)$ represents the sigmoid function applied to the logit $z_{n,c}$.

Furthermore, for the purpose of simplicity and standardization, consistent with J. Yang et al. \cite{medmnistv2}, we employ the same evaluation metrics including accuracy (ACC) for a fixed operating point of $0.5$ and the area under the receiver operating characteristic curve (AUC) to assess the model's ability to differentiate between classes. Our choice of AUC over metrics such as sensitivity, specificity, and false positive rates is driven by its suitability for comparing different architectural choices and training paradigms. Unlike sensitivity and specificity, which require setting a fixed operating point, a process that is application-dependent and not trivial, AUC provides a holistic measure of a model's discriminatory power across all operating points. Additionally, we include accuracy due to its prevalence in the field and ease of interpretation.

To quantify the level of class imbalance within each dataset, we report Shannon’s Equitability for each data split. Shannon’s Equitability (EH) is a normalized measure of class distribution uniformity, derived from the Shannon Diversity Index \cite{Shannon1948}. It is computed as:
\begin{equation}
    EH = \frac{H}{H_{\text{max}}} = \frac{- \sum_{i=1}^{S} p_i \ln{p_i}}{\ln{S}},
\end{equation}
where $H$ is the Shannon Diversity Index, $S$ represents the total number of classes, $p_i$ denotes the proportion of samples belonging to class $i$ and $\ln$ stands for the natural logarithm. The maximum possible diversity, $H_{\text{max}} = \ln{S}$, occurs when all classes are equally represented. Shannon’s Equitability ranges from 0 (total imbalance) to 1 (perfect balance), allowing for a standardized comparison of class distribution across datasets and splits.
\section*{Data availability}
The used datasets are licensed under Creative Commons Attribution 4.0 International (CC BY 4.0), except DermaMNIST under Creative Commons Attribution-NonCommercial 4.0 International (CC BY-NC 4.0) and publicly available here: Yang, J. et al. [MedMNIST+] 18x Standardized Datasets for 2D and 3D Biomedical Image Classification with Multiple Size Options: 28 (MNIST-Like), 64, 128, and 224. \href{https://zenodo.org/records/10519652}{https://zenodo.org/records/10519652} (2024).

\section*{Code availability}
The source code is available at \href{https://github.com/sdoerrich97/rethinking-model-prototyping-MedMNISTPlus}{https://github.com/sdoerrich97/rethinking-model-prototyping-MedMNISTPlus}.

\bibliography{references}

\section*{Acknowledgements}
Funded through the Hightech Agenda Bayern (HTA) of the Free State of Bavaria, Germany.

\section*{Author contributions statement}
S.D., F.D.S., J.B., and C.L. worked on the methods and did the study design. S.D., F.D.S., and J.B. implemented the algorithms. J.B. analyzed the individual data sets in detail. S.D. and F.D.S. conceived and conducted the experiments. S.D. analyzed the results and created all images. C.L. supervised the study. S.D., F.D.S., and C.L. wrote the manuscript. All authors reviewed, corrected, and approved the paper.

\section*{Additional information}
\noindent\textbf{Compliance with ethical standards}:
This research study was conducted retrospectively using human subject data made available in open access. Ethical approval was not required as confirmed by the license attached with the open-access data.

\noindent\textbf{Competing interests}:
The authors declare no competing interests.

\appendix
\setcounter{table}{0}
\renewcommand{\tablename}{Supplementary Table}
\renewcommand{\thetable}{S\arabic{table}}

\section*{Appendix}

\begin{table}[!htpb]
\setlength{\tabcolsep}{4.9pt}
\renewcommand{\arraystretch}{1.1}
\caption{Benchmark outcomes summarizing the mean and standard deviation of accuracy (ACC), for a fixed operating point of $0.5$, and area under the receiver operating characteristic curve (AUC) for the BloodMNIST dataset across all training scheme-model-image resolution combinations, derived from three independent random seeds. Notably, the \mbox{$k$-NN} algorithm, devoid of a training phase, remains unaffected by the stochasticity inherent in model training, thus reporting only the total ACC value without standard deviation. Moreover, owing to its direct utilization of embeddings and labels for classification, \mbox{$k$-NN} does not furnish a reliable AUC score. The overall best result across all training schemes, models, and resolutions is highlighted with a \colorbox{bg}{background color}; the best result per resolution across all training schemes and models is highlighted with \underline{underline}; and the best result per training scheme and resolution is highlighted in \textbf{bold}.}
\label{tab:bloodmnist benchmark}
\centering
\begin{tabular}{lcccccccccc}
& \\
\toprule
\multicolumn{10}{c}{\textbf{BloodMNIST}} \\
\toprule
\multirow{2.5}{*}{Methods} & \multicolumn{4}{c}{Accuracy (ACC)} &  & \multicolumn{4}{c}{Area Under the ROC Curve (AUC)} \\
\cmidrule(r){2-5}\cmidrule(l){7-10}
& $28 \times 28$ & $64 \times 64$ & $128 \times 128$ & $224 \times 224$ &  & $28 \times 28$ & $64 \times 64$ & $128 \times 128$ & $224 \times 224$\\ 
\midrule
{\textsc{End-to-End}} & & & & & \\        
\; VGG16           & \underline{\textbf{94.85{\scriptsize$\pm$0.47}}} & \underline{\textbf{98.20{\scriptsize$\pm$0.16}}} & 98.52{\scriptsize$\pm$0.25} & 98.77{\scriptsize$\pm$0.31} &  & \underline{\textbf{99.66{\scriptsize$\pm$0.06}}} & \underline{\textbf{99.91{\scriptsize$\pm$0.00}}} & \underline{\textbf{99.93{\scriptsize$\pm$0.01}}} & \cellcolor{bg}\underline{\textbf{99.95{\scriptsize$\pm$0.01}}} \\ 
\; AlexNet         & 90.39{\scriptsize$\pm$0.46} & 96.41{\scriptsize$\pm$0.50} & 97.74{\scriptsize$\pm$0.29} & 98.18{\scriptsize$\pm$0.39} &  & 99.12{\scriptsize$\pm$0.04} & 99.83{\scriptsize$\pm$0.00} & 99.89{\scriptsize$\pm$0.01} & 99.92{\scriptsize$\pm$0.00} \\ 
\; ResNet-18       & 91.93{\scriptsize$\pm$0.37} & 97.06{\scriptsize$\pm$0.11} & 98.34{\scriptsize$\pm$0.26} & 98.94{\scriptsize$\pm$0.08} &  & 99.24{\scriptsize$\pm$0.04} & 99.82{\scriptsize$\pm$0.01} & 99.92{\scriptsize$\pm$0.01} & 99.92{\scriptsize$\pm$0.01} \\ 
\; DenseNet-121    & 93.87{\scriptsize$\pm$0.40} & 97.99{\scriptsize$\pm$0.28} & \underline{\textbf{98.90{\scriptsize$\pm$0.09}}} & \cellcolor{bg}\underline{\textbf{99.02{\scriptsize$\pm$0.14}}} & & 99.58{\scriptsize$\pm$0.03} & \underline{\textbf{99.91{\scriptsize$\pm$0.02}}} & 99.92{\scriptsize$\pm$0.01} & \cellcolor{bg}\underline{\textbf{99.95{\scriptsize$\pm$0.01}}} \\ 
\; EfficientNet-B4 & 78.06{\scriptsize$\pm$1.34} & 90.61{\scriptsize$\pm$0.39} & 96.00{\scriptsize$\pm$0.19} & 97.40{\scriptsize$\pm$0.27} & & 96.62{\scriptsize$\pm$0.26} & 99.26{\scriptsize$\pm$0.05} & 99.78{\scriptsize$\pm$0.03} & 99.90{\scriptsize$\pm$0.00} \\ 
\; ViT-B/16        & 90.59{\scriptsize$\pm$0.56} & 96.42{\scriptsize$\pm$0.26} & 97.54{\scriptsize$\pm$0.21} & 98.43{\scriptsize$\pm$0.09} & &  99.21{\scriptsize$\pm$0.08} & 99.80{\scriptsize$\pm$0.00} & 99.91{\scriptsize$\pm$0.01} & 99.93{\scriptsize$\pm$0.01} \\ 
\; CLIP ViT-B/16   & 89.92{\scriptsize$\pm$0.25} & 93.56{\scriptsize$\pm$0.90} & 95.60{\scriptsize$\pm$0.42} & 96.82{\scriptsize$\pm$0.04} &  & 99.08{\scriptsize$\pm$0.03} & 99.56{\scriptsize$\pm$0.04} & 99.78{\scriptsize$\pm$0.03} & 99.87{\scriptsize$\pm$0.01} \\ 
\; EVA-02 ViT-B/16 & 91.24{\scriptsize$\pm$0.63} & 95.65{\scriptsize$\pm$0.26} & 97.94{\scriptsize$\pm$0.59} & 98.42{\scriptsize$\pm$0.12} &  & 99.15{\scriptsize$\pm$0.10} & 99.75{\scriptsize$\pm$0.03} & 99.92{\scriptsize$\pm$0.02} & 99.93{\scriptsize$\pm$0.01} \\ 
\; DINO ViT-B/16   & 89.87{\scriptsize$\pm$0.47} & 95.69{\scriptsize$\pm$0.12} & 97.54{\scriptsize$\pm$0.54} & 97.98{\scriptsize$\pm$0.46} &  & 99.13{\scriptsize$\pm$0.09} & 99.78{\scriptsize$\pm$0.01} & 99.89{\scriptsize$\pm$0.03} & 99.93{\scriptsize$\pm$0.01} \\ 
\; SAM ViT-B/16    & 91.19{\scriptsize$\pm$0.54} & 95.89{\scriptsize$\pm$0.12} & 97.33{\scriptsize$\pm$0.32} & 98.55{\scriptsize$\pm$0.26} &  & 99.16{\scriptsize$\pm$0.07} & 99.72{\scriptsize$\pm$0.01} & 99.86{\scriptsize$\pm$0.04} & 99.92{\scriptsize$\pm$0.02} \\ 
\midrule
{\textsc{Linear Probing}} & & & & & \\         
\; VGG16           & 74.95{\scriptsize$\pm$0.02} & 84.53{\scriptsize$\pm$0.04} & 88.86{\scriptsize$\pm$0.09} & 93.77{\scriptsize$\pm$0.09} & & 95.66{\scriptsize$\pm$0.00} & 98.24{\scriptsize$\pm$0.00} & 99.10{\scriptsize$\pm$0.00} & 99.64{\scriptsize$\pm$0.00} \\ 
\; AlexNet         & 66.67{\scriptsize$\pm$0.25} & 81.97{\scriptsize$\pm$0.22} & 89.61{\scriptsize$\pm$0.12} & 91.51{\scriptsize$\pm$0.02} & & 93.84{\scriptsize$\pm$0.01} & 97.69{\scriptsize$\pm$0.05} & 99.16{\scriptsize$\pm$0.02} & 99.39{\scriptsize$\pm$0.00} \\
\; ResNet-18       & 63.40{\scriptsize$\pm$0.13} & 67.15{\scriptsize$\pm$0.06} & 80.58{\scriptsize$\pm$0.10} & 91.10{\scriptsize$\pm$0.10} & 
& 90.58{\scriptsize$\pm$0.05} & 93.67{\scriptsize$\pm$0.01} & 97.15{\scriptsize$\pm$0.00} & 99.25{\scriptsize$\pm$0.00} \\ 
\; DenseNet-121    & 71.34{\scriptsize$\pm$0.12} & 84.03{\scriptsize$\pm$0.07} & 93.36{\scriptsize$\pm$0.02} & 95.65{\scriptsize$\pm$0.06} & & 94.08{\scriptsize$\pm$0.03} & 97.99{\scriptsize$\pm$0.01} & 99.55{\scriptsize$\pm$0.00} & 99.81{\scriptsize$\pm$0.00} \\ 
\; EfficientNet-B4 & 59.46{\scriptsize$\pm$0.72} & 69.54{\scriptsize$\pm$0.24} & 84.10{\scriptsize$\pm$0.17} & 90.19{\scriptsize$\pm$0.10} & & 88.94{\scriptsize$\pm$0.20} & 93.97{\scriptsize$\pm$0.06} & 97.91{\scriptsize$\pm$0.03} & 99.11{\scriptsize$\pm$0.04} \\ 
\; ViT-B/16        & 80.67{\scriptsize$\pm$0.07} & 93.21{\scriptsize$\pm$0.12} & 97.07{\scriptsize$\pm$0.01} & 97.95{\scriptsize$\pm$0.04} & & 97.13{\scriptsize$\pm$0.01} & 99.50{\scriptsize$\pm$0.00} & 99.88{\scriptsize$\pm$0.00} & 99.92{\scriptsize$\pm$0.00} \\ 
\; CLIP ViT-B/16   & 83.05{\scriptsize$\pm$0.20} & 92.21{\scriptsize$\pm$0.04} & 96.00{\scriptsize$\pm$0.02} & 96.13{\scriptsize$\pm$0.07} & & 97.80{\scriptsize$\pm$0.01} & 99.38{\scriptsize$\pm$0.01} & 99.77{\scriptsize$\pm$0.00} & 99.86{\scriptsize$\pm$0.00} \\ 
\; EVA-02 ViT-B/16 & 82.35{\scriptsize$\pm$0.01} & 89.89{\scriptsize$\pm$0.00} & 93.10{\scriptsize$\pm$0.00} & 94.09{\scriptsize$\pm$0.01} & & 97.58{\scriptsize$\pm$0.00} & 99.03{\scriptsize$\pm$0.00} & 99.53{\scriptsize$\pm$0.00} & 99.66{\scriptsize$\pm$0.00} \\ 
\; DINO ViT-B/16   & \textbf{88.79{\scriptsize$\pm$0.04}} & \textbf{97.15{\scriptsize$\pm$0.09}} & \textbf{98.25{\scriptsize$\pm$0.10}} & \textbf{98.70{\scriptsize$\pm$0.01}} & & \textbf{98.81{\scriptsize$\pm$0.00}} & \textbf{99.86{\scriptsize$\pm$0.00}} & \textbf{99.92{\scriptsize$\pm$0.00}} & \textbf{99.94{\scriptsize$\pm$0.00}} \\ 
\; SAM ViT-B/16    & 20.00{\scriptsize$\pm$0.04} & 24.61{\scriptsize$\pm$0.11} & 34.73{\scriptsize$\pm$0.17} & 51.70{\scriptsize$\pm$0.10} & & 76.78{\scriptsize$\pm$0.16} & 81.19{\scriptsize$\pm$0.18} & 87.54{\scriptsize$\pm$0.03} & 88.98{\scriptsize$\pm$0.03} \\ 
\midrule
{\textsc{$k$-NN ($k = 11$)}} & & & & & \\       
\; VGG16           & 71.76 & 75.97 & 79.22 & 81.38 & & - & - & - & - \\ 
\; AlexNet         & 68.34 & 82.23 & 87.55 & 89.94 & & - & - & - & - \\  
\; ResNet-18       & 75.04 & 76.18 & 80.27 & 89.04 & & - & - & - & - \\ 
\; DenseNet-121    & 72.29 & 80.94 & 84.10 & 88.54 & & - & - & - & - \\ 
\; EfficientNet-B4 & 73.63 & 78.49 & 81.00 & 84.48 & & - & - & - & - \\ 
\; ViT-B/16        & 70.21 & 78.95 & 92.17 & 94.80 & & - & - & - & - \\ 
\; CLIP ViT-B/16   & 68.66 & 82.87 & 91.38 & 93.19 & & - & - & - & - \\ 
\; EVA-02 ViT-B/16 & 77.70 & 84.07 & 89.94 & 92.43 & & - & - & - & - \\ 
\; DINO ViT-B/16   & \textbf{85.24} & \textbf{93.04} & \textbf{96.67} & \textbf{96.81} & & - & - & - & - \\ 
\; SAM ViT-B/16    & 62.50 & 64.40 & 76.15 & 78.22 & & - & - & - & - \\ 
\bottomrule
\end{tabular}
\end{table}

\begin{table}[!htpb]
\setlength{\tabcolsep}{4.9pt}
\renewcommand{\arraystretch}{1.1}
\caption{Benchmark outcomes summarizing the mean and standard deviation of accuracy (ACC), for a fixed operating point of $0.5$, and area under the receiver operating characteristic curve (AUC) for the BreastMNIST dataset across all training scheme-model-image resolution combinations, derived from three independent random seeds. Notably, the \mbox{$k$-NN} algorithm, devoid of a training phase, remains unaffected by the stochasticity inherent in model training, thus reporting only the total ACC value without standard deviation. Moreover, owing to its direct utilization of embeddings and labels for classification, \mbox{$k$-NN} does not furnish a reliable AUC score. The overall best result across all training schemes, models, and resolutions is highlighted with a \colorbox{bg}{background color}; the best result per resolution across all training schemes and models is highlighted with \underline{underline}; and the best result per training scheme and resolution is highlighted in \textbf{bold}.}
\label{tab:breastmnist benchmark}
\centering
\begin{tabular}{lcccccccccc}
& \\
\toprule
\multicolumn{10}{c}{\textbf{BreastMNIST}} \\
\toprule
\multirow{2.5}{*}{Methods} & \multicolumn{4}{c}{Accuracy (ACC)} &  & \multicolumn{4}{c}{Area Under the ROC Curve (AUC)} \\
\cmidrule(r){2-5}\cmidrule(l){7-10}
& $28 \times 28$ & $64 \times 64$ & $128 \times 128$ & $224 \times 224$ &  & $28 \times 28$ & $64 \times 64$ & $128 \times 128$ & $224 \times 224$\\ 
\midrule
{\textsc{End-to-End}} & & & & & \\        
\; VGG16           & 85.26{\scriptsize$\pm$1.05} & \underline{\textbf{87.82{\scriptsize$\pm$2.09}}} & 89.32{\scriptsize$\pm$0.80} & 88.03{\scriptsize$\pm$0.80} &  & 87.48{\scriptsize$\pm$0.96} & 90.21{\scriptsize$\pm$0.48} & 92.40{\scriptsize$\pm$0.68} & 91.11{\scriptsize$\pm$0.44} \\ 
\; AlexNet         & \underline{\textbf{86.54{\scriptsize$\pm$1.89}}} & 87.61{\scriptsize$\pm$1.21} & \cellcolor{bg}\underline{\textbf{90.60{\scriptsize$\pm$1.09}}} & \textbf{88.46{\scriptsize$\pm$0.52}} &  & \underline{\textbf{89.08{\scriptsize$\pm$2.41}}} & 89.67{\scriptsize$\pm$1.38} & \cellcolor{bg}\underline{\textbf{93.97{\scriptsize$\pm$1.36}}} & \underline{\textbf{93.86{\scriptsize$\pm$0.48}}} \\ 
\; ResNet-18       & 83.97{\scriptsize$\pm$1.81} & 83.33{\scriptsize$\pm$0.52} & 85.47{\scriptsize$\pm$1.51} & 87.18{\scriptsize$\pm$0.52} &  & 87.61{\scriptsize$\pm$1.16} & 87.14{\scriptsize$\pm$1.77} & 87.78{\scriptsize$\pm$0.63} & 89.94{\scriptsize$\pm$0.57} \\ 
\; DenseNet-121    & 83.33{\scriptsize$\pm$2.91} & 85.90{\scriptsize$\pm$0.91} & 85.68{\scriptsize$\pm$0.30} & 87.39{\scriptsize$\pm$0.80} &  & 86.18{\scriptsize$\pm$3.25} & \textbf{90.56{\scriptsize$\pm$0.32}} & 86.63{\scriptsize$\pm$0.55} & 89.67{\scriptsize$\pm$0.57} \\ 
\; EfficientNet-B4 & 76.50{\scriptsize$\pm$2.47} & 74.57{\scriptsize$\pm$2.88} & 76.50{\scriptsize$\pm$2.47} & 74.57{\scriptsize$\pm$1.32} &  & 75.62{\scriptsize$\pm$1.69} & 74.09{\scriptsize$\pm$4.58} & 76.73{\scriptsize$\pm$2.92} & 70.84{\scriptsize$\pm$5.18} \\ 
\; ViT-B/16        & 82.05{\scriptsize$\pm$0.52} & 81.62{\scriptsize$\pm$2.58} & 82.48{\scriptsize$\pm$0.80} & 83.76{\scriptsize$\pm$1.09} &  & 84.83{\scriptsize$\pm$0.96} & 84.49{\scriptsize$\pm$4.67} & 83.63{\scriptsize$\pm$1.49} & 86.18{\scriptsize$\pm$0.26} \\ 
\; CLIP ViT-B/16   & 77.56{\scriptsize$\pm$3.43} & 76.71{\scriptsize$\pm$0.60} & 79.06{\scriptsize$\pm$0.80} & 80.13{\scriptsize$\pm$1.81} &  & 75.19{\scriptsize$\pm$10.6} & 77.58{\scriptsize$\pm$2.42} & 78.38{\scriptsize$\pm$0.73} & 77.53{\scriptsize$\pm$2.64} \\ 
\; EVA-02 ViT-B/16 & 74.79{\scriptsize$\pm$5.50} & 73.08{\scriptsize$\pm$0.00} & 72.44{\scriptsize$\pm$1.38} & 82.91{\scriptsize$\pm$3.86} &  & 74.88{\scriptsize$\pm$6.72} & 73.85{\scriptsize$\pm$1.78} & 78.36{\scriptsize$\pm$3.35} & 83.08{\scriptsize$\pm$5.07} \\ 
\; DINO ViT-B/16   & 84.83{\scriptsize$\pm$3.79} & 81.20{\scriptsize$\pm$0.80} & 79.70{\scriptsize$\pm$2.36} & 84.40{\scriptsize$\pm$1.68} &  & 88.39{\scriptsize$\pm$2.64} & 81.95{\scriptsize$\pm$1.58} & 83.58{\scriptsize$\pm$0.79} & 86.31{\scriptsize$\pm$2.70} \\ 
\; SAM ViT-B/16    & 82.05{\scriptsize$\pm$0.52} & 77.35{\scriptsize$\pm$6.04} & 82.91{\scriptsize$\pm$2.63} & 81.62{\scriptsize$\pm$2.12} &  & 80.16{\scriptsize$\pm$4.38} & 79.16{\scriptsize$\pm$5.18} & 82.07{\scriptsize$\pm$7.69} & 78.51{\scriptsize$\pm$1.13} \\ 
\midrule
{\textsc{Linear Probing}} & & & & & \\         
\; VGG16           & 78.63{\scriptsize$\pm$0.30} & 77.78{\scriptsize$\pm$0.60} & 84.62{\scriptsize$\pm$0.00} & 80.98{\scriptsize$\pm$0.60} &  & 83.10{\scriptsize$\pm$0.24} & 81.86{\scriptsize$\pm$0.22} & 89.00{\scriptsize$\pm$0.08} & 85.06{\scriptsize$\pm$0.17} \\ 
\; AlexNet         & 77.56{\scriptsize$\pm$1.05} & 81.62{\scriptsize$\pm$0.60} & 84.40{\scriptsize$\pm$0.30} & 86.11{\scriptsize$\pm$0.80} &  & 77.39{\scriptsize$\pm$0.43} & 82.64{\scriptsize$\pm$0.38} & 90.38{\scriptsize$\pm$0.35} & \textbf{93.47{\scriptsize$\pm$0.11}} \\
\; ResNet-18       & 73.08{\scriptsize$\pm$0.00} & 73.08{\scriptsize$\pm$0.00} & 73.08{\scriptsize$\pm$0.00} & 73.08{\scriptsize$\pm$0.00} &  & 64.77{\scriptsize$\pm$1.42} & 54.92{\scriptsize$\pm$1.76} & 61.72{\scriptsize$\pm$8.75} & 68.33{\scriptsize$\pm$1.75} \\ 
\; DenseNet-121    & 75.00{\scriptsize$\pm$1.81} & 78.42{\scriptsize$\pm$0.80} & 78.21{\scriptsize$\pm$0.91} & 80.77{\scriptsize$\pm$0.52} &  & 64.96{\scriptsize$\pm$10.86} & 73.84{\scriptsize$\pm$2.94} & 78.43{\scriptsize$\pm$1.87} & 79.82{\scriptsize$\pm$0.95} \\ 
\; EfficientNet-B4 & 62.18{\scriptsize$\pm$4.99} & 63.25{\scriptsize$\pm$5.16} & 61.11{\scriptsize$\pm$2.18} & 59.62{\scriptsize$\pm$6.80} &  & 54.69{\scriptsize$\pm$5.15} & 52.17{\scriptsize$\pm$0.82} & 53.73{\scriptsize$\pm$6.10} & 52.49{\scriptsize$\pm$2.50} \\ 
\; ViT-B/16        & 77.14{\scriptsize$\pm$0.30} & 80.34{\scriptsize$\pm$0.60} & 85.04{\scriptsize$\pm$1.51} & 84.40{\scriptsize$\pm$1.32} &  & 78.22{\scriptsize$\pm$1.36} & 81.53{\scriptsize$\pm$1.96} & 90.23{\scriptsize$\pm$0.19} & 91.15{\scriptsize$\pm$0.71} \\ 
\; CLIP ViT-B/16   & 79.27{\scriptsize$\pm$0.30} & 81.20{\scriptsize$\pm$1.32} & 85.26{\scriptsize$\pm$0.91} & 83.55{\scriptsize$\pm$0.30} &  & 75.48{\scriptsize$\pm$0.31} & 84.40{\scriptsize$\pm$1.65} & 89.73{\scriptsize$\pm$0.47} & 86.52{\scriptsize$\pm$0.73} \\ 
\; EVA-02 ViT-B/16 & 76.07{\scriptsize$\pm$0.30} & 78.63{\scriptsize$\pm$0.30} & 80.98{\scriptsize$\pm$0.30} & 82.05{\scriptsize$\pm$0.00} &  & 76.68{\scriptsize$\pm$0.04} & 80.95{\scriptsize$\pm$0.11} & 86.86{\scriptsize$\pm$0.07} & 85.46{\scriptsize$\pm$0.03} \\ 
\; DINO ViT-B/16   & \textbf{83.33{\scriptsize$\pm$0.52}} & \textbf{85.68{\scriptsize$\pm$0.60}} & \textbf{88.89{\scriptsize$\pm$0.80}} & \underline{\textbf{88.68{\scriptsize$\pm$1.09}}} &  & \textbf{86.77{\scriptsize$\pm$0.44}} & \underline{\textbf{91.70{\scriptsize$\pm$0.27}}} & \textbf{93.25{\scriptsize$\pm$0.89}} & 93.43{\scriptsize$\pm$0.82} \\ 
\; SAM ViT-B/16    & 73.08{\scriptsize$\pm$0.00} & 73.08{\scriptsize$\pm$0.00} & 73.08{\scriptsize$\pm$0.00} & 73.08{\scriptsize$\pm$0.00} &  & 48.26{\scriptsize$\pm$2.40} & 61.65{\scriptsize$\pm$3.48} & 70.64{\scriptsize$\pm$0.64} & 73.75{\scriptsize$\pm$0.40} \\ 
\midrule
{\textsc{$k$-NN ($k = 11$)}} & & & & & \\       
\; VGG16           & 74.36 & 80.77 & 80.77 & 79.49 &  & - & - & - & - \\ 
\; AlexNet         & 81.41 & 81.41 & 82.69 & 82.69 &  & - & - & - & - \\  
\; ResNet-18       & 78.21 & 81.41 & \textbf{85.90} & 80.77 &  & - & - & - & - \\ 
\; DenseNet-121    & 78.85 & 75.64 & 83.33 & 84.62 &  & - & - & - & - \\ 
\; EfficientNet-B4 & \textbf{83.33} & 81.41 & 83.97 & 86.54 &  & - & - & - & - \\ 
\; ViT-B/16        & 75.64 & 79.49 & 83.97 & 81.41 &  & - & - & - & - \\ 
\; CLIP ViT-B/16   & 78.21 & 78.21 & 80.13 & 80.13 &  & - & - & - & - \\ 
\; EVA-02 ViT-B/16 & 75.64 & 82.69 & 82.69 & 81.41 &  & - & - & - & - \\ 
\; DINO ViT-B/16   & 78.21 & \textbf{86.54} & 85.26 & \textbf{87.18} &  & - & - & - & - \\ 
\; SAM ViT-B/16    & 74.36 & 82.69 & 77.56 & 78.85 &  & - & - & - & - \\ 
\bottomrule
\end{tabular}
\end{table}

\begin{table}[!htpb]
\setlength{\tabcolsep}{4.9pt}
\renewcommand{\arraystretch}{1.1}
\caption{Benchmark outcomes summarizing the mean and standard deviation of accuracy (ACC), for a fixed operating point of $0.5$ and area under the receiver operating characteristic curve (AUC) for the ChestMNIST dataset across all training scheme-model-image resolution combinations, derived from three independent random seeds. Notably, the \mbox{$k$-NN} algorithm, devoid of a training phase, remains unaffected by the stochasticity inherent in model training, thus reporting only the total ACC value without standard deviation. Moreover, owing to its direct utilization of embeddings and labels for classification, \mbox{$k$-NN} does not furnish a reliable AUC score. The overall best result across all training schemes, models, and resolutions is highlighted with a \colorbox{bg}{background color}; the best result per resolution across all training schemes and models is highlighted with \underline{underline}; and the best result per training scheme and resolution is highlighted in \textbf{bold}.}
\label{tab:chestmnist benchmark}
\centering
\begin{tabular}{lcccccccccc}
& \\
\toprule
\multicolumn{10}{c}{\textbf{ChestMNIST}} \\
\toprule
\multirow{2.5}{*}{Methods} & \multicolumn{4}{c}{Accuracy (ACC)} &  & \multicolumn{4}{c}{Area Under the ROC Curve (AUC)} \\
\cmidrule(r){2-5}\cmidrule(l){7-10}
& $28 \times 28$ & $64 \times 64$ & $128 \times 128$ & $224 \times 224$ &  & $28 \times 28$ & $64 \times 64$ & $128 \times 128$ & $224 \times 224$\\ 
\midrule
{\textsc{End-to-End}} & & & & & \\        
\; VGG16           & \underline{\textbf{94.79{\scriptsize$\pm$0.02}}} & \underline{\textbf{94.82{\scriptsize$\pm$0.01}}} & 94.79{\scriptsize$\pm$0.06} & 94.83{\scriptsize$\pm$0.01} &  & \underline{\textbf{76.16{\scriptsize$\pm$0.31}}} & 78.59{\scriptsize$\pm$0.11} & 80.07{\scriptsize$\pm$0.26} & 80.92{\scriptsize$\pm$0.11}\\ 
\; AlexNet         & 94.75{\scriptsize$\pm$0.01} & 94.77{\scriptsize$\pm$0.00} & \underline{\textbf{94.80{\scriptsize$\pm$0.02}}} & 94.79{\scriptsize$\pm$0.01} &  & 72.31{\scriptsize$\pm$0.20} & 74.75{\scriptsize$\pm$0.29} & 77.21{\scriptsize$\pm$0.32} & 78.87{\scriptsize$\pm$0.22}\\ 
\; ResNet-18       & 94.75{\scriptsize$\pm$0.01} & 94.77{\scriptsize$\pm$0.00} & 94.79{\scriptsize$\pm$0.02} & 94.79{\scriptsize$\pm$0.01} &  & 73.82{\scriptsize$\pm$0.12} & 75.59{\scriptsize$\pm$0.24} & 78.24{\scriptsize$\pm$0.32} & 78.96{\scriptsize$\pm$0.30}\\ 
\; DenseNet-121    & 94.77{\scriptsize$\pm$0.02} & 94.79{\scriptsize$\pm$0.02} & \underline{\textbf{94.80{\scriptsize$\pm$0.03}}} & \cellcolor{bg}\underline{\textbf{94.84{\scriptsize$\pm$0.03}}} &  & 75.93{\scriptsize$\pm$0.09} & \underline{\textbf{78.70{\scriptsize$\pm$0.38}}} & \underline{\textbf{81.21{\scriptsize$\pm$0.10}}} & \cellcolor{bg}\underline{\textbf{82.22{\scriptsize$\pm$0.14}}} \\ 
\; EfficientNet-B4 & 94.37{\scriptsize$\pm$0.32} & 94.31{\scriptsize$\pm$0.27} & 94.76{\scriptsize$\pm$0.02} & 94.76{\scriptsize$\pm$0.03} &  & 67.44{\scriptsize$\pm$1.87} & 71.63{\scriptsize$\pm$0.64} & 77.65{\scriptsize$\pm$0.35} & 78.25{\scriptsize$\pm$0.31}\\ 
\; ViT-B/16        & 94.74{\scriptsize$\pm$0.00} & 94.74{\scriptsize$\pm$0.03} & 94.78{\scriptsize$\pm$0.02} & 94.79{\scriptsize$\pm$0.05} &  & 73.74{\scriptsize$\pm$0.05} & 76.21{\scriptsize$\pm$0.09} & 78.05{\scriptsize$\pm$0.18} & 80.08{\scriptsize$\pm$0.14}\\ 
\; CLIP ViT-B/16   & 94.74{\scriptsize$\pm$0.01} & 94.75{\scriptsize$\pm$0.00} & 94.76{\scriptsize$\pm$0.01} & 94.74{\scriptsize$\pm$0.06} &  & 72.44{\scriptsize$\pm$0.35} & 73.20{\scriptsize$\pm$0.42} & 74.18{\scriptsize$\pm$0.23} & 77.77{\scriptsize$\pm$0.42}\\ 
\; EVA-02 ViT-B/16 & 94.73{\scriptsize$\pm$0.01} & 94.75{\scriptsize$\pm$0.01} & 94.76{\scriptsize$\pm$0.01} & 94.76{\scriptsize$\pm$0.01} &  & 72.44{\scriptsize$\pm$0.33} & 74.13{\scriptsize$\pm$0.28} & 75.23{\scriptsize$\pm$0.68} & 76.52{\scriptsize$\pm$0.40}\\ 
\; DINO ViT-B/16   & 94.74{\scriptsize$\pm$0.01} & 94.73{\scriptsize$\pm$0.03} & 94.76{\scriptsize$\pm$0.01} & 94.78{\scriptsize$\pm$0.02} &  & 73.23{\scriptsize$\pm$0.21} & 75.13{\scriptsize$\pm$0.27} & 76.81{\scriptsize$\pm$0.26} & 78.99{\scriptsize$\pm$0.22}\\ 
\; SAM ViT-B/16    & 94.73{\scriptsize$\pm$0.01} & 94.74{\scriptsize$\pm$0.01} & 94.73{\scriptsize$\pm$0.01} & 94.77{\scriptsize$\pm$0.00} &  & 72.68{\scriptsize$\pm$0.19} & 72.27{\scriptsize$\pm$0.72} & 74.13{\scriptsize$\pm$0.31} & 75.44{\scriptsize$\pm$0.23} \\ 
\midrule
{\textsc{Linear Probing}} & & & & & \\         
\; VGG16           & \textbf{94.75{\scriptsize$\pm$0.00}} & 94.75{\scriptsize$\pm$0.00} & 94.75{\scriptsize$\pm$0.00} & \textbf{94.76{\scriptsize$\pm$0.00}} &  & 69.04{\scriptsize$\pm$0.02} & 70.45{\scriptsize$\pm$0.03} & 73.69{\scriptsize$\pm$0.04} & 73.91{\scriptsize$\pm$0.06} \\ 
\; AlexNet         & 94.74{\scriptsize$\pm$0.00} & 94.74{\scriptsize$\pm$0.00} & 94.75{\scriptsize$\pm$0.00} & \textbf{94.76{\scriptsize$\pm$0.00}} &  & 61.61{\scriptsize$\pm$0.36} & 66.60{\scriptsize$\pm$0.31} & 71.49{\scriptsize$\pm$0.18} & 74.37{\scriptsize$\pm$0.04} \\
\; ResNet-18       & 94.74{\scriptsize$\pm$0.00} & 94.74{\scriptsize$\pm$0.00} & 94.74{\scriptsize$\pm$0.00} & 94.74{\scriptsize$\pm$0.00} &  & 61.51{\scriptsize$\pm$0.07} & 62.56{\scriptsize$\pm$0.36} & 66.63{\scriptsize$\pm$0.18} & 75.12{\scriptsize$\pm$0.01} \\ 
\; DenseNet-121    & 94.74{\scriptsize$\pm$0.00} & 94.74{\scriptsize$\pm$0.00} & 94.73{\scriptsize$\pm$0.00} & 94.75{\scriptsize$\pm$0.00} &  & 63.14{\scriptsize$\pm$0.07} & 68.16{\scriptsize$\pm$0.13} & 71.66{\scriptsize$\pm$0.04} & 76.26{\scriptsize$\pm$0.02} \\ 
\; EfficientNet-B4 & 94.68{\scriptsize$\pm$0.02} & 94.69{\scriptsize$\pm$0.03} & 94.72{\scriptsize$\pm$0.01} & 94.74{\scriptsize$\pm$0.00} &  & 55.93{\scriptsize$\pm$2.16} & 58.77{\scriptsize$\pm$4.36} & 68.20{\scriptsize$\pm$0.30} & 74.89{\scriptsize$\pm$0.01} \\ 
\; ViT-B/16        & 94.74{\scriptsize$\pm$0.00} & 94.73{\scriptsize$\pm$0.00} & 94.74{\scriptsize$\pm$0.00} & 94.73{\scriptsize$\pm$0.00} &  & 68.04{\scriptsize$\pm$0.03} & 71.37{\scriptsize$\pm$0.11} & 74.67{\scriptsize$\pm$0.18} & 76.09{\scriptsize$\pm$0.07} \\ 
\; CLIP ViT-B/16   & \textbf{94.75{\scriptsize$\pm$0.00}} & 94.72{\scriptsize$\pm$0.00} & 94.74{\scriptsize$\pm$0.00} & 94.75{\scriptsize$\pm$0.00} &  & 70.65{\scriptsize$\pm$0.03} & 71.22{\scriptsize$\pm$0.01} & 74.41{\scriptsize$\pm$0.15} & 76.29{\scriptsize$\pm$0.15} \\ 
\; EVA-02 ViT-B/16 & 94.74{\scriptsize$\pm$0.00} & 94.74{\scriptsize$\pm$0.00} & 94.73{\scriptsize$\pm$0.00} & 94.74{\scriptsize$\pm$0.00} &  & 68.87{\scriptsize$\pm$0.01} & 70.89{\scriptsize$\pm$0.01} & 70.65{\scriptsize$\pm$0.22} & 72.11{\scriptsize$\pm$0.22} \\ 
\; DINO ViT-B/16   & \textbf{94.75{\scriptsize$\pm$0.00}} & \textbf{94.76{\scriptsize$\pm$0.00}} & \textbf{94.76{\scriptsize$\pm$0.00}} & \textbf{94.76{\scriptsize$\pm$0.01}} &  & \textbf{71.69{\scriptsize$\pm$0.20}} & \textbf{74.41{\scriptsize$\pm$0.03}} & \textbf{77.90{\scriptsize$\pm$0.06}} & \textbf{78.88{\scriptsize$\pm$0.05}} \\ 
\; SAM ViT-B/16    & 94.74{\scriptsize$\pm$0.00} & 94.74{\scriptsize$\pm$0.00} & 94.74{\scriptsize$\pm$0.00} & 94.74{\scriptsize$\pm$0.00} &  & 60.42{\scriptsize$\pm$0.03} & 60.24{\scriptsize$\pm$0.01} & 61.11{\scriptsize$\pm$0.04} & 63.71{\scriptsize$\pm$0.02} \\ 
\midrule
{\textsc{$k$-NN ($k = 11$)}} & & & & & \\       
\; VGG16           & \textbf{94.67} & 94.66 & 94.66 & 94.67 &  & - & - & - & - \\ 
\; AlexNet         & 94.66 & 94.68 & 94.67 & 94.66 &  & - & - & - & - \\  
\; ResNet-18       & 94.66 & 94.68 & 94.66 & 94.66 &  & - & - & - & - \\ 
\; DenseNet-121    & 94.66 & 94.67 & 94.66 & 94.66 &  & - & - & - & - \\ 
\; EfficientNet-B4 & 94.65 & 94.67 & 94.68 & 94.67 &  & - & - & - & - \\ 
\; ViT-B/16        & 94.66 & 94.66 & \textbf{94.69} & \textbf{94.68} &  & - & - & - & - \\ 
\; CLIP ViT-B/16   & 94.66 & 94.66 & 94.67 & 94.67 &  & - & - & - & - \\ 
\; EVA-02 ViT-B/16 & 94.64 & 94.67 & 94.68 & 94.66 &  & - & - & - & - \\ 
\; DINO ViT-B/16   & 94.63 & 94.66 & 94.67 & 94.64 &  & - & - & - & - \\ 
\; SAM ViT-B/16    & 94.64 & \textbf{94.69} & \textbf{94.69} & \textbf{94.68} &  & - & - & - & - \\ 
\bottomrule
\end{tabular}
\end{table}

\begin{table}[!htpb]
\setlength{\tabcolsep}{4.9pt}
\renewcommand{\arraystretch}{1.1}
\caption{Benchmark outcomes summarizing the mean and standard deviation of accuracy (ACC), for a fixed operating point of $0.5$ and area under the receiver operating characteristic curve (AUC) for the DermaMNIST dataset across all training scheme-model-image resolution combinations, derived from three independent random seeds. Notably, the \mbox{$k$-NN} algorithm, devoid of a training phase, remains unaffected by the stochasticity inherent in model training, thus reporting only the total ACC value without standard deviation. Moreover, owing to its direct utilization of embeddings and labels for classification, \mbox{$k$-NN} does not furnish a reliable AUC score. The overall best result across all training schemes, models, and resolutions is highlighted with a \colorbox{bg}{background color}; the best result per resolution across all training schemes and models is highlighted with \underline{underline}; and the best result per training scheme and resolution is highlighted in \textbf{bold}.}
\label{tab:dermamnist benchmark}
\centering
\begin{tabular}{lcccccccccc}
& \\
\toprule
\multicolumn{10}{c}{\textbf{DermaMNIST}} \\
\toprule
\multirow{2.5}{*}{Methods} & \multicolumn{4}{c}{Accuracy (ACC)} &  & \multicolumn{4}{c}{Area Under the ROC Curve (AUC)} \\
\cmidrule(r){2-5}\cmidrule(l){7-10}
& $28 \times 28$ & $64 \times 64$ & $128 \times 128$ & $224 \times 224$ &  & $28 \times 28$ & $64 \times 64$ & $128 \times 128$ & $224 \times 224$\\ 
\midrule
{\textsc{End-to-End}} & & & & & \\        
\; VGG16           & \underline{\textbf{76.58{\scriptsize$\pm$0.26}}} & \textbf{79.53{\scriptsize$\pm$1.19}} & \underline{\textbf{81.80{\scriptsize$\pm$1.29}}} & 81.55{\scriptsize$\pm$1.89} &  & \underline{\textbf{92.69{\scriptsize$\pm$0.59}}} & \textbf{94.36{\scriptsize$\pm$0.40}} & 95.71{\scriptsize$\pm$0.46} & 95.51{\scriptsize$\pm$0.69} \\ 
\; AlexNet         & 76.13{\scriptsize$\pm$0.35} & 78.69{\scriptsize$\pm$0.18} & 80.42{\scriptsize$\pm$0.75} & 82.04{\scriptsize$\pm$1.19} &  & 92.23{\scriptsize$\pm$0.21} & 94.34{\scriptsize$\pm$0.06} & 95.13{\scriptsize$\pm$0.25} & 96.10{\scriptsize$\pm$0.09} \\ 
\; ResNet-18       & 73.38{\scriptsize$\pm$0.61} & 76.36{\scriptsize$\pm$0.22} & 79.19{\scriptsize$\pm$0.51} & 82.33{\scriptsize$\pm$0.36} &  & 88.35{\scriptsize$\pm$0.28} & 91.70{\scriptsize$\pm$0.76} & 93.46{\scriptsize$\pm$0.39} & 95.27{\scriptsize$\pm$0.27} \\ 
\; DenseNet-121    & 74.16{\scriptsize$\pm$0.65} & 76.16{\scriptsize$\pm$0.58} & 81.76{\scriptsize$\pm$0.25} & \cellcolor{bg}\underline{\textbf{84.74{\scriptsize$\pm$0.51}}} &  & 91.11{\scriptsize$\pm$0.61} & 93.05{\scriptsize$\pm$0.24} & 95.71{\scriptsize$\pm$0.19} & 96.26{\scriptsize$\pm$0.06} \\ 
\; EfficientNet-B4 & 68.74{\scriptsize$\pm$0.75} & 71.45{\scriptsize$\pm$0.10} & 73.83{\scriptsize$\pm$0.30} & 76.38{\scriptsize$\pm$1.29} &  & 83.87{\scriptsize$\pm$0.66} & 87.58{\scriptsize$\pm$0.27} & 89.32{\scriptsize$\pm$0.16} & 91.88{\scriptsize$\pm$0.85} \\ 
\; ViT-B/16        & 74.40{\scriptsize$\pm$1.51} & 77.01{\scriptsize$\pm$1.45} & 80.81{\scriptsize$\pm$0.89} & 82.31{\scriptsize$\pm$1.36} &  & 90.62{\scriptsize$\pm$2.21} & 93.77{\scriptsize$\pm$0.80} & \textbf{95.89{\scriptsize$\pm$0.30}} & 96.28{\scriptsize$\pm$0.67} \\ 
\; CLIP ViT-B/16   & 72.97{\scriptsize$\pm$0.15} & 72.44{\scriptsize$\pm$0.19} & 74.73{\scriptsize$\pm$0.42} & 75.31{\scriptsize$\pm$0.28} &  & 90.32{\scriptsize$\pm$0.12} & 91.55{\scriptsize$\pm$0.45} & 92.51{\scriptsize$\pm$0.14} & 92.59{\scriptsize$\pm$0.26} \\ 
\; EVA-02 ViT-B/16 & 73.48{\scriptsize$\pm$0.65} & 75.28{\scriptsize$\pm$0.98} & 76.41{\scriptsize$\pm$0.46} & 77.94{\scriptsize$\pm$0.29} &  & 90.17{\scriptsize$\pm$0.34} & 91.47{\scriptsize$\pm$0.34} & 92.63{\scriptsize$\pm$0.33} & 93.23{\scriptsize$\pm$0.33} \\ 
\; DINO ViT-B/16   & 74.40{\scriptsize$\pm$0.80} & 76.87{\scriptsize$\pm$0.12} & 79.22{\scriptsize$\pm$1.56} & 81.31{\scriptsize$\pm$1.05} &  & 91.60{\scriptsize$\pm$0.32} & 93.57{\scriptsize$\pm$0.17} & 95.34{\scriptsize$\pm$0.41} & \textbf{96.50{\scriptsize$\pm$0.51}} \\ 
\; SAM ViT-B/16    & 73.08{\scriptsize$\pm$0.65} & 74.68{\scriptsize$\pm$0.10} & 76.71{\scriptsize$\pm$1.06} & 77.42{\scriptsize$\pm$0.17} &  & 87.97{\scriptsize$\pm$0.65} & 88.33{\scriptsize$\pm$0.58} & 90.66{\scriptsize$\pm$1.36} & 92.60{\scriptsize$\pm$0.65} \\ 
\midrule
{\textsc{Linear Probing}} & & & & & \\         
\; VGG16           & 72.15{\scriptsize$\pm$0.05} & 73.77{\scriptsize$\pm$0.11} & 75.38{\scriptsize$\pm$0.16} & 75.99{\scriptsize$\pm$0.13} &  & 87.61{\scriptsize$\pm$0.01} & 88.71{\scriptsize$\pm$0.01} & 90.53{\scriptsize$\pm$0.01} & 92.03{\scriptsize$\pm$0.01} \\ 
\; AlexNet         & 72.40{\scriptsize$\pm$0.12} & 75.41{\scriptsize$\pm$0.11} & 77.11{\scriptsize$\pm$0.15} & 78.90{\scriptsize$\pm$0.23} &  & 88.77{\scriptsize$\pm$0.02} & 91.33{\scriptsize$\pm$0.04} & 92.81{\scriptsize$\pm$0.03} & 94.01{\scriptsize$\pm$0.06} \\
\; ResNet-18       & 67.83{\scriptsize$\pm$0.00} & 68.94{\scriptsize$\pm$0.10} & 70.57{\scriptsize$\pm$0.14} & 71.19{\scriptsize$\pm$0.12} &  & 84.09{\scriptsize$\pm$0.05} & 85.41{\scriptsize$\pm$0.08} & 86.44{\scriptsize$\pm$0.01} & 88.38{\scriptsize$\pm$0.02} \\ 
\; DenseNet-121    & 71.70{\scriptsize$\pm$0.06} & 74.26{\scriptsize$\pm$0.15} & 77.39{\scriptsize$\pm$0.12} & 77.06{\scriptsize$\pm$0.18} &  & 89.80{\scriptsize$\pm$0.03} & 90.94{\scriptsize$\pm$0.08} & 92.15{\scriptsize$\pm$0.05} & 93.18{\scriptsize$\pm$0.04} \\ 
\; EfficientNet-B4 & 69.99{\scriptsize$\pm$0.02} & 72.77{\scriptsize$\pm$0.04} & 72.97{\scriptsize$\pm$0.04} & 73.23{\scriptsize$\pm$0.06} &  & 84.83{\scriptsize$\pm$0.04} & 88.53{\scriptsize$\pm$0.04} & 89.70{\scriptsize$\pm$0.01} & 90.51{\scriptsize$\pm$0.01} \\ 
\; ViT-B/16        & 72.15{\scriptsize$\pm$0.15} & 77.64{\scriptsize$\pm$0.24} & 80.90{\scriptsize$\pm$0.25} & 82.01{\scriptsize$\pm$0.02} &  & 89.89{\scriptsize$\pm$0.04} & 93.58{\scriptsize$\pm$0.04} & 95.05{\scriptsize$\pm$0.02} & 95.88{\scriptsize$\pm$0.03} \\ 
\; CLIP ViT-B/16   & 74.15{\scriptsize$\pm$0.34} & 77.22{\scriptsize$\pm$0.08} & 80.28{\scriptsize$\pm$0.31} & 81.93{\scriptsize$\pm$0.31} &  & 90.08{\scriptsize$\pm$0.05} & 93.40{\scriptsize$\pm$0.05} & 94.89{\scriptsize$\pm$0.03} & 95.92{\scriptsize$\pm$0.06} \\ 
\; EVA-02 ViT-B/16 & 73.47{\scriptsize$\pm$0.12} & 75.54{\scriptsize$\pm$0.12} & 77.17{\scriptsize$\pm$0.10} & 79.29{\scriptsize$\pm$0.02} &  & 90.50{\scriptsize$\pm$0.05} & 92.76{\scriptsize$\pm$0.04} & 93.88{\scriptsize$\pm$0.05} & 94.69{\scriptsize$\pm$0.03} \\ 
\; DINO ViT-B/16   & \textbf{75.78{\scriptsize$\pm$0.17}} & \underline{\textbf{79.88{\scriptsize$\pm$0.68}}} & \textbf{81.65{\scriptsize$\pm$0.66}} & \textbf{84.42{\scriptsize$\pm$0.23}} &  & \textbf{91.87{\scriptsize$\pm$0.10}} & \underline{\textbf{95.40{\scriptsize$\pm$0.20}}} & \underline{\textbf{95.95{\scriptsize$\pm$0.17}}} & \cellcolor{bg}\underline{\textbf{96.83{\scriptsize$\pm$0.07}}} \\ 
\; SAM ViT-B/16    & 66.88{\scriptsize$\pm$0.00} & 66.88{\scriptsize$\pm$0.00} & 66.88{\scriptsize$\pm$0.00} & 66.88{\scriptsize$\pm$0.00} &  & 66.73{\scriptsize$\pm$0.88} & 70.87{\scriptsize$\pm$0.37} & 72.64{\scriptsize$\pm$0.46} & 69.34{\scriptsize$\pm$0.33} \\ 
\midrule
{\textsc{$k$-NN ($k = 11$)}} & & & & & \\       
\; VGG16           & 70.27 & 72.27 & 72.57 & 71.67 &  & - & - & - & - \\ 
\; AlexNet         & 70.62 & 73.92 & 74.16 & 74.66 &  & - & - & - & - \\  
\; ResNet-18       & 70.72 & 71.52 & 71.37 & 73.07 &  & - & - & - & - \\ 
\; DenseNet-121    & 69.33 & 70.62 & 72.67 & 73.17 &  & - & - & - & - \\ 
\; EfficientNet-B4 & 69.48 & 71.62 & 71.87 & 71.72 &  & - & - & - & - \\ 
\; ViT-B/16        & 69.43 & 70.87 & 72.62 & 75.21 &  & - & - & - & - \\ 
\; CLIP ViT-B/16   & 72.12 & 71.17 & 73.52 & 74.46 &  & - & - & - & - \\ 
\; EVA-02 ViT-B/16 & 73.22 & 73.77 & 74.11 & 75.61 &  & - & - & - & - \\ 
\; DINO ViT-B/16   & \textbf{73.97} & \textbf{75.91} & \textbf{76.51} & \textbf{78.35} &  & - & - & - & - \\ 
\; SAM ViT-B/16    & 69.73 & 70.42 & 70.22 & 68.38 &  & - & - & - & - \\ 
\bottomrule
\end{tabular}
\end{table}

\begin{table}[!htpb]
\setlength{\tabcolsep}{4.9pt}
\renewcommand{\arraystretch}{1.1}
\caption{Benchmark outcomes summarizing the mean and standard deviation of accuracy (ACC), for a fixed operating point of $0.5$ and area under the receiver operating characteristic curve (AUC) for the OctMNIST dataset across all training scheme-model-image resolution combinations, derived from three independent random seeds. Notably, the \mbox{$k$-NN} algorithm, devoid of a training phase, remains unaffected by the stochasticity inherent in model training, thus reporting only the total ACC value without standard deviation. Moreover, owing to its direct utilization of embeddings and labels for classification, \mbox{$k$-NN} does not furnish a reliable AUC score. The overall best result across all training schemes, models, and resolutions is highlighted with a \colorbox{bg}{background color}; the best result per resolution across all training schemes and models is highlighted with \underline{underline}; and the best result per training scheme and resolution is highlighted in \textbf{bold}.}
\label{tab:octmnist benchmark}
\centering
\begin{tabular}{lcccccccccc}
& \\
\toprule
\multicolumn{10}{c}{\textbf{OctMNIST}} \\
\toprule
\multirow{2.5}{*}{Methods} & \multicolumn{4}{c}{Accuracy (ACC)} &  & \multicolumn{4}{c}{Area Under the ROC Curve (AUC)} \\
\cmidrule(r){2-5}\cmidrule(l){7-10}
& $28 \times 28$ & $64 \times 64$ & $128 \times 128$ & $224 \times 224$ &  & $28 \times 28$ & $64 \times 64$ & $128 \times 128$ & $224 \times 224$\\ 
\midrule
{\textsc{End-to-End}} & & & & & \\        
\; VGG16           & \underline{\textbf{77.50{\scriptsize$\pm$2.86}}} & 81.93{\scriptsize$\pm$2.44} & \cellcolor{bg}\underline{\textbf{90.50{\scriptsize$\pm$0.85}}} & \underline{\textbf{90.30{\scriptsize$\pm$2.62}}} &  & \underline{\textbf{95.84{\scriptsize$\pm$1.00}}} & 97.50{\scriptsize$\pm$0.79} & 98.73{\scriptsize$\pm$0.26} & 99.17{\scriptsize$\pm$0.13} \\ 
\; AlexNet         & 66.40{\scriptsize$\pm$1.13} & 75.47{\scriptsize$\pm$2.28} & 80.57{\scriptsize$\pm$1.90} & 84.63{\scriptsize$\pm$1.73} &  & 90.96{\scriptsize$\pm$0.39} & 95.09{\scriptsize$\pm$0.76} & 97.92{\scriptsize$\pm$0.22} & 98.01{\scriptsize$\pm$0.48} \\ 
\; ResNet-18       & 69.07{\scriptsize$\pm$0.83} & 80.70{\scriptsize$\pm$2.01} & 84.03{\scriptsize$\pm$1.97} & 85.10{\scriptsize$\pm$1.12} &  & 92.10{\scriptsize$\pm$0.36} & 97.53{\scriptsize$\pm$0.26} & 97.77{\scriptsize$\pm$0.58} & 98.81{\scriptsize$\pm$0.28} \\ 
\; DenseNet-121    & 72.87{\scriptsize$\pm$2.17} & \underline{\textbf{84.40{\scriptsize$\pm$1.93}}} & 89.83{\scriptsize$\pm$3.65} & 86.63{\scriptsize$\pm$0.42} &  & 94.35{\scriptsize$\pm$0.83} & \underline{\textbf{98.00{\scriptsize$\pm$0.40}}} & 98.60{\scriptsize$\pm$0.61} & \cellcolor{bg}\underline{\textbf{99.26{\scriptsize$\pm$0.24}}} \\ 
\; EfficientNet-B4 & 59.93{\scriptsize$\pm$2.79} & 75.87{\scriptsize$\pm$1.55} & 80.73{\scriptsize$\pm$2.00} & 82.57{\scriptsize$\pm$4.67} &  & 88.65{\scriptsize$\pm$0.74} & 95.61{\scriptsize$\pm$0.08} & 98.09{\scriptsize$\pm$0.12} & 98.93{\scriptsize$\pm$0.24} \\ 
\; ViT-B/16        & 64.00{\scriptsize$\pm$0.33} & 80.23{\scriptsize$\pm$0.58} & 87.47{\scriptsize$\pm$2.21} & 90.13{\scriptsize$\pm$0.39} &  & 88.59{\scriptsize$\pm$0.47} & 96.51{\scriptsize$\pm$0.22} & \underline{\textbf{98.95{\scriptsize$\pm$0.22}}} & 99.12{\scriptsize$\pm$0.15} \\ 
\; CLIP ViT-B/16   & 61.53{\scriptsize$\pm$0.70} & 77.73{\scriptsize$\pm$1.78} & 83.43{\scriptsize$\pm$3.42} & 86.80{\scriptsize$\pm$3.07} &  & 87.46{\scriptsize$\pm$0.13} & 95.68{\scriptsize$\pm$0.03} & 98.11{\scriptsize$\pm$0.58} & 98.88{\scriptsize$\pm$0.41} \\ 
\; EVA-02 ViT-B/16 & 60.63{\scriptsize$\pm$0.98} & 73.00{\scriptsize$\pm$2.79} & 84.33{\scriptsize$\pm$1.32} & 87.43{\scriptsize$\pm$1.11} &  & 86.60{\scriptsize$\pm$0.94} & 94.24{\scriptsize$\pm$0.79} & 98.13{\scriptsize$\pm$0.63} & 98.93{\scriptsize$\pm$0.18} \\ 
\; DINO ViT-B/16   & 64.53{\scriptsize$\pm$1.03} & 78.40{\scriptsize$\pm$1.49} & 84.03{\scriptsize$\pm$1.27} & 85.07{\scriptsize$\pm$2.23} &  & 89.26{\scriptsize$\pm$0.70} & 96.39{\scriptsize$\pm$0.17} & 98.22{\scriptsize$\pm$0.51} & 98.57{\scriptsize$\pm$0.29} \\ 
\; SAM ViT-B/16    & 64.87{\scriptsize$\pm$2.38} & 80.07{\scriptsize$\pm$0.58} & 87.50{\scriptsize$\pm$1.19} & 87.30{\scriptsize$\pm$0.88} &  & 89.26{\scriptsize$\pm$0.24} & 96.83{\scriptsize$\pm$0.52} & 98.87{\scriptsize$\pm$0.32} & 99.19{\scriptsize$\pm$0.13} \\ 
\midrule
{\textsc{Linear Probing}} & & & & & \\         
\; VGG16           & 50.03{\scriptsize$\pm$0.05} & 58.90{\scriptsize$\pm$0.08} & 70.77{\scriptsize$\pm$0.25} & 67.30{\scriptsize$\pm$0.36} &  & 84.59{\scriptsize$\pm$0.01} & 89.79{\scriptsize$\pm$0.02} & 94.66{\scriptsize$\pm$0.01} & 95.81{\scriptsize$\pm$0.02} \\ 
\; AlexNet         & 47.07{\scriptsize$\pm$0.05} & 56.57{\scriptsize$\pm$0.19} & 62.50{\scriptsize$\pm$0.24} & 68.30{\scriptsize$\pm$0.29} &  & 82.24{\scriptsize$\pm$0.02} & 84.84{\scriptsize$\pm$0.06} & 90.71{\scriptsize$\pm$0.05} & 94.31{\scriptsize$\pm$0.09} \\
\; ResNet-18       & 46.73{\scriptsize$\pm$0.05} & 53.40{\scriptsize$\pm$0.08} & 68.60{\scriptsize$\pm$0.00} & 72.00{\scriptsize$\pm$0.20} &  & 83.04{\scriptsize$\pm$0.01} & 88.34{\scriptsize$\pm$0.02} & 96.39{\scriptsize$\pm$0.01} & 97.65{\scriptsize$\pm$0.01} \\ 
\; DenseNet-121    & 56.17{\scriptsize$\pm$0.12} & 66.07{\scriptsize$\pm$0.24} & 71.10{\scriptsize$\pm$0.33} & 78.47{\scriptsize$\pm$0.34} &  & 88.87{\scriptsize$\pm$0.01} & 94.37{\scriptsize$\pm$0.01} & 97.27{\scriptsize$\pm$0.01} & 98.72{\scriptsize$\pm$0.01} \\ 
\; EfficientNet-B4 & 54.17{\scriptsize$\pm$0.05} & 66.23{\scriptsize$\pm$0.05} & 72.17{\scriptsize$\pm$0.05} & 76.53{\scriptsize$\pm$0.05} &  & 88.94{\scriptsize$\pm$0.01} & 93.56{\scriptsize$\pm$0.00} & 96.44{\scriptsize$\pm$0.01} & 97.73{\scriptsize$\pm$0.00} \\ 
\; ViT-B/16        & 54.43{\scriptsize$\pm$0.05} & 66.80{\scriptsize$\pm$0.33} & 76.87{\scriptsize$\pm$0.12} & \textbf{83.57{\scriptsize$\pm$0.12}} &  & 86.31{\scriptsize$\pm$0.01} & 94.94{\scriptsize$\pm$0.01} & 97.22{\scriptsize$\pm$0.04} & \textbf{98.96{\scriptsize$\pm$0.03}} \\ 
\; CLIP ViT-B/16   & 58.20{\scriptsize$\pm$0.16} & 63.37{\scriptsize$\pm$0.26} & \textbf{77.23{\scriptsize$\pm$0.17}} & 81.13{\scriptsize$\pm$0.17} &  & 89.47{\scriptsize$\pm$0.01} & 93.00{\scriptsize$\pm$0.05} & 97.66{\scriptsize$\pm$0.02} & 98.66{\scriptsize$\pm$0.00} \\ 
\; EVA-02 ViT-B/16 & 53.33{\scriptsize$\pm$0.17} & 60.27{\scriptsize$\pm$0.05} & 63.87{\scriptsize$\pm$0.19} & 68.50{\scriptsize$\pm$0.24} &  & 87.44{\scriptsize$\pm$0.02} & 92.72{\scriptsize$\pm$0.02} & 95.95{\scriptsize$\pm$0.01} & 96.47{\scriptsize$\pm$0.01} \\ 
\; DINO ViT-B/16   & \textbf{62.63{\scriptsize$\pm$0.09}} & \textbf{71.20{\scriptsize$\pm$0.49}} & 75.47{\scriptsize$\pm$0.21} & 73.73{\scriptsize$\pm$0.99} &  & \textbf{92.30{\scriptsize$\pm$0.04}} & \textbf{95.99{\scriptsize$\pm$0.11}} & \textbf{97.83{\scriptsize$\pm$0.04}} & 98.35{\scriptsize$\pm$0.08} \\ 
\; SAM ViT-B/16    & 26.20{\scriptsize$\pm$0.00} & 28.87{\scriptsize$\pm$0.05} & 34.50{\scriptsize$\pm$0.00} & 40.80{\scriptsize$\pm$0.00} &  & 66.00{\scriptsize$\pm$0.05} & 72.17{\scriptsize$\pm$0.06} & 71.19{\scriptsize$\pm$0.03} & 80.36{\scriptsize$\pm$0.03} \\ 
\midrule
{\textsc{$k$-NN ($k = 11$)}} & & & & & \\       
\; VGG16           & 46.90 & 46.50 & 52.30 & 61.30 &  & - & - & - & - \\ 
\; AlexNet         & 42.60 & 49.90 & 50.60 & 54.60 &  & - & - & - & - \\  
\; ResNet-18       & 46.80 & 46.40 & 56.00 & 68.90 &  & - & - & - & - \\ 
\; DenseNet-121    & \textbf{49.20} & 50.30 & 55.30 & 63.20 &  & - & - & - & - \\ 
\; EfficientNet-B4 & 48.20 & 54.20 & 63.60 & 65.80 &  & - & - & - & - \\ 
\; ViT-B/16        & 46.70 & 48.70 & 58.90 & 65.70 &  & - & - & - & - \\ 
\; CLIP ViT-B/16   & 47.30 & 52.30 & 61.40 & 58.90 &  & - & - & - & - \\ 
\; EVA-02 ViT-B/16 & 46.00 & 51.20 & 55.50 & 58.90 &  & - & - & - & - \\ 
\; DINO ViT-B/16   & 46.50 & \textbf{61.10} & \textbf{72.20} & \textbf{74.10} &  & - & - & - & - \\ 
\; SAM ViT-B/16    & 39.30 & 39.10 & 40.50 & 44.00 &  & - & - & - & - \\ 
\bottomrule
\end{tabular}
\end{table}

\begin{table}[!htpb]
\setlength{\tabcolsep}{4.9pt}
\renewcommand{\arraystretch}{1.1}
\caption{Benchmark outcomes summarizing the mean and standard deviation of accuracy (ACC), for a fixed operating point of $0.5$ and area under the receiver operating characteristic curve (AUC) for the OrganAMNIST dataset across all training scheme-model-image resolution combinations, derived from three independent random seeds. Notably, the \mbox{$k$-NN} algorithm, devoid of a training phase, remains unaffected by the stochasticity inherent in model training, thus reporting only the total ACC value without standard deviation. Moreover, owing to its direct utilization of embeddings and labels for classification, \mbox{$k$-NN} does not furnish a reliable AUC score. The overall best result across all training schemes, models, and resolutions is highlighted with a \colorbox{bg}{background color}; the best result per resolution across all training schemes and models is highlighted with \underline{underline}; and the best result per training scheme and resolution is highlighted in \textbf{bold}.}
\label{tab:organamnist benchmark}
\centering
\begin{tabular}{lcccccccccc}
& \\
\toprule
\multicolumn{10}{c}{\textbf{OrganAMNIST}} \\
\toprule
\multirow{2.5}{*}{Methods} & \multicolumn{4}{c}{Accuracy (ACC)} &  & \multicolumn{4}{c}{Area Under the ROC Curve (AUC)} \\
\cmidrule(r){2-5}\cmidrule(l){7-10}
& $28 \times 28$ & $64 \times 64$ & $128 \times 128$ & $224 \times 224$ &  & $28 \times 28$ & $64 \times 64$ & $128 \times 128$ & $224 \times 224$\\ 
\midrule
{\textsc{End-to-End}} & & & & & \\   
\; VGG16           & \underline{\textbf{93.19{\scriptsize$\pm$0.49}}} & \underline{\textbf{96.44{\scriptsize$\pm$0.35}}} & 94.24{\scriptsize$\pm$1.94} & 94.61{\scriptsize$\pm$0.64} &  & 99.47{\scriptsize$\pm$0.05} & 99.79{\scriptsize$\pm$0.04} & 99.75{\scriptsize$\pm$0.12} & 99.75{\scriptsize$\pm$0.05} \\ 
\; AlexNet         & 90.87{\scriptsize$\pm$0.21} & 95.23{\scriptsize$\pm$0.38} & 95.34{\scriptsize$\pm$0.33} & 95.93{\scriptsize$\pm$0.29} &  & 99.37{\scriptsize$\pm$0.03} & 99.78{\scriptsize$\pm$0.06} & 99.79{\scriptsize$\pm$0.03} & 99.85{\scriptsize$\pm$0.04} \\ 
\; ResNet-18       & 92.04{\scriptsize$\pm$0.04} & 95.49{\scriptsize$\pm$0.12} & 96.21{\scriptsize$\pm$0.30} & 96.01{\scriptsize$\pm$0.11} &  & 99.30{\scriptsize$\pm$0.12} & 99.76{\scriptsize$\pm$0.05} & 99.84{\scriptsize$\pm$0.02} & 99.70{\scriptsize$\pm$0.02} \\ 
\; DenseNet-121    & 91.62{\scriptsize$\pm$0.64} & 96.16{\scriptsize$\pm$0.29} & \underline{\textbf{96.51{\scriptsize$\pm$0.14}}} & 96.72{\scriptsize$\pm$0.29} &  & \underline{\textbf{99.52{\scriptsize$\pm$0.02}}} & \underline{\textbf{99.84{\scriptsize$\pm$0.02}}} & \underline{\textbf{99.88{\scriptsize$\pm$0.01}}} & 99.84{\scriptsize$\pm$0.05} \\ 
\; EfficientNet-B4 & 85.81{\scriptsize$\pm$1.30} & 93.52{\scriptsize$\pm$0.67} & 95.58{\scriptsize$\pm$0.23} & 95.11{\scriptsize$\pm$0.27} &  & 98.76{\scriptsize$\pm$0.09} & 99.65{\scriptsize$\pm$0.08} & 99.85{\scriptsize$\pm$0.02} & 99.79{\scriptsize$\pm$0.01} \\ 
\; ViT-B/16        & 90.29{\scriptsize$\pm$0.62} & 94.78{\scriptsize$\pm$0.94} & 96.27{\scriptsize$\pm$0.55} & \cellcolor{bg}\underline{\textbf{96.93{\scriptsize$\pm$0.43}}} &  & 99.20{\scriptsize$\pm$0.07} & 99.80{\scriptsize$\pm$0.08} & 99.86{\scriptsize$\pm$0.03} & \cellcolor{bg}\underline{\textbf{99.91{\scriptsize$\pm$0.02}}} \\ 
\; CLIP ViT-B/16   & 88.10{\scriptsize$\pm$0.23} & 93.90{\scriptsize$\pm$0.34} & 94.83{\scriptsize$\pm$0.43} & 95.25{\scriptsize$\pm$0.39} &  & 98.97{\scriptsize$\pm$0.06} & 99.70{\scriptsize$\pm$0.06} & 99.81{\scriptsize$\pm$0.03} & 99.81{\scriptsize$\pm$0.05} \\ 
\; EVA-02 ViT-B/16 & 88.54{\scriptsize$\pm$0.47} & 93.66{\scriptsize$\pm$1.30} & 95.93{\scriptsize$\pm$0.56} & 96.12{\scriptsize$\pm$0.22} &  & 98.94{\scriptsize$\pm$0.14} & 99.57{\scriptsize$\pm$0.15} & 99.84{\scriptsize$\pm$0.03} & 99.88{\scriptsize$\pm$0.01} \\ 
\; DINO ViT-B/16   & 89.98{\scriptsize$\pm$0.05} & 94.99{\scriptsize$\pm$0.37} & 95.96{\scriptsize$\pm$0.57} & 96.13{\scriptsize$\pm$0.32} &  & 99.31{\scriptsize$\pm$0.05} & 99.81{\scriptsize$\pm$0.04} & 99.87{\scriptsize$\pm$0.03} & 99.90{\scriptsize$\pm$0.02} \\ 
\; SAM ViT-B/16    & 90.41{\scriptsize$\pm$0.32} & 94.42{\scriptsize$\pm$0.39} & 95.96{\scriptsize$\pm$0.29} & 95.60{\scriptsize$\pm$0.19} &  & 99.00{\scriptsize$\pm$0.08} & 99.59{\scriptsize$\pm$0.07} & 99.75{\scriptsize$\pm$0.03} & 99.78{\scriptsize$\pm$0.02} \\ 
\midrule
{\textsc{Linear Probing}} & & & & & \\
\; VGG16           & 79.36{\scriptsize$\pm$0.09} & 85.62{\scriptsize$\pm$0.37} & 89.09{\scriptsize$\pm$0.08} & 91.46{\scriptsize$\pm$0.04} &  & 97.49{\scriptsize$\pm$0.02} & 98.75{\scriptsize$\pm$0.06} & 99.25{\scriptsize$\pm$0.01} & 99.53{\scriptsize$\pm$0.00} \\ 
\; AlexNet         & 79.87{\scriptsize$\pm$0.50} & 90.03{\scriptsize$\pm$0.09} & 92.32{\scriptsize$\pm$0.16} & 93.32{\scriptsize$\pm$0.08} &  & 97.52{\scriptsize$\pm$0.08} & 99.36{\scriptsize$\pm$0.00} & 99.58{\scriptsize$\pm$0.02} & 99.67{\scriptsize$\pm$0.00} \\
\; ResNet-18       & 70.37{\scriptsize$\pm$0.04} & 84.38{\scriptsize$\pm$0.06} & 89.30{\scriptsize$\pm$0.06} & 90.20{\scriptsize$\pm$0.05} &  & 94.85{\scriptsize$\pm$0.00} & 98.43{\scriptsize$\pm$0.00} & 99.20{\scriptsize$\pm$0.00} & 99.35{\scriptsize$\pm$0.00} \\ 
\; DenseNet-121    & 81.99{\scriptsize$\pm$0.08} & 90.71{\scriptsize$\pm$0.04} & 92.73{\scriptsize$\pm$0.05} & 93.63{\scriptsize$\pm$0.10} &  & 98.03{\scriptsize$\pm$0.02} & 99.42{\scriptsize$\pm$0.01} & 99.67{\scriptsize$\pm$0.00} & 99.73{\scriptsize$\pm$0.01} \\ 
\; EfficientNet-B4 & 74.23{\scriptsize$\pm$0.02} & 86.98{\scriptsize$\pm$0.03} & 90.13{\scriptsize$\pm$0.15} & 90.79{\scriptsize$\pm$0.04} &  & 96.15{\scriptsize$\pm$0.00} & 98.95{\scriptsize$\pm$0.01} & 99.35{\scriptsize$\pm$0.01} & 99.38{\scriptsize$\pm$0.01} \\ 
\; ViT-B/16        & 81.45{\scriptsize$\pm$0.43} & 90.21{\scriptsize$\pm$0.15} & 92.14{\scriptsize$\pm$0.06} & 93.06{\scriptsize$\pm$0.36} &  & 97.49{\scriptsize$\pm$0.08} & 99.42{\scriptsize$\pm$0.02} & 99.59{\scriptsize$\pm$0.00} & 99.65{\scriptsize$\pm$0.03} \\ 
\; CLIP ViT-B/16   & 80.36{\scriptsize$\pm$0.06} & 88.20{\scriptsize$\pm$0.11} & 90.19{\scriptsize$\pm$0.12} & 90.96{\scriptsize$\pm$0.08} &  & 97.62{\scriptsize$\pm$0.01} & 99.19{\scriptsize$\pm$0.02} & 99.42{\scriptsize$\pm$0.01} & 99.45{\scriptsize$\pm$0.01} \\ 
\; EVA-02 ViT-B/16 & 81.68{\scriptsize$\pm$0.29} & 87.12{\scriptsize$\pm$0.05} & 88.50{\scriptsize$\pm$0.10} & 89.97{\scriptsize$\pm$0.20} &  & 97.92{\scriptsize$\pm$0.03} & 98.93{\scriptsize$\pm$0.01} & 99.18{\scriptsize$\pm$0.01} & 99.37{\scriptsize$\pm$0.03} \\ 
\; DINO ViT-B/16   & \textbf{89.74{\scriptsize$\pm$0.37}} & \textbf{93.91{\scriptsize$\pm$0.16}} & \textbf{94.97{\scriptsize$\pm$0.13}} & \textbf{94.96{\scriptsize$\pm$0.02}} &  & \textbf{99.27{\scriptsize$\pm$0.04}} & \textbf{99.74{\scriptsize$\pm$0.01}} & \textbf{99.79{\scriptsize$\pm$0.01}} & \textbf{99.78{\scriptsize$\pm$0.01}} \\ 
\; SAM ViT-B/16    & 22.54{\scriptsize$\pm$0.07} & 39.10{\scriptsize$\pm$0.25} & 61.65{\scriptsize$\pm$0.08} & 71.18{\scriptsize$\pm$0.05} &  & 79.07{\scriptsize$\pm$0.05} & 90.11{\scriptsize$\pm$0.02} & 94.50{\scriptsize$\pm$0.01} & 95.29{\scriptsize$\pm$0.01} \\ 
\midrule
{\textsc{$k$-NN ($k = 11$)}} & & & & & \\       
\; VGG16           & 70.42 & 80.40 & 82.90 & 84.55 &  & - & - & - & - \\ 
\; AlexNet         & 72.71 & 82.87 & 86.54 & 88.38 &  & - & - & - & - \\  
\; ResNet-18       & 69.68 & 81.25 & 86.08 & 86.69 &  & - & - & - & - \\ 
\; DenseNet-121    & 69.48 & 81.93 & 86.73 & 87.28 &  & - & - & - & - \\ 
\; EfficientNet-B4 & 69.92 & 81.16 & 83.19 & 82.15 &  & - & - & - & - \\ 
\; ViT-B/16        & 66.71 & 80.67 & 83.13 & 83.86 &  & - & - & - & - \\ 
\; CLIP ViT-B/16   & 66.65 & 79.68 & 81.29 & 82.77 &  & - & - & - & - \\ 
\; EVA-02 ViT-B/16 & 73.33 & 80.74 & 82.71 & 83.59 &  & - & - & - & - \\ 
\; DINO ViT-B/16   & \textbf{84.59} & \textbf{90.75} & \textbf{91.25} & \textbf{90.69} &  & - & - & - & - \\ 
\; SAM ViT-B/16    & 70.08 & 82.34 & 83.67 & 83.14 &  & - & - & - & - \\ 
\bottomrule
\end{tabular}
\end{table}

\begin{table}[!htpb]
\setlength{\tabcolsep}{4.9pt}
\renewcommand{\arraystretch}{1.1}
\caption{Benchmark outcomes summarizing the mean and standard deviation of accuracy (ACC), for a fixed operating point of $0.5$ and area under the receiver operating characteristic curve (AUC) for the OrganCMNIST dataset across all training scheme-model-image resolution combinations, derived from three independent random seeds. Notably, the \mbox{$k$-NN} algorithm, devoid of a training phase, remains unaffected by the stochasticity inherent in model training, thus reporting only the total ACC value without standard deviation. Moreover, owing to its direct utilization of embeddings and labels for classification, \mbox{$k$-NN} does not furnish a reliable AUC score. The overall best result across all training schemes, models, and resolutions is highlighted with a \colorbox{bg}{background color}; the best result per resolution across all training schemes and models is highlighted with \underline{underline}; and the best result per training scheme and resolution is highlighted in \textbf{bold}.}
\label{tab:Organcmnist benchmark}
\centering
\begin{tabular}{lcccccccccc}
& \\
\toprule
\multicolumn{10}{c}{\textbf{OrganCMNIST}} \\
\toprule
\multirow{2.5}{*}{Methods} & \multicolumn{4}{c}{Accuracy (ACC)} &  & \multicolumn{4}{c}{Area Under the ROC Curve (AUC)} \\
\cmidrule(r){2-5}\cmidrule(l){7-10}
& $28 \times 28$ & $64 \times 64$ & $128 \times 128$ & $224 \times 224$ &  & $28 \times 28$ & $64 \times 64$ & $128 \times 128$ & $224 \times 224$\\ 
\midrule
{\textsc{End-to-End}} & & & & & \\        
\; VGG16           & 91.35{\scriptsize$\pm$0.41} & 93.51{\scriptsize$\pm$0.72} & 92.54{\scriptsize$\pm$0.88} & 93.37{\scriptsize$\pm$0.34} &  & \underline{\textbf{99.32{\scriptsize$\pm$0.08}}} & 99.64{\scriptsize$\pm$0.05} & 99.61{\scriptsize$\pm$0.03} & 99.65{\scriptsize$\pm$0.02} \\ 
\; AlexNet         & 87.76{\scriptsize$\pm$1.73} & 92.82{\scriptsize$\pm$0.25} & 93.84{\scriptsize$\pm$0.20} & 93.17{\scriptsize$\pm$0.16} &  & 99.13{\scriptsize$\pm$0.16} & 99.61{\scriptsize$\pm$0.02} & 99.72{\scriptsize$\pm$0.03} & 99.71{\scriptsize$\pm$0.03} \\ 
\; ResNet-18       & 90.48{\scriptsize$\pm$0.42} & 93.25{\scriptsize$\pm$0.27} & 93.94{\scriptsize$\pm$0.31} & 93.20{\scriptsize$\pm$0.12} &  & 99.14{\scriptsize$\pm$0.03} & 99.60{\scriptsize$\pm$0.06} & 99.68{\scriptsize$\pm$0.02} & 99.61{\scriptsize$\pm$0.07} \\ 
\; DenseNet-121    & \underline{\textbf{91.52{\scriptsize$\pm$0.33}}} & \underline{\textbf{93.99{\scriptsize$\pm$0.32}}} & \cellcolor{bg}\underline{\textbf{94.42{\scriptsize$\pm$0.37}}} & 93.72{\scriptsize$\pm$0.52} &  & 99.27{\scriptsize$\pm$0.05} & \underline{\textbf{99.69{\scriptsize$\pm$0.02}}} & 99.65{\scriptsize$\pm$0.05} & 99.67{\scriptsize$\pm$0.02} \\ 
\; EfficientNet-B4 & 84.25{\scriptsize$\pm$1.79} & 89.15{\scriptsize$\pm$2.09} & 90.93{\scriptsize$\pm$0.70} & 90.60{\scriptsize$\pm$0.16} &  & 98.39{\scriptsize$\pm$0.17} & 99.17{\scriptsize$\pm$0.23} & 99.50{\scriptsize$\pm$0.05} & 99.46{\scriptsize$\pm$0.01} \\ 
\; ViT-B/16        & 89.34{\scriptsize$\pm$1.00} & 93.20{\scriptsize$\pm$0.53} & 93.34{\scriptsize$\pm$0.46} & \underline{\textbf{94.02{\scriptsize$\pm$0.45}}} &  & 99.13{\scriptsize$\pm$0.08} & 99.71{\scriptsize$\pm$0.02} & 99.72{\scriptsize$\pm$0.05} & \cellcolor{bg}\underline{\textbf{99.78{\scriptsize$\pm$0.04}}} \\ 
\; CLIP ViT-B/16   & 87.52{\scriptsize$\pm$0.25} & 90.88{\scriptsize$\pm$0.84} & 91.54{\scriptsize$\pm$1.63} & 92.39{\scriptsize$\pm$0.43} &  & 98.94{\scriptsize$\pm$0.03} & 99.40{\scriptsize$\pm$0.02} & 99.53{\scriptsize$\pm$0.15} & 99.61{\scriptsize$\pm$0.08} \\ 
\; EVA-02 ViT-B/16 & 88.77{\scriptsize$\pm$0.56} & 91.86{\scriptsize$\pm$0.59} & 93.93{\scriptsize$\pm$0.34} & \underline{\textbf{94.02{\scriptsize$\pm$0.84}}} &  & 98.77{\scriptsize$\pm$0.16} & 99.33{\scriptsize$\pm$0.08} & 99.61{\scriptsize$\pm$0.09} & 99.61{\scriptsize$\pm$0.05} \\ 
\; DINO ViT-B/16   & 89.51{\scriptsize$\pm$0.75} & 92.36{\scriptsize$\pm$0.09} & 93.78{\scriptsize$\pm$1.39} & 93.68{\scriptsize$\pm$0.14} &  & 99.24{\scriptsize$\pm$0.01} & 99.63{\scriptsize$\pm$0.02} & \underline{\textbf{99.75{\scriptsize$\pm$0.09}}} & \cellcolor{bg}\underline{\textbf{99.78{\scriptsize$\pm$0.02}}} \\ 
\; SAM ViT-B/16    & 89.04{\scriptsize$\pm$0.54} & 92.94{\scriptsize$\pm$0.05} & 92.12{\scriptsize$\pm$1.13} & 92.71{\scriptsize$\pm$0.51} &  & 98.63{\scriptsize$\pm$0.17} & 99.31{\scriptsize$\pm$0.09} & 99.33{\scriptsize$\pm$0.13} & 99.39{\scriptsize$\pm$0.10} \\ 
\midrule
{\textsc{Linear Probing}} & & & & & \\         
\; VGG16           & 75.62{\scriptsize$\pm$0.28} & 81.49{\scriptsize$\pm$0.21} & 84.30{\scriptsize$\pm$0.36} & 85.11{\scriptsize$\pm$0.11} &  & 96.67{\scriptsize$\pm$0.05} & 98.03{\scriptsize$\pm$0.03} & 98.58{\scriptsize$\pm$0.05} & 98.81{\scriptsize$\pm$0.01} \\ 
\; AlexNet         & 78.15{\scriptsize$\pm$0.50} & 87.39{\scriptsize$\pm$0.13} & 88.66{\scriptsize$\pm$0.12} & 89.74{\scriptsize$\pm$0.14} &  & 97.32{\scriptsize$\pm$0.10} & 98.97{\scriptsize$\pm$0.02} & 99.16{\scriptsize$\pm$0.01} & 99.20{\scriptsize$\pm$0.01} \\
\; ResNet-18       & 60.70{\scriptsize$\pm$0.02} & 75.86{\scriptsize$\pm$0.12} & 81.65{\scriptsize$\pm$0.03} & 82.21{\scriptsize$\pm$0.01} &  & 92.56{\scriptsize$\pm$0.01} & 97.04{\scriptsize$\pm$0.00} & 98.20{\scriptsize$\pm$0.00} & 98.33{\scriptsize$\pm$0.00} \\ 
\; DenseNet-121    & 76.22{\scriptsize$\pm$0.10} & 85.80{\scriptsize$\pm$0.09} & 87.70{\scriptsize$\pm$0.10} & 88.09{\scriptsize$\pm$0.13} &  & 97.01{\scriptsize$\pm$0.00} & 98.88{\scriptsize$\pm$0.00} & 99.14{\scriptsize$\pm$0.01} & 99.21{\scriptsize$\pm$0.01} \\ 
\; EfficientNet-B4 & 65.08{\scriptsize$\pm$0.03} & 78.61{\scriptsize$\pm$0.06} & 81.67{\scriptsize$\pm$0.07} & 83.85{\scriptsize$\pm$0.07} &  & 93.77{\scriptsize$\pm$0.00} & 97.70{\scriptsize$\pm$0.00} & 98.27{\scriptsize$\pm$0.00} & 98.59{\scriptsize$\pm$0.00} \\ 
\; ViT-B/16        & 77.36{\scriptsize$\pm$0.30} & 85.46{\scriptsize$\pm$0.14} & 90.82{\scriptsize$\pm$5.27} & 88.20{\scriptsize$\pm$0.05}  &  & 97.22{\scriptsize$\pm$0.04} & 98.92{\scriptsize$\pm$0.01} & 99.06{\scriptsize$\pm$0.02} & 99.19{\scriptsize$\pm$0.00} \\ 
\; CLIP ViT-B/16   & 76.78{\scriptsize$\pm$0.11} & 83.04{\scriptsize$\pm$0.20} & 83.93{\scriptsize$\pm$0.13} & 84.12{\scriptsize$\pm$0.19} &  & 96.93{\scriptsize$\pm$0.02} & 98.54{\scriptsize$\pm$0.03} & 98.68{\scriptsize$\pm$0.00} & 98.72{\scriptsize$\pm$0.03} \\ 
\; EVA-02 ViT-B/16 & 78.51{\scriptsize$\pm$0.04} & 81.63{\scriptsize$\pm$0.29} & 83.60{\scriptsize$\pm$0.34} & 83.85{\scriptsize$\pm$0.10} &  & 97.41{\scriptsize$\pm$0.00} & 98.36{\scriptsize$\pm$0.05} & 98.66{\scriptsize$\pm$0.04} & 98.66{\scriptsize$\pm$0.00} \\ 
\; DINO ViT-B/16   & \textbf{88.34{\scriptsize$\pm$0.06}} & \textbf{92.21{\scriptsize$\pm$0.12}} & \textbf{91.82{\scriptsize$\pm$0.22}} & \textbf{92.46{\scriptsize$\pm$1.70}} &  & \textbf{99.08{\scriptsize$\pm$0.01}} & \textbf{99.62{\scriptsize$\pm$0.01}} & \textbf{99.60{\scriptsize$\pm$0.02}} & \textbf{99.54{\scriptsize$\pm$0.01}} \\ 
\; SAM ViT-B/16    & 22.33{\scriptsize$\pm$0.00} & 22.33{\scriptsize$\pm$0.00} & 32.01{\scriptsize$\pm$0.11} & 58.60{\scriptsize$\pm$0.10} &  & 62.32{\scriptsize$\pm$3.37} & 87.56{\scriptsize$\pm$0.08} & 93.09{\scriptsize$\pm$0.02} & 93.31{\scriptsize$\pm$0.00} \\ 
\midrule
{\textsc{$k$-NN ($k = 11$)}} & & & & & \\       
\; VGG16           & 67.70 & 75.21 & 77.31 & 75.65 &  & - & - & - & - \\ 
\; AlexNet         & 72.47 & 80.22 & 81.86 & 83.00 &  & - & - & - & - \\  
\; ResNet-18       & 66.20 & 76.56 & 80.00 & 80.08 &  & - & - & - & - \\ 
\; DenseNet-121    & 62.97 & 73.03 & 78.44 & 79.99 &  & - & - & - & - \\ 
\; EfficientNet-B4 & 63.83 & 74.25 & 72.69 & 73.47 &  & - & - & - & - \\ 
\; ViT-B/16        & 64.74 & 72.59 & 76.06 & 78.25 &  & - & - & - & - \\ 
\; CLIP ViT-B/16   & 59.30 & 70.87 & 73.69 & 74.66 &  & - & - & - & - \\ 
\; EVA-02 ViT-B/16 & 72.48 & 75.23 & 76.64 & 77.57 &  & - & - & - & - \\ 
\; DINO ViT-B/16   & \textbf{83.63} & \textbf{87.52} & \textbf{87.13} & \textbf{86.00} &  & - & - & - & - \\ 
\; SAM ViT-B/16    & 71.15 & 82.05 & 84.07 & 83.80 &  & - & - & - & - \\ 
\bottomrule
\end{tabular}
\end{table}

\begin{table}[!htpb]
\setlength{\tabcolsep}{4.9pt}
\renewcommand{\arraystretch}{1.1}
\caption{Benchmark outcomes summarizing the mean and standard deviation of accuracy (ACC), for a fixed operating point of $0.5$ and area under the receiver operating characteristic curve (AUC) for the OrganSMNIST dataset across all training scheme-model-image resolution combinations, derived from three independent random seeds. Notably, the \mbox{$k$-NN} algorithm, devoid of a training phase, remains unaffected by the stochasticity inherent in model training, thus reporting only the total ACC value without standard deviation. Moreover, owing to its direct utilization of embeddings and labels for classification, \mbox{$k$-NN} does not furnish a reliable AUC score. The overall best result across all training schemes, models, and resolutions is highlighted with a \colorbox{bg}{background color}; the best result per resolution across all training schemes and models is highlighted with \underline{underline}; and the best result per training scheme and resolution is highlighted in \textbf{bold}.}
\label{tab:organsmnist benchmark}
\centering
\begin{tabular}{lcccccccccc}
& \\
\toprule
\multicolumn{10}{c}{\textbf{OrganSMNIST}} \\
\toprule
\multirow{2.5}{*}{Methods} & \multicolumn{4}{c}{Accuracy (ACC)} &  & \multicolumn{4}{c}{Area Under the ROC Curve (AUC)} \\
\cmidrule(r){2-5}\cmidrule(l){7-10}
& $28 \times 28$ & $64 \times 64$ & $128 \times 128$ & $224 \times 224$ &  & $28 \times 28$ & $64 \times 64$ & $128 \times 128$ & $224 \times 224$\\ 
\midrule
{\textsc{End-to-End}} & & & & & \\        
\; VGG16           & \underline{\textbf{78.87{\scriptsize$\pm$0.68}}} & 81.60{\scriptsize$\pm$0.21} & 82.36{\scriptsize$\pm$0.08} & 82.06{\scriptsize$\pm$0.48} &  & \underline{\textbf{97.64{\scriptsize$\pm$0.09}}} & 98.06{\scriptsize$\pm$0.24} & 98.11{\scriptsize$\pm$0.06} & 97.89{\scriptsize$\pm$0.18} \\ 
\; AlexNet         & 76.75{\scriptsize$\pm$0.44} & 80.58{\scriptsize$\pm$0.90} & 81.71{\scriptsize$\pm$0.17} & 81.56{\scriptsize$\pm$0.36} &  & 97.30{\scriptsize$\pm$0.13} & 97.87{\scriptsize$\pm$0.17} & 98.07{\scriptsize$\pm$0.06} & 98.17{\scriptsize$\pm$0.03} \\ 
\; ResNet-18       & 76.24{\scriptsize$\pm$0.36} & 80.34{\scriptsize$\pm$0.82} & 82.30{\scriptsize$\pm$0.29} & 81.24{\scriptsize$\pm$0.42} &  & 97.23{\scriptsize$\pm$0.15} & 97.91{\scriptsize$\pm$0.07} & 98.19{\scriptsize$\pm$0.13} & 97.86{\scriptsize$\pm$0.15} \\ 
\; DenseNet-121    & 77.70{\scriptsize$\pm$0.61} & \cellcolor{bg}\underline{\textbf{83.47{\scriptsize$\pm$0.24}}} & \underline{\textbf{83.18{\scriptsize$\pm$0.19}}} & 81.80{\scriptsize$\pm$0.66} &  & 97.17{\scriptsize$\pm$0.19} & 97.93{\scriptsize$\pm$0.05} & 97.94{\scriptsize$\pm$0.13} & 98.11{\scriptsize$\pm$0.16} \\ 
\; EfficientNet-B4 & 67.97{\scriptsize$\pm$0.79} & 76.01{\scriptsize$\pm$0.73} & 77.33{\scriptsize$\pm$0.38} & 76.37{\scriptsize$\pm$0.42} &  & 95.07{\scriptsize$\pm$0.20} & 97.17{\scriptsize$\pm$0.08} & 97.50{\scriptsize$\pm$0.04} & 97.33{\scriptsize$\pm$0.03} \\ 
\; ViT-B/16        & 76.45{\scriptsize$\pm$0.83} & 81.43{\scriptsize$\pm$1.28} & 82.94{\scriptsize$\pm$0.33} & \underline{\textbf{82.50{\scriptsize$\pm$0.60}}} &  & 97.12{\scriptsize$\pm$0.03} & \underline{\textbf{98.19{\scriptsize$\pm$0.19}}} & \cellcolor{bg}\underline{\textbf{98.50{\scriptsize$\pm$0.08}}} & 98.31{\scriptsize$\pm$0.18} \\ 
\; CLIP ViT-B/16   & 73.35{\scriptsize$\pm$0.31} & 78.08{\scriptsize$\pm$2.76} & 79.65{\scriptsize$\pm$1.72} & 78.69{\scriptsize$\pm$1.71} &  & 96.53{\scriptsize$\pm$0.19} & 97.56{\scriptsize$\pm$0.33} & 97.96{\scriptsize$\pm$0.15} & 97.76{\scriptsize$\pm$0.11} \\ 
\; EVA-02 ViT-B/16 & 71.91{\scriptsize$\pm$3.07} & 78.77{\scriptsize$\pm$1.99} & 81.19{\scriptsize$\pm$1.35} & 81.62{\scriptsize$\pm$0.80} &  & 95.80{\scriptsize$\pm$0.65} & 97.49{\scriptsize$\pm$0.24} & 98.02{\scriptsize$\pm$0.15} & 98.09{\scriptsize$\pm$0.11} \\ 
\; DINO ViT-B/16   & 76.10{\scriptsize$\pm$0.85} & 79.38{\scriptsize$\pm$2.39} & 82.72{\scriptsize$\pm$0.33} & 81.72{\scriptsize$\pm$0.57} &  & 97.09{\scriptsize$\pm$0.18} & 98.00{\scriptsize$\pm$0.12} & 98.33{\scriptsize$\pm$0.05} & \underline{\textbf{98.33{\scriptsize$\pm$0.05}}} \\ 
\; SAM ViT-B/16    & 76.71{\scriptsize$\pm$0.41} & 80.18{\scriptsize$\pm$0.19} & 80.61{\scriptsize$\pm$0.54} & 81.00{\scriptsize$\pm$0.63} &  & 96.11{\scriptsize$\pm$0.14} & 97.34{\scriptsize$\pm$0.15} & 97.48{\scriptsize$\pm$0.21} & 97.70{\scriptsize$\pm$0.10} \\ 
\midrule
{\textsc{Linear Probing}} & & & & & \\         
\; VGG16           & 63.33{\scriptsize$\pm$0.23} & 68.83{\scriptsize$\pm$0.28} & 72.08{\scriptsize$\pm$0.23} & 72.85{\scriptsize$\pm$0.07} &  & 93.70{\scriptsize$\pm$0.07} & 95.48{\scriptsize$\pm$0.06} & 96.37{\scriptsize$\pm$0.01} & 96.62{\scriptsize$\pm$0.00} \\ 
\; AlexNet         & 63.75{\scriptsize$\pm$0.77} & 72.14{\scriptsize$\pm$0.38} & 74.91{\scriptsize$\pm$0.04} & 75.65{\scriptsize$\pm$0.18} &  & 94.44{\scriptsize$\pm$0.16} & 96.73{\scriptsize$\pm$0.06} & 97.24{\scriptsize$\pm$0.00} & 97.12{\scriptsize$\pm$0.00} \\
\; ResNet-18       & 53.65{\scriptsize$\pm$0.06} & 64.95{\scriptsize$\pm$0.11} & 70.19{\scriptsize$\pm$0.04} & 70.12{\scriptsize$\pm$0.06} &  & 89.55{\scriptsize$\pm$0.00} & 94.31{\scriptsize$\pm$0.01} & 95.78{\scriptsize$\pm$0.01} & 95.84{\scriptsize$\pm$0.00} \\ 
\; DenseNet-121    & 66.28{\scriptsize$\pm$0.03} & 74.96{\scriptsize$\pm$0.06} & 76.10{\scriptsize$\pm$0.12} & 76.56{\scriptsize$\pm$0.19} &  & 94.86{\scriptsize$\pm$0.00} & 96.93{\scriptsize$\pm$0.02} & 97.32{\scriptsize$\pm$0.02} & 97.41{\scriptsize$\pm$0.02} \\ 
\; EfficientNet-B4 & 58.50{\scriptsize$\pm$0.05} & 70.32{\scriptsize$\pm$0.03} & 71.35{\scriptsize$\pm$0.03} & 72.06{\scriptsize$\pm$0.07} &  & 91.96{\scriptsize$\pm$0.00} & 95.72{\scriptsize$\pm$0.01} & 96.16{\scriptsize$\pm$0.01} & 96.44{\scriptsize$\pm$0.00} \\ 
\; ViT-B/16        & 65.44{\scriptsize$\pm$0.15} & 74.63{\scriptsize$\pm$0.28} & 77.41{\scriptsize$\pm$0.24} & 77.45{\scriptsize$\pm$0.22} &  & 94.46{\scriptsize$\pm$0.01} & 97.05{\scriptsize$\pm$0.02} & 97.48{\scriptsize$\pm$0.05} & 97.55{\scriptsize$\pm$0.01} \\ 
\; CLIP ViT-B/16   & 65.39{\scriptsize$\pm$0.07} & 72.82{\scriptsize$\pm$0.07} & 74.67{\scriptsize$\pm$0.15} & 74.77{\scriptsize$\pm$0.13} &  & 94.23{\scriptsize$\pm$0.01} & 96.61{\scriptsize$\pm$0.00} & 96.87{\scriptsize$\pm$0.04} & 96.93{\scriptsize$\pm$0.02} \\ 
\; EVA-02 ViT-B/16 & 66.42{\scriptsize$\pm$0.13} & 69.52{\scriptsize$\pm$0.05} & 73.08{\scriptsize$\pm$0.09} & 74.24{\scriptsize$\pm$0.09} &  & 95.01{\scriptsize$\pm$0.02} & 96.21{\scriptsize$\pm$0.02} & 96.54{\scriptsize$\pm$0.02} & 96.91{\scriptsize$\pm$0.01} \\ 
\; DINO ViT-B/16   & \textbf{74.25{\scriptsize$\pm$0.36}} & \textbf{78.70{\scriptsize$\pm$0.06}} & \textbf{80.79{\scriptsize$\pm$0.24}} & \textbf{79.27{\scriptsize$\pm$0.17}} &  & \textbf{97.03{\scriptsize$\pm$0.02}} & \textbf{97.92{\scriptsize$\pm$0.00}} & \textbf{98.09{\scriptsize$\pm$0.00}} & \textbf{97.89{\scriptsize$\pm$0.01}} \\ 
\; SAM ViT-B/16    & 23.54{\scriptsize$\pm$0.00} & 23.54{\scriptsize$\pm$0.00} & 35.64{\scriptsize$\pm$0.02} & 49.05{\scriptsize$\pm$0.03} &  & 55.66{\scriptsize$\pm$5.45} & 56.44{\scriptsize$\pm$3.28} & 89.56{\scriptsize$\pm$0.02} & 89.66{\scriptsize$\pm$0.02} \\ 
\midrule
{\textsc{$k$-NN ($k = 11$)}} & & & & & \\       
\; VGG16           & 55.25 & 63.80 & 64.98 & 65.80 &  & - & - & - & - \\ 
\; AlexNet         & 60.67 & 66.50 & 68.29 & 68.97 &  & - & - & - & - \\  
\; ResNet-18       & 56.67 & 65.83 & 70.26 & 70.08 &  & - & - & - & - \\ 
\; DenseNet-121    & 55.61 & 64.96 & 67.96 & 69.56 &  & - & - & - & - \\ 
\; EfficientNet-B4 & 55.84 & 66.22 & 64.81 & 63.92 &  & - & - & - & - \\ 
\; ViT-B/16        & 52.02 & 66.04 & 69.30 & 69.81 &  & - & - & - & - \\ 
\; CLIP ViT-B/16   & 50.91 & 64.35 & 66.65 & 67.27 &  & - & - & - & - \\ 
\; EVA-02 ViT-B/16 & 60.61 & 65.29 & 65.62 & 68.81 &  & - & - & - & - \\ 
\; DINO ViT-B/16   & \textbf{70.69} & \textbf{76.64} & \textbf{77.70} & \textbf{74.82} &  & - & - & - & - \\ 
\; SAM ViT-B/16    & 56.72 & 67.13 & 70.62 & 69.77 &  & - & - & - & - \\ 
\bottomrule
\end{tabular}
\end{table}

\begin{table}[!htpb]
\setlength{\tabcolsep}{4.9pt}
\renewcommand{\arraystretch}{1.1}
\caption{Benchmark outcomes summarizing the mean and standard deviation of accuracy (ACC), for a fixed operating point of $0.5$ and area under the receiver operating characteristic curve (AUC) for the PathMNIST dataset across all training scheme-model-image resolution combinations, derived from three independent random seeds. Notably, the \mbox{$k$-NN} algorithm, devoid of a training phase, remains unaffected by the stochasticity inherent in model training, thus reporting only the total ACC value without standard deviation. Moreover, owing to its direct utilization of embeddings and labels for classification, \mbox{$k$-NN} does not furnish a reliable AUC score. The overall best result across all training schemes, models, and resolutions is highlighted with a \colorbox{bg}{background color}; the best result per resolution across all training schemes and models is highlighted with \underline{underline}; and the best result per training scheme and resolution is highlighted in \textbf{bold}.}
\label{tab:pathmnist benchmark}
\centering
\begin{tabular}{lcccccccccc}
& \\
\toprule
\multicolumn{10}{c}{\textbf{PathMNIST}} \\
\toprule
\multirow{2.5}{*}{Methods} & \multicolumn{4}{c}{Accuracy (ACC)} &  & \multicolumn{4}{c}{Area Under the ROC Curve (AUC)} \\
\cmidrule(r){2-5}\cmidrule(l){7-10}
& $28 \times 28$ & $64 \times 64$ & $128 \times 128$ & $224 \times 224$ &  & $28 \times 28$ & $64 \times 64$ & $128 \times 128$ & $224 \times 224$\\ 
\midrule
{\textsc{End-to-End}} & & & & & \\        
\; VGG16           & \underline{\textbf{88.93{\scriptsize$\pm$0.53}}} & \textbf{93.83{\scriptsize$\pm$1.12}} & 94.85{\scriptsize$\pm$0.61} & 94.73{\scriptsize$\pm$0.31} &  & \underline{\textbf{98.57{\scriptsize$\pm$0.14}}} & 99.34{\scriptsize$\pm$0.12} & 99.42{\scriptsize$\pm$0.18} & 99.33{\scriptsize$\pm$0.06} \\ 
\; AlexNet         & 80.75{\scriptsize$\pm$1.63} & 89.00{\scriptsize$\pm$1.35} & 93.00{\scriptsize$\pm$0.65} & 94.19{\scriptsize$\pm$0.99} &  & 96.67{\scriptsize$\pm$0.41} & 98.78{\scriptsize$\pm$0.17} & 99.35{\scriptsize$\pm$0.04} & 99.41{\scriptsize$\pm$0.11} \\ 
\; ResNet-18       & 85.54{\scriptsize$\pm$0.87} & 93.36{\scriptsize$\pm$0.65} & 95.27{\scriptsize$\pm$0.23} & 93.82{\scriptsize$\pm$1.19} &  & 98.24{\scriptsize$\pm$0.14} & 99.40{\scriptsize$\pm$0.11} & 99.62{\scriptsize$\pm$0.10} & 99.30{\scriptsize$\pm$0.15} \\ 
\; DenseNet-121    & 85.26{\scriptsize$\pm$0.36} & 92.34{\scriptsize$\pm$0.84} & 94.69{\scriptsize$\pm$0.31} & 95.74{\scriptsize$\pm$0.58} &  & 98.18{\scriptsize$\pm$0.14} & \textbf{99.43{\scriptsize$\pm$0.06}} & \underline{\textbf{99.70{\scriptsize$\pm$0.06}}} & \cellcolor{bg}\underline{\textbf{99.79{\scriptsize$\pm$0.06}}} \\ 
\; EfficientNet-B4 & 76.35{\scriptsize$\pm$3.04} & 88.39{\scriptsize$\pm$1.94} & 94.20{\scriptsize$\pm$0.64} & 92.60{\scriptsize$\pm$0.57} &  & 96.27{\scriptsize$\pm$0.61} & 98.90{\scriptsize$\pm$0.25} & 99.57{\scriptsize$\pm$0.02} & 99.45{\scriptsize$\pm$0.13} \\ 
\; ViT-B/16        & 82.82{\scriptsize$\pm$0.29} & 91.97{\scriptsize$\pm$0.39} & 94.61{\scriptsize$\pm$0.90} & 95.82{\scriptsize$\pm$0.25} &  & 97.81{\scriptsize$\pm$0.31} & 99.28{\scriptsize$\pm$0.11} & 99.67{\scriptsize$\pm$0.08} & 99.64{\scriptsize$\pm$0.10} \\ 
\; CLIP ViT-B/16   & 81.41{\scriptsize$\pm$0.75} & 87.47{\scriptsize$\pm$1.96} & 92.71{\scriptsize$\pm$0.34} & 92.51{\scriptsize$\pm$1.56} &  & 97.79{\scriptsize$\pm$0.22} & 98.79{\scriptsize$\pm$0.13} & 99.49{\scriptsize$\pm$0.01} & 99.47{\scriptsize$\pm$0.11} \\ 
\; EVA-02 ViT-B/16 & 82.33{\scriptsize$\pm$1.71} & 90.54{\scriptsize$\pm$1.68} & 94.88{\scriptsize$\pm$1.02} & 95.97{\scriptsize$\pm$0.85} &  & 97.69{\scriptsize$\pm$0.16} & 99.17{\scriptsize$\pm$0.19} & 99.69{\scriptsize$\pm$0.10} & 99.75{\scriptsize$\pm$0.10} \\ 
\; DINO ViT-B/16   & 82.23{\scriptsize$\pm$1.32} & 90.29{\scriptsize$\pm$1.80} & 94.24{\scriptsize$\pm$0.23} & 94.33{\scriptsize$\pm$0.90} &  & 97.56{\scriptsize$\pm$0.38} & 99.13{\scriptsize$\pm$0.07} & 99.68{\scriptsize$\pm$0.03} & 99.54{\scriptsize$\pm$0.22} \\ 
\; SAM ViT-B/16    & 84.45{\scriptsize$\pm$0.53} & 91.82{\scriptsize$\pm$1.60} & \underline{\textbf{95.67{\scriptsize$\pm$1.27}}} & \textbf{96.07{\scriptsize$\pm$0.41}} &  & 98.10{\scriptsize$\pm$0.23} & 99.37{\scriptsize$\pm$0.10} & 99.66{\scriptsize$\pm$0.02} & 99.76{\scriptsize$\pm$0.05} \\ 
\midrule
{\textsc{Linear Probing}} & & & & & \\         
\; VGG16           & 80.04{\scriptsize$\pm$0.01} & 84.88{\scriptsize$\pm$0.05} & 87.00{\scriptsize$\pm$0.04} & 87.84{\scriptsize$\pm$0.06} &  & 96.79{\scriptsize$\pm$0.00} & 98.21{\scriptsize$\pm$0.00} & 98.50{\scriptsize$\pm$0.00} & 98.76{\scriptsize$\pm$0.01} \\ 
\; AlexNet         & 76.95{\scriptsize$\pm$0.04} & 81.48{\scriptsize$\pm$0.06} & 86.86{\scriptsize$\pm$0.14} & 88.21{\scriptsize$\pm$0.14} &  & 95.33{\scriptsize$\pm$0.00} & 96.87{\scriptsize$\pm$0.02} & 98.59{\scriptsize$\pm$0.01} & 98.79{\scriptsize$\pm$0.03} \\
\; ResNet-18       & 73.53{\scriptsize$\pm$0.01} & 84.42{\scriptsize$\pm$0.02} & 88.12{\scriptsize$\pm$0.02} & 88.94{\scriptsize$\pm$0.02} &  & 95.42{\scriptsize$\pm$0.00} & 98.05{\scriptsize$\pm$0.00} & 98.57{\scriptsize$\pm$0.00} & 98.88{\scriptsize$\pm$0.00} \\ 
\; DenseNet-121    & 81.82{\scriptsize$\pm$0.01} & 90.03{\scriptsize$\pm$0.05} & 92.02{\scriptsize$\pm$0.02} & 91.28{\scriptsize$\pm$0.06} &  & 97.80{\scriptsize$\pm$0.01} & 99.11{\scriptsize$\pm$0.00} & 99.40{\scriptsize$\pm$0.00} & 99.21{\scriptsize$\pm$0.01} \\ 
\; EfficientNet-B4 & 80.51{\scriptsize$\pm$0.03} & 86.45{\scriptsize$\pm$0.02} & 87.62{\scriptsize$\pm$0.02} & 89.67{\scriptsize$\pm$0.02} &  & 97.33{\scriptsize$\pm$0.00} & 98.74{\scriptsize$\pm$0.00} & 98.84{\scriptsize$\pm$0.00} & 98.88{\scriptsize$\pm$0.00} \\ 
\; ViT-B/16        & 83.90{\scriptsize$\pm$0.04} & 91.43{\scriptsize$\pm$0.02} & 92.70{\scriptsize$\pm$0.02} & 93.54{\scriptsize$\pm$0.08} &  & 97.90{\scriptsize$\pm$0.00} & 99.33{\scriptsize$\pm$0.00} & 99.50{\scriptsize$\pm$0.00} & 99.63{\scriptsize$\pm$0.01} \\ 
\; CLIP ViT-B/16   & 83.78{\scriptsize$\pm$0.02} & 90.83{\scriptsize$\pm$0.08} & 91.12{\scriptsize$\pm$0.17} & 91.82{\scriptsize$\pm$0.07} &  & \textbf{98.01{\scriptsize$\pm$0.00}} & 99.17{\scriptsize$\pm$0.00} & 99.33{\scriptsize$\pm$0.00} & 99.32{\scriptsize$\pm$0.01} \\ 
\; EVA-02 ViT-B/16 & 82.53{\scriptsize$\pm$0.05} & 90.07{\scriptsize$\pm$0.04} & 91.73{\scriptsize$\pm$0.03} & 87.37{\scriptsize$\pm$0.03} &  & 97.76{\scriptsize$\pm$0.00} & 99.03{\scriptsize$\pm$0.00} & 99.34{\scriptsize$\pm$0.00} & 99.02{\scriptsize$\pm$0.00} \\ 
\; DINO ViT-B/16   & \textbf{85.05{\scriptsize$\pm$0.11}} & \underline{\textbf{93.90{\scriptsize$\pm$0.11}}} & \textbf{94.27{\scriptsize$\pm$0.07}} & \cellcolor{bg}\underline{\textbf{96.14{\scriptsize$\pm$0.05}}} &  & 97.93{\scriptsize$\pm$0.02} & \underline{\textbf{99.59{\scriptsize$\pm$0.01}}} & \textbf{99.64{\scriptsize$\pm$0.01}} & \textbf{99.75{\scriptsize$\pm$0.00}} \\ 
\; SAM ViT-B/16    & 31.28{\scriptsize$\pm$0.05} & 56.31{\scriptsize$\pm$0.01} & 64.34{\scriptsize$\pm$0.05} & 75.75{\scriptsize$\pm$0.01} &  & 77.04{\scriptsize$\pm$0.02} & 88.94{\scriptsize$\pm$0.01} & 91.06{\scriptsize$\pm$0.00} & 96.37{\scriptsize$\pm$0.00} \\ 
\midrule
{\textsc{$k$-NN ($k = 11$)}} & & & & & \\       
\; VGG16           & 70.32 & 76.82 & 78.84 & 82.19 &  & - & - & - & - \\ 
\; AlexNet         & 71.52 & 74.55 & 81.23 & 84.21 &  & - & - & - & - \\  
\; ResNet-18       & 68.89 & 79.25 & 81.85 & 83.47 &  & - & - & - & - \\ 
\; DenseNet-121    & 72.90 & 82.16 & 86.16 & 85.86 &  & - & - & - & - \\ 
\; EfficientNet-B4 & 73.84 & 80.45 & 81.25 & 80.45 &  & - & - & - & - \\ 
\; ViT-B/16        & 71.96 & 81.50 & 86.57 & 88.04 &  & - & - & - & - \\ 
\; CLIP ViT-B/16   & 73.48 & 83.02 & 85.58 & 86.49 &  & - & - & - & - \\ 
\; EVA-02 ViT-B/16 & 76.17 & 83.58 & 86.49 & 79.94 &  & - & - & - & - \\ 
\; DINO ViT-B/16   & \textbf{80.54} & \textbf{90.39} & \textbf{93.72} & \textbf{94.32} &  & - & - & - & - \\ 
\; SAM ViT-B/16    & 63.91 & 76.99 & 78.02 & 77.28 &  & - & - & - & - \\ 
\bottomrule
\end{tabular}
\end{table}

\begin{table}[!htpb]
\setlength{\tabcolsep}{4.9pt}
\renewcommand{\arraystretch}{1.1}
\caption{Benchmark outcomes summarizing the mean and standard deviation of accuracy (ACC), for a fixed operating point of $0.5$ and area under the receiver operating characteristic curve (AUC) for the PneumoniaMNIST dataset across all training scheme-model-image resolution combinations, derived from three independent random seeds. Notably, the \mbox{$k$-NN} algorithm, devoid of a training phase, remains unaffected by the stochasticity inherent in model training, thus reporting only the total ACC value without standard deviation. Moreover, owing to its direct utilization of embeddings and labels for classification, \mbox{$k$-NN} does not furnish a reliable AUC score. The overall best result across all training schemes, models, and resolutions is highlighted with a \colorbox{bg}{background color}; the best result per resolution across all training schemes and models is highlighted with \underline{underline}; and the best result per training scheme and resolution is highlighted in \textbf{bold}.}
\label{tab:pneumoniamnist benchmark}
\centering
\begin{tabular}{lcccccccccc}
& \\
\toprule
\multicolumn{10}{c}{\textbf{PneumoniaMNIST}} \\
\toprule
\multirow{2.5}{*}{Methods} & \multicolumn{4}{c}{Accuracy (ACC)} &  & \multicolumn{4}{c}{Area Under the ROC Curve (AUC)} \\
\cmidrule(r){2-5}\cmidrule(l){7-10}
& $28 \times 28$ & $64 \times 64$ & $128 \times 128$ & $224 \times 224$ &  & $28 \times 28$ & $64 \times 64$ & $128 \times 128$ & $224 \times 224$\\ 
\midrule
{\textsc{End-to-End}} & & & & & \\        
\; VGG16           & 84.78{\scriptsize$\pm$2.55} & \underline{\textbf{89.26{\scriptsize$\pm$0.39}}} & 87.13{\scriptsize$\pm$1.93} & 87.39{\scriptsize$\pm$1.00} &  & \textbf{97.04{\scriptsize$\pm$0.54}} & \underline{\textbf{98.69{\scriptsize$\pm$0.16}}} & 98.55{\scriptsize$\pm$0.44} & \textbf{98.47{\scriptsize$\pm$0.16}} \\ 
\; AlexNet         & \textbf{85.10{\scriptsize$\pm$0.73}} & 88.03{\scriptsize$\pm$1.64} & 87.18{\scriptsize$\pm$1.67} & 87.23{\scriptsize$\pm$0.77} &  & 95.88{\scriptsize$\pm$0.42} & 96.71{\scriptsize$\pm$0.64} & 97.85{\scriptsize$\pm$0.32} & 98.25{\scriptsize$\pm$0.13} \\ 
\; ResNet-18       & 83.12{\scriptsize$\pm$1.10} & 86.00{\scriptsize$\pm$1.44} & 89.85{\scriptsize$\pm$1.37} & \textbf{91.13{\scriptsize$\pm$0.96}} &  & 95.01{\scriptsize$\pm$0.30} & 94.34{\scriptsize$\pm$0.71} & 97.91{\scriptsize$\pm$0.23} & 98.04{\scriptsize$\pm$0.30} \\ 
\; DenseNet-121    & 83.55{\scriptsize$\pm$2.21} & 85.42{\scriptsize$\pm$1.36} & 89.80{\scriptsize$\pm$0.79} & 88.94{\scriptsize$\pm$2.30} &  & 96.06{\scriptsize$\pm$0.24} & 97.19{\scriptsize$\pm$0.29} & \textbf{98.75{\scriptsize$\pm$0.23}} & 97.41{\scriptsize$\pm$0.80} \\ 
\; EfficientNet-B4 & 79.01{\scriptsize$\pm$1.44} & 82.69{\scriptsize$\pm$0.13} & 85.15{\scriptsize$\pm$0.93} & 87.13{\scriptsize$\pm$1.14} &  & 90.69{\scriptsize$\pm$0.46} & 95.34{\scriptsize$\pm$0.52} & 96.91{\scriptsize$\pm$0.21} & 96.71{\scriptsize$\pm$0.72} \\ 
\; ViT-B/16        & 83.65{\scriptsize$\pm$2.77} & 85.84{\scriptsize$\pm$1.32} & 83.76{\scriptsize$\pm$2.17} & 86.11{\scriptsize$\pm$3.57} &  & 95.60{\scriptsize$\pm$0.27} & 96.41{\scriptsize$\pm$0.10} & 96.81{\scriptsize$\pm$0.73} & 96.33{\scriptsize$\pm$0.51} \\ 
\; CLIP ViT-B/16   & 84.56{\scriptsize$\pm$1.96} & 84.62{\scriptsize$\pm$1.04} & 84.62{\scriptsize$\pm$1.73} & 83.81{\scriptsize$\pm$1.83} &  & 94.66{\scriptsize$\pm$0.36} & 94.63{\scriptsize$\pm$0.19} & 94.61{\scriptsize$\pm$0.75} & 95.59{\scriptsize$\pm$1.30} \\ 
\; EVA-02 ViT-B/16 & 85.04{\scriptsize$\pm$1.52} & 86.59{\scriptsize$\pm$1.14} & 83.17{\scriptsize$\pm$0.32} & 82.75{\scriptsize$\pm$2.53} &  & 93.88{\scriptsize$\pm$1.04} & 94.99{\scriptsize$\pm$1.14} & 93.81{\scriptsize$\pm$1.21} & 93.54{\scriptsize$\pm$1.82} \\ 
\; DINO ViT-B/16   & 84.13{\scriptsize$\pm$2.06} & 85.04{\scriptsize$\pm$2.16} & \textbf{90.06{\scriptsize$\pm$0.65}} & 84.08{\scriptsize$\pm$2.13} &  & 95.52{\scriptsize$\pm$0.62} & 96.00{\scriptsize$\pm$0.28} & 97.96{\scriptsize$\pm$0.71} & 96.67{\scriptsize$\pm$0.95} \\ 
\; SAM ViT-B/16    & 83.71{\scriptsize$\pm$2.08} & 86.65{\scriptsize$\pm$0.98} & 86.16{\scriptsize$\pm$2.69} & 83.81{\scriptsize$\pm$2.85} &  & 91.94{\scriptsize$\pm$1.45} & 93.68{\scriptsize$\pm$2.74} & 95.88{\scriptsize$\pm$1.61} & 94.53{\scriptsize$\pm$2.71} \\ 
\midrule
{\textsc{Linear Probing}} & & & & & \\         
\; VGG16           & 81.57{\scriptsize$\pm$0.26} & 84.13{\scriptsize$\pm$0.32} & 83.33{\scriptsize$\pm$0.23} & 86.22{\scriptsize$\pm$0.13} &  & 91.93{\scriptsize$\pm$0.03} & 95.59{\scriptsize$\pm$0.10} & 96.15{\scriptsize$\pm$0.03} & 96.75{\scriptsize$\pm$0.04} \\ 
\; AlexNet         & 78.10{\scriptsize$\pm$0.50} & 83.92{\scriptsize$\pm$0.40} & 85.90{\scriptsize$\pm$0.26} & 88.09{\scriptsize$\pm$0.15} &  & 91.14{\scriptsize$\pm$0.07} & 94.89{\scriptsize$\pm$0.05} & 97.51{\scriptsize$\pm$0.06} & 97.57{\scriptsize$\pm$0.01} \\
\; ResNet-18       & 74.36{\scriptsize$\pm$0.26} & 78.90{\scriptsize$\pm$0.20} & 80.40{\scriptsize$\pm$0.20} & 83.17{\scriptsize$\pm$0.13} &  & 87.69{\scriptsize$\pm$0.08} & 94.21{\scriptsize$\pm$0.02} & 94.40{\scriptsize$\pm$0.05} & 96.91{\scriptsize$\pm$0.01} \\ 
\; DenseNet-121    & 81.94{\scriptsize$\pm$0.20} & 82.85{\scriptsize$\pm$0.13} & 84.88{\scriptsize$\pm$0.27} & 86.59{\scriptsize$\pm$0.20} &  & 92.68{\scriptsize$\pm$0.14} & 96.01{\scriptsize$\pm$0.06} & 96.53{\scriptsize$\pm$0.05} & 97.34{\scriptsize$\pm$0.05} \\ 
\; EfficientNet-B4 & 82.64{\scriptsize$\pm$0.08} & 84.19{\scriptsize$\pm$0.08} & 84.40{\scriptsize$\pm$0.15} & 87.07{\scriptsize$\pm$0.08} &  & 93.66{\scriptsize$\pm$0.03} & 96.29{\scriptsize$\pm$0.02} & 95.41{\scriptsize$\pm$0.02} & 97.21{\scriptsize$\pm$0.01} \\ 
\; ViT-B/16        & 83.12{\scriptsize$\pm$0.20} & 84.51{\scriptsize$\pm$0.72} & 87.50{\scriptsize$\pm$0.35} & 88.30{\scriptsize$\pm$0.26} &  & 94.28{\scriptsize$\pm$0.02} & 95.93{\scriptsize$\pm$0.11} & 97.33{\scriptsize$\pm$0.03} & 97.63{\scriptsize$\pm$0.07} \\ 
\; CLIP ViT-B/16   & 84.88{\scriptsize$\pm$0.20} & 85.04{\scriptsize$\pm$0.20} & 84.67{\scriptsize$\pm$0.20} & 84.99{\scriptsize$\pm$0.27} &  & 94.48{\scriptsize$\pm$0.19} & 96.54{\scriptsize$\pm$0.05} & 96.64{\scriptsize$\pm$0.11} & 97.17{\scriptsize$\pm$0.07} \\ 
\; EVA-02 ViT-B/16 & 83.60{\scriptsize$\pm$0.08} & 79.22{\scriptsize$\pm$0.15} & 81.52{\scriptsize$\pm$0.08} & 83.65{\scriptsize$\pm$0.13} &  & 94.10{\scriptsize$\pm$0.03} & 94.30{\scriptsize$\pm$0.01} & 95.42{\scriptsize$\pm$0.03} & 96.34{\scriptsize$\pm$0.05} \\ 
\; DINO ViT-B/16   & \underline{\textbf{86.59{\scriptsize$\pm$0.54}}} & \textbf{86.43{\scriptsize$\pm$0.65}} & \textbf{90.33{\scriptsize$\pm$0.15}} & \cellcolor{bg}\underline{\textbf{91.56{\scriptsize$\pm$0.27}}} &  & \underline{\textbf{97.29{\scriptsize$\pm$0.07}}} & \textbf{97.56{\scriptsize$\pm$0.10}} & \underline{\textbf{98.89{\scriptsize$\pm$0.03}}} & \cellcolor{bg}\underline{\textbf{98.92{\scriptsize$\pm$0.08}}} \\ 
\; SAM ViT-B/16    & 62.50{\scriptsize$\pm$0.00} & 62.50{\scriptsize$\pm$0.00} & 62.50{\scriptsize$\pm$0.00} & 62.50{\scriptsize$\pm$0.00} &  & 81.25{\scriptsize$\pm$0.50} & 87.16{\scriptsize$\pm$0.25} & 92.52{\scriptsize$\pm$0.05} & 90.13{\scriptsize$\pm$0.14} \\ 
\midrule
{\textsc{$k$-NN ($k = 11$)}} & & & & & \\       
\; VGG16           & 76.12 & 83.01 & 81.73 & 81.57 &  & - & - & - & - \\ 
\; AlexNet         & 81.89 & 81.25 & 83.65 & 84.62 &  & - & - & - & - \\  
\; ResNet-18       & 81.09 & 86.06 & 83.97 & 87.34 &  & - & - & - & - \\ 
\; DenseNet-121    & 82.05 & 82.37 & 84.29 & 86.06 &  & - & - & - & - \\ 
\; EfficientNet-B4 & 85.58 & 83.81 & 84.78 & 84.62 &  & - & - & - & - \\ 
\; ViT-B/16        & 83.01 & 77.88 & 84.94 & 87.82 &  & - & - & - & - \\ 
\; CLIP ViT-B/16   & 85.26 & 83.01 & 86.38 & 87.50 &  & - & - & - & - \\ 
\; EVA-02 ViT-B/16 & 84.29 & 79.81 & 83.97 & 84.94 &  & - & - & - & - \\ 
\; DINO ViT-B/16   & \textbf{85.74} & \textbf{87.82} & \underline{\textbf{90.54}} & \textbf{89.74} &  & - & - & - & - \\ 
\; SAM ViT-B/16    & 80.93 & 83.33 & 84.13 & 86.06 &  & - & - & - & - \\ 
\bottomrule
\end{tabular}
\end{table}

\begin{table}[!htpb]
\setlength{\tabcolsep}{4.9pt}
\renewcommand{\arraystretch}{1.1}
\caption{Benchmark outcomes summarizing the mean and standard deviation of accuracy (ACC), for a fixed operating point of $0.5$ and area under the receiver operating characteristic curve (AUC) for the RetinaMNIST dataset across all training scheme-model-image resolution combinations, derived from three independent random seeds. Notably, the \mbox{$k$-NN} algorithm, devoid of a training phase, remains unaffected by the stochasticity inherent in model training, thus reporting only the total ACC value without standard deviation. Moreover, owing to its direct utilization of embeddings and labels for classification, \mbox{$k$-NN} does not furnish a reliable AUC score. The overall best result across all training schemes, models, and resolutions is highlighted with a \colorbox{bg}{background color}; the best result per resolution across all training schemes and models is highlighted with \underline{underline}; and the best result per training scheme and resolution is highlighted in \textbf{bold}.}
\label{tab:retinamnist benchmark}
\centering
\begin{tabular}{lcccccccccc}
& \\
\toprule
\multicolumn{10}{c}{\textbf{RetinaMNIST}} \\
\toprule
\multirow{2.5}{*}{Methods} & \multicolumn{4}{c}{Accuracy (ACC)} &  & \multicolumn{4}{c}{Area Under the ROC Curve (AUC)} \\
\cmidrule(r){2-5}\cmidrule(l){7-10}
& $28 \times 28$ & $64 \times 64$ & $128 \times 128$ & $224 \times 224$ &  & $28 \times 28$ & $64 \times 64$ & $128 \times 128$ & $224 \times 224$\\ 
\midrule
{\textsc{End-to-End}} & & & & & \\        
\; VGG16           & \textbf{54.17{\scriptsize$\pm$0.77}} & \textbf{55.75{\scriptsize$\pm$4.48}} & \underline{\textbf{62.00{\scriptsize$\pm$1.02}}} & \cellcolor{bg}\underline{\textbf{64.17{\scriptsize$\pm$2.42}}} &  & \underline{\textbf{75.35{\scriptsize$\pm$0.98}}} & \underline{\textbf{80.63{\scriptsize$\pm$0.68}}} & \underline{\textbf{85.26{\scriptsize$\pm$0.60}}} & \cellcolor{bg}\underline{\textbf{87.82{\scriptsize$\pm$0.71}}} \\ 
\; AlexNet         & 51.83{\scriptsize$\pm$0.66} & 52.58{\scriptsize$\pm$0.42} & 58.25{\scriptsize$\pm$1.74} & 59.42{\scriptsize$\pm$0.96} &  & 72.80{\scriptsize$\pm$0.64} & 75.08{\scriptsize$\pm$0.38} & 80.02{\scriptsize$\pm$0.95} & 83.20{\scriptsize$\pm$1.02} \\ 
\; ResNet-18       & 52.33{\scriptsize$\pm$2.37} & 53.08{\scriptsize$\pm$1.48} & 59.25{\scriptsize$\pm$0.41} & 61.50{\scriptsize$\pm$1.34} &  & 70.41{\scriptsize$\pm$0.55} & 74.40{\scriptsize$\pm$1.90} & 80.81{\scriptsize$\pm$0.61} & 83.16{\scriptsize$\pm$0.65} \\ 
\; DenseNet-121    & 48.67{\scriptsize$\pm$0.51} & 53.25{\scriptsize$\pm$2.15} & 61.75{\scriptsize$\pm$0.20} & 61.75{\scriptsize$\pm$1.08} &  & 71.17{\scriptsize$\pm$1.16} & 74.30{\scriptsize$\pm$0.74} & 81.45{\scriptsize$\pm$0.51} & 82.90{\scriptsize$\pm$0.57} \\ 
\; EfficientNet-B4 & 47.25{\scriptsize$\pm$2.70} & 50.67{\scriptsize$\pm$0.77} & 53.92{\scriptsize$\pm$1.36} & 52.42{\scriptsize$\pm$1.66} &  & 64.12{\scriptsize$\pm$2.59} & 70.83{\scriptsize$\pm$0.30} & 73.61{\scriptsize$\pm$0.65} & 73.83{\scriptsize$\pm$0.76} \\ 
\; ViT-B/16        & 49.83{\scriptsize$\pm$1.53} & 54.08{\scriptsize$\pm$0.96} & 54.00{\scriptsize$\pm$2.27} & 55.08{\scriptsize$\pm$2.05} &  & 71.72{\scriptsize$\pm$0.89} & 73.58{\scriptsize$\pm$1.65} & 73.71{\scriptsize$\pm$0.81} & 78.58{\scriptsize$\pm$2.32} \\ 
\; CLIP ViT-B/16   & 52.50{\scriptsize$\pm$1.06} & 51.58{\scriptsize$\pm$0.85} & 50.58{\scriptsize$\pm$1.01} & 50.33{\scriptsize$\pm$0.62} &  & 72.80{\scriptsize$\pm$0.93} & 72.69{\scriptsize$\pm$1.72} & 71.43{\scriptsize$\pm$0.62} & 70.84{\scriptsize$\pm$1.46} \\ 
\; EVA-02 ViT-B/16 & 51.25{\scriptsize$\pm$1.22} & 51.67{\scriptsize$\pm$0.24} & 47.67{\scriptsize$\pm$3.37} & 54.42{\scriptsize$\pm$1.53} &  & 71.23{\scriptsize$\pm$0.91} & 71.15{\scriptsize$\pm$1.19} & 69.91{\scriptsize$\pm$2.11} & 74.73{\scriptsize$\pm$2.80} \\ 
\; DINO ViT-B/16   & 52.33{\scriptsize$\pm$1.64} & 50.83{\scriptsize$\pm$1.90} & 50.33{\scriptsize$\pm$2.01} & 54.25{\scriptsize$\pm$2.41} &  & 73.32{\scriptsize$\pm$0.28} & 71.91{\scriptsize$\pm$0.70} & 71.96{\scriptsize$\pm$0.60} & 78.32{\scriptsize$\pm$3.37} \\ 
\; SAM ViT-B/16    & 50.00{\scriptsize$\pm$1.95} & 50.67{\scriptsize$\pm$2.71} & 51.17{\scriptsize$\pm$1.05} & 51.33{\scriptsize$\pm$1.45} &  & 71.25{\scriptsize$\pm$1.60} & 71.64{\scriptsize$\pm$1.66} & 71.90{\scriptsize$\pm$0.96} & 71.72{\scriptsize$\pm$1.39} \\ 
\midrule
{\textsc{Linear Probing}} & & & & & \\         
\; VGG16           & 50.58{\scriptsize$\pm$0.24} & 53.42{\scriptsize$\pm$1.01} & 57.33{\scriptsize$\pm$0.24} & 61.08{\scriptsize$\pm$0.42} &  & 71.84{\scriptsize$\pm$0.31} & 75.50{\scriptsize$\pm$0.17} & 81.06{\scriptsize$\pm$0.18} & 84.86{\scriptsize$\pm$0.27} \\ 
\; AlexNet         & 51.25{\scriptsize$\pm$0.35} & 54.25{\scriptsize$\pm$0.74} & 56.17{\scriptsize$\pm$0.47} & 58.08{\scriptsize$\pm$0.12} &  & 70.94{\scriptsize$\pm$0.17} & 74.07{\scriptsize$\pm$0.32} & 78.31{\scriptsize$\pm$0.1} & 81.33{\scriptsize$\pm$0.20} \\
\; ResNet-18       & 43.50{\scriptsize$\pm$0.00} & 46.58{\scriptsize$\pm$0.24} & 47.50{\scriptsize$\pm$0.20} & 49.50{\scriptsize$\pm$0.20} &  & 68.91{\scriptsize$\pm$0.52} & 71.18{\scriptsize$\pm$0.35} & 75.52{\scriptsize$\pm$0.16} & 79.45{\scriptsize$\pm$0.21} \\ 
\; DenseNet-121    & 52.08{\scriptsize$\pm$0.77} & 54.83{\scriptsize$\pm$0.59} & \textbf{60.17{\scriptsize$\pm$0.31}} & \textbf{62.67{\scriptsize$\pm$0.92}} &  & \textbf{73.41{\scriptsize$\pm$0.14}} & 77.01{\scriptsize$\pm$0.22} & \textbf{83.46{\scriptsize$\pm$0.24}} & \textbf{85.93{\scriptsize$\pm$0.23}} \\ 
\; EfficientNet-B4 & 51.83{\scriptsize$\pm$0.42} & \underline{\textbf{57.42{\scriptsize$\pm$0.42}}} & 58.58{\scriptsize$\pm$0.12} & 58.75{\scriptsize$\pm$0.20} &  & 72.51{\scriptsize$\pm$0.14} & 76.23{\scriptsize$\pm$0.07} & 80.23{\scriptsize$\pm$0.07} & 81.32{\scriptsize$\pm$0.07} \\ 
\; ViT-B/16        & \underline{\textbf{54.25{\scriptsize$\pm$1.08}}} & 55.75{\scriptsize$\pm$1.27} & 59.00{\scriptsize$\pm$0.54} & 61.17{\scriptsize$\pm$0.72} &  & 73.09{\scriptsize$\pm$0.07} & 75.15{\scriptsize$\pm$0.32} & 82.47{\scriptsize$\pm$0.52} & 85.16{\scriptsize$\pm$0.55} \\ 
\; CLIP ViT-B/16   & 54.17{\scriptsize$\pm$1.45} & 56.00{\scriptsize$\pm$1.41} & 59.33{\scriptsize$\pm$0.66} & 61.25{\scriptsize$\pm$0.54} &  & 73.10{\scriptsize$\pm$0.29} & \textbf{77.77{\scriptsize$\pm$0.55}} & 82.29{\scriptsize$\pm$0.33} & 85.35{\scriptsize$\pm$0.09} \\ 
\; EVA-02 ViT-B/16 & 49.58{\scriptsize$\pm$1.03} & 53.00{\scriptsize$\pm$0.89} & 53.83{\scriptsize$\pm$0.12} & 53.33{\scriptsize$\pm$0.31} &  & 72.45{\scriptsize$\pm$0.32} & 76.86{\scriptsize$\pm$0.14} & 79.75{\scriptsize$\pm$0.04} & 81.05{\scriptsize$\pm$0.15} \\ 
\; DINO ViT-B/16   & 52.08{\scriptsize$\pm$0.12} & 55.92{\scriptsize$\pm$1.36} & 58.42{\scriptsize$\pm$0.31} & 62.58{\scriptsize$\pm$1.90} &  & 71.99{\scriptsize$\pm$0.63} & 77.18{\scriptsize$\pm$0.69} & 82.17{\scriptsize$\pm$0.13} & 85.57{\scriptsize$\pm$0.57} \\ 
\; SAM ViT-B/16    & 43.50{\scriptsize$\pm$0.00} & 43.50{\scriptsize$\pm$0.00} & 43.50{\scriptsize$\pm$0.00} & 43.50{\scriptsize$\pm$0.00} &  & 56.00{\scriptsize$\pm$3.62} & 66.13{\scriptsize$\pm$1.83} & 65.07{\scriptsize$\pm$0.27} & 63.22{\scriptsize$\pm$0.33} \\ 
\midrule
{\textsc{$k$-NN ($k = 11$)}} & & & & & \\       
\; VGG16           & 47.75 & 51.25 & 53.25 & 55.75 &  & - & - & - & - \\ 
\; AlexNet         & 46.75 & 48.25 & 52.75 & 54.75 &  & - & - & - & - \\  
\; ResNet-18       & 47.50 & 49.00 & 51.00 & 53.50 &  & - & - & - & - \\ 
\; DenseNet-121    & 49.50 & 48.75 & \textbf{55.00} & 58.00 &  & - & - & - & - \\ 
\; EfficientNet-B4 & 49.75 & 52.00 & 54.75 & 51.00 &  & - & - & - & - \\ 
\; ViT-B/16        & 48.50 & 48.75 & 50.25 & 56.25 &  & - & - & - & - \\ 
\; CLIP ViT-B/16   & \textbf{52.25} & 48.75 & 50.00 & 52.75 &  & - & - & - & - \\ 
\; EVA-02 ViT-B/16 & 51.75 & 49.75 & 50.75 & 54.50 &  & - & - & - & - \\ 
\; DINO ViT-B/16   & 48.00 & \textbf{52.50} & 51.00 & \textbf{59.00} &  & - & - & - & - \\ 
\; SAM ViT-B/16    & 49.25 & 52.00 & 49.75 & 52.25 &  & - & - & - & - \\ 
\bottomrule
\end{tabular}
\end{table}

\begin{table}[!htpb]
\setlength{\tabcolsep}{4.9pt}
\renewcommand{\arraystretch}{1.1}
\caption{Benchmark outcomes summarizing the mean and standard deviation of accuracy (ACC), for a fixed operating point of $0.5$ and area under the receiver operating characteristic curve (AUC) for the TissueMNIST dataset across all training scheme-model-image resolution combinations, derived from three independent random seeds. Notably, the \mbox{$k$-NN} algorithm, devoid of a training phase, remains unaffected by the stochasticity inherent in model training, thus reporting only the total ACC value without standard deviation. Moreover, owing to its direct utilization of embeddings and labels for classification, \mbox{$k$-NN} does not furnish a reliable AUC score. The overall best result across all training schemes, models, and resolutions is highlighted with a \colorbox{bg}{background color}; the best result per resolution across all training schemes and models is highlighted with \underline{underline}; and the best result per training scheme and resolution is highlighted in \textbf{bold}.}
\label{tab:tissuemnist benchmark}
\centering
\begin{tabular}{lcccccccccc}
& \\
\toprule
\multicolumn{10}{c}{\textbf{TissueMNIST}} \\
\toprule
\multirow{2.5}{*}{Methods} & \multicolumn{4}{c}{Accuracy (ACC)} &  & \multicolumn{4}{c}{Area Under the ROC Curve (AUC)} \\
\cmidrule(r){2-5}\cmidrule(l){7-10}
& $28 \times 28$ & $64 \times 64$ & $128 \times 128$ & $224 \times 224$ &  & $28 \times 28$ & $64 \times 64$ & $128 \times 128$ & $224 \times 224$\\ 
\midrule
{\textsc{End-to-End}} & & & & & \\        
\; VGG16           &  \underline{\textbf{67.75{\scriptsize$\pm$0.46}}} & 71.29{\scriptsize$\pm$0.81} & 71.62{\scriptsize$\pm$0.16} & 70.57{\scriptsize$\pm$0.35} &  & \underline{\textbf{92.67{\scriptsize$\pm$0.12}}} & 94.12{\scriptsize$\pm$0.23} & 94.37{\scriptsize$\pm$0.10} & 94.05{\scriptsize$\pm$0.06} \\ 
\; AlexNet         & 59.80{\scriptsize$\pm$0.43} & 64.12{\scriptsize$\pm$0.25} & 67.06{\scriptsize$\pm$0.06} & 69.25{\scriptsize$\pm$0.33} &  & 88.79{\scriptsize$\pm$0.11} & 91.16{\scriptsize$\pm$0.04} & 92.56{\scriptsize$\pm$0.07} & 93.50{\scriptsize$\pm$0.08} \\ 
\; ResNet-18       & 63.02{\scriptsize$\pm$0.10} & 67.36{\scriptsize$\pm$0.20} & 70.17{\scriptsize$\pm$0.73} & 69.35{\scriptsize$\pm$0.67} &  & 90.59{\scriptsize$\pm$0.08} & 92.65{\scriptsize$\pm$0.09} & 93.65{\scriptsize$\pm$0.21} & 93.57{\scriptsize$\pm$0.13} \\ 
\; DenseNet-121    & 66.53{\scriptsize$\pm$0.37} & \underline{\textbf{71.54{\scriptsize$\pm$0.71}}} & \cellcolor{bg}\underline{\textbf{74.25{\scriptsize$\pm$0.39}}} & \underline{\textbf{74.08{\scriptsize$\pm$0.32}}} &  & 92.47{\scriptsize$\pm$0.04} & \underline{\textbf{94.46{\scriptsize$\pm$0.24}}} & \cellcolor{bg}\underline{\textbf{95.35{\scriptsize$\pm$0.07}}} & \underline{\textbf{95.25{\scriptsize$\pm$0.04}}} \\ 
\; EfficientNet-B4 & 59.92{\scriptsize$\pm$0.55} & 65.16{\scriptsize$\pm$1.63} & 71.35{\scriptsize$\pm$0.24} & 69.31{\scriptsize$\pm$1.50} &  & 88.97{\scriptsize$\pm$0.50} & 91.59{\scriptsize$\pm$0.71} & 94.15{\scriptsize$\pm$0.06} & 93.35{\scriptsize$\pm$0.53} \\ 
\; ViT-B/16        & 60.55{\scriptsize$\pm$0.56} & 66.72{\scriptsize$\pm$0.71} & 71.29{\scriptsize$\pm$0.40} & 72.89{\scriptsize$\pm$0.70} &  & 88.97{\scriptsize$\pm$0.18} & 92.62{\scriptsize$\pm$0.31} & 94.32{\scriptsize$\pm$0.17} & 94.84{\scriptsize$\pm$0.16} \\ 
\; CLIP ViT-B/16   & 56.63{\scriptsize$\pm$0.53} & 62.97{\scriptsize$\pm$0.67} & 66.47{\scriptsize$\pm$0.29} & 66.25{\scriptsize$\pm$0.28} &  & 86.44{\scriptsize$\pm$0.27} & 90.55{\scriptsize$\pm$0.35} & 92.32{\scriptsize$\pm$0.25} & 92.20{\scriptsize$\pm$0.11} \\ 
\; EVA-02 ViT-B/16 & 57.60{\scriptsize$\pm$1.00} & 64.42{\scriptsize$\pm$0.67} & 70.42{\scriptsize$\pm$0.87} & 70.34{\scriptsize$\pm$0.93} &  & 87.31{\scriptsize$\pm$0.58} & 91.21{\scriptsize$\pm$0.29} & 93.92{\scriptsize$\pm$0.33} & 93.97{\scriptsize$\pm$0.35} \\ 
\; DINO ViT-B/16   & 59.44{\scriptsize$\pm$0.31} & 65.80{\scriptsize$\pm$1.01} & 69.35{\scriptsize$\pm$1.72} & 70.38{\scriptsize$\pm$0.99} &  & 88.55{\scriptsize$\pm$0.27} & 91.98{\scriptsize$\pm$0.44} & 93.48{\scriptsize$\pm$0.77} & 94.02{\scriptsize$\pm$0.37} \\ 
\; SAM ViT-B/16    & 58.86{\scriptsize$\pm$0.20} & 66.19{\scriptsize$\pm$0.70} & 69.42{\scriptsize$\pm$0.60} & 71.47{\scriptsize$\pm$0.41} &  & 88.07{\scriptsize$\pm$0.32} & 92.25{\scriptsize$\pm$0.33} & 93.70{\scriptsize$\pm$0.23} & 94.40{\scriptsize$\pm$0.13} \\ 
\midrule
{\textsc{Linear Probing}} & & & & & \\         
\; VGG16           & 53.19{\scriptsize$\pm$0.00} & 53.53{\scriptsize$\pm$0.02} & 55.50{\scriptsize$\pm$0.08} & 58.12{\scriptsize$\pm$0.13} &  & 83.95{\scriptsize$\pm$0.00} & 84.03{\scriptsize$\pm$0.00} & 86.36{\scriptsize$\pm$0.03} & 87.89{\scriptsize$\pm$0.04} \\ 
\; AlexNet         & 49.10{\scriptsize$\pm$0.10} & 53.80{\scriptsize$\pm$0.06} & 55.75{\scriptsize$\pm$0.06} & 59.57{\scriptsize$\pm$0.05} &  & 80.35{\scriptsize$\pm$0.13} & 84.74{\scriptsize$\pm$0.06} & 86.47{\scriptsize$\pm$0.03} & 88.98{\scriptsize$\pm$0.01} \\
\; ResNet-18       & 51.06{\scriptsize$\pm$0.00} & 53.43{\scriptsize$\pm$0.02} & 54.60{\scriptsize$\pm$0.02} & 56.46{\scriptsize$\pm$0.01} &  & 82.44{\scriptsize$\pm$0.00} & 84.76{\scriptsize$\pm$0.00} & 85.65{\scriptsize$\pm$0.00} & 86.97{\scriptsize$\pm$0.00} \\ 
\; DenseNet-121    & 55.93{\scriptsize$\pm$0.01} & 59.46{\scriptsize$\pm$0.02} & 60.86{\scriptsize$\pm$0.02} & 61.09{\scriptsize$\pm$0.02} &  & 86.45{\scriptsize$\pm$0.00} & 88.64{\scriptsize$\pm$0.00} & 89.38{\scriptsize$\pm$0.01} & 89.64{\scriptsize$\pm$0.01} \\ 
\; EfficientNet-B4 & 54.24{\scriptsize$\pm$0.01} & 56.91{\scriptsize$\pm$0.00} & 57.88{\scriptsize$\pm$0.02} & 58.47{\scriptsize$\pm$0.01} &  & 85.29{\scriptsize$\pm$0.00} & 87.17{\scriptsize$\pm$0.00} & 87.63{\scriptsize$\pm$0.00} & 88.33{\scriptsize$\pm$0.00} \\ 
\; ViT-B/16        & 53.91{\scriptsize$\pm$0.06} & 60.69{\scriptsize$\pm$0.09} & 62.79{\scriptsize$\pm$0.02} & 63.80{\scriptsize$\pm$0.08} &  & 84.95{\scriptsize$\pm$0.03} & 89.27{\scriptsize$\pm$0.03} & 90.39{\scriptsize$\pm$0.01} & 90.88{\scriptsize$\pm$0.06} \\ 
\; CLIP ViT-B/16   & 55.28{\scriptsize$\pm$0.03} & 59.39{\scriptsize$\pm$0.04} & 61.03{\scriptsize$\pm$0.08} & 61.50{\scriptsize$\pm$0.06} &  & 86.02{\scriptsize$\pm$0.01} & 88.60{\scriptsize$\pm$0.02} & 89.62{\scriptsize$\pm$0.01} & 89.76{\scriptsize$\pm$0.08} \\ 
\; EVA-02 ViT-B/16 & 54.23{\scriptsize$\pm$0.01} & 58.35{\scriptsize$\pm$0.02} & 59.71{\scriptsize$\pm$0.00} & 60.51{\scriptsize$\pm$0.02} &  & 85.14{\scriptsize$\pm$0.00} & 87.93{\scriptsize$\pm$0.00} & 88.68{\scriptsize$\pm$0.00} & 89.24{\scriptsize$\pm$0.00} \\ 
\; DINO ViT-B/16   & \textbf{57.46{\scriptsize$\pm$0.22}} & \textbf{63.09{\scriptsize$\pm$0.02}} & \textbf{63.92{\scriptsize$\pm$0.05}} & \textbf{64.04{\scriptsize$\pm$0.12}} &  & \textbf{87.22{\scriptsize$\pm$0.10}} & \textbf{90.51{\scriptsize$\pm$0.02}} & \textbf{91.02{\scriptsize$\pm$0.01}} & \textbf{90.98{\scriptsize$\pm$0.03}} \\ 
\; SAM ViT-B/16    & 37.70{\scriptsize$\pm$0.00} & 41.70{\scriptsize$\pm$0.00} & 46.05{\scriptsize$\pm$0.01} & 46.66{\scriptsize$\pm$0.01} &  & 68.59{\scriptsize$\pm$0.01} & 73.73{\scriptsize$\pm$0.01} & 76.61{\scriptsize$\pm$0.00} & 76.89{\scriptsize$\pm$0.00} \\ 
\midrule
{\textsc{$k$-NN ($k = 11$)}} & & & & & \\       
\; VGG16           & 47.30 & 47.20 & 48.55 & 51.30 &  & - & - & - & - \\ 
\; AlexNet         & 45.97 & 49.96 & 50.79 & 54.11 &  & - & - & - & - \\  
\; ResNet-18       & 48.25 & 48.89 & 49.05 & 51.54 &  & - & - & - & - \\ 
\; DenseNet-121    & 48.66 & 50.10 & 51.04 & 52.67 &  & - & - & - & - \\ 
\; EfficientNet-B4 & 49.79 & 51.14 & 50.35 & 51.26 &  & - & - & - & - \\ 
\; ViT-B/16        & 47.47 & 50.73 & 52.84 & 54.33 &  & - & - & - & - \\ 
\; CLIP ViT-B/16   & 48.04 & 50.48 & 53.29 & 53.46 &  & - & - & - & - \\ 
\; EVA-02 ViT-B/16 & 49.74 & 50.92 & 52.52 & 53.86 &  & - & - & - & - \\ 
\; DINO ViT-B/16   & \textbf{51.56} & \textbf{56.10} & \textbf{57.39} & \textbf{57.12} &  & - & - & - & - \\ 
\; SAM ViT-B/16    & 48.00 & 49.67 & 49.64 & 46.97 &  & - & - & - & - \\ 
\bottomrule
\end{tabular}
\end{table}

\end{document}